\documentclass[twocolumn,twocolappendix]{aastex701}

\usepackage{amsmath}
\usepackage{afterpage}
\usepackage{makecell}

\newcommand{\cloudy}{\texttt{CLOUDY}}

\newcommand{\changa}{{\sc ChaNGa}}
\newcommand{\MM}{{\sc M+M}}
\newcommand{\gizmo}{{\sc GIZMO}}

\definecolor{pink}{HTML}{de02ab}

\newcommand{\OVI}{O{\sc ~vi}}
\newcommand{\HI}{H{\sc ~i}}
\newcommand{\HII}{H{\sc ~ii}}

\newcommand{\HeI}{He{\sc ~i}}
\newcommand{\HeII}{He{\sc ~ii}}
\newcommand{\HeIII}{He{\sc ~iii}}

\newcommand{\CII}{C{\sc ~ii}}

\newcommand{\CIV}{C{\sc ~iv}}

\newcommand{\SiII}{Si{\sc ~ii}}
\newcommand{\SiIII}{Si{\sc ~iii}}
\newcommand{\SiIV}{Si{\sc ~iv}}

\newcommand{\NOVI}{$N_{\text{OVI}}$}

\newcommand{\OI}{O{\sc ~i}}\newcommand{\OII}{O{\sc ~ii}}
\newcommand{\OIII}{O{\sc ~iii}}
\newcommand{\OIV}{O{\sc ~iv}}
\newcommand{\OV}{O{\sc ~v}} 
\newcommand{\OVII}{O{\sc ~vii}}
\newcommand{\OVIII}{O{\sc ~viii}}
\newcommand{\OIX}{O{\sc ~IX}}

\newcommand{\Mstar}{$M_{*}$}
\newcommand{\Mvir}{$M_{vir}$}
\newcommand{\MOVI}{$M_{\rm OVI}$}
\newcommand{\MishraMOVI}{$M^{M24}_{\rm OVI}$}
\newcommand{\MOform}{$M_{\rm O,form}$}
\defcitealias{Mishra_24}{M24}
\defcitealias{HM2012}{HM12}
\defcitealias{FaucherGiguère2009}{FG09}

\begin{document}

\title{The Simulated Oxygen Shortage (SOS): Mapping the Missing \OVI\, in Simulated Dwarf Galaxies to Subgrid Physics}




\author[orcid=0000-0003-0200-6986]{Daniel R. Piacitelli}
\affiliation{Rutgers University, Department of Physics and Astronomy, Piscataway, NJ 08854, USA}\email[show]{piacitelli.danielr@gmail.com}

\author{Alyson M. Brooks}\affiliation{Rutgers University, Department of Physics and Astronomy, Piscataway, NJ 08854, USA}\affiliation{Center for Computational Astrophysics, Flatiron Institute, 162 Fifth Ave, New York, NY 10010, USA}\email{abrooks@physics.rutgers.edu}

\author[0000-0001-7589-6188]{N. Nicole Sanchez}\affiliation{The Observatories of the Carnegie Institution for Science, 813 Santa Barbara Street, Pasadena, CA 91101, USA}\affiliation{Cahill Center for Astronomy and Astrophysics, California Institute of Technology, MC249-17, Pasadena, CA 91125, USA}\email{nnicolesanchez@gmail.com}

\author[0009-0006-1680-8540]{Hetvi Khatri}\affiliation{Rutgers University, Department of Physics and Astronomy, Piscataway, NJ 08854, USA}\email{}

\author[0000-0001-6779-3429]{Charlotte Christensen}\affiliation{Physics Department, Grinnell College, 1116 Eighth Avenue, Grinnell, IA 50112, USA }\email{christenc@grinnell.edu}

\author[0000-0002-3817-8133]{Cameron Hummels}\affiliation{California Institute of Technology, Pasadena, CA 91125, USA}\email{}

\author[0000-0002-9141-9792]{Nishant Mishra}\affiliation{Department of Astronomy, University of Michigan, 1085 S. University, Ann Arbor, MI 48109, USA}\affiliation{The Observatories of the Carnegie Institution for Science, 813 Santa Barbara Street, Pasadena, CA 91101, USA}\email{}


\author[0000-0001-7831-4892]{Akaxia Cruz}
\affiliation{Center for Computational Astrophysics, Flatiron Institute, 162 Fifth Ave, New York, NY 10010, USA}\affiliation{Department of Physics, Princeton University, Princeton, NJ 08544, USA}\affiliation{Department of Astrophysical Sciences, Princeton University, Princeton, NJ 08544, USA}\email{acruz@flatironinstitute.org}


\author[0000-0002-9642-7193]{Ben Keller}
\affiliation{Department of Physics and Materials Science, University of Memphis, 3720 Alumni Avenue, Memphis, TN 38152, USA}\email{bkeller1@memphis.edu}

\author[0000-0001-5510-2803]{Thomas R. Quinn}
\affiliation{Department of Astronomy, University of Washington, Seattle, WA 98195, USA}\email{trq@astro.washington.edu}

\author{Sijing Shen}\affiliation{Institute of Theoretical Astrophysics, University of Oslo, PO Box 1029, Blindern 0315, Oslo, Norway}\email{sijing.shen@astro.uio.no}

\author{James Wadsley}
\affiliation{Department of Physics \& Astronomy, McMaster University, ABB-241, 1280 Main Street West, Hamilton, Ontario, L8S 4M1, Canada}\email{wadsley@mcmaster.ca}

\begin{abstract}
 Observations reveal extended \OVI\, reservoirs in the circumgalactic medium (CGM) of dwarf galaxies, yet current simulations systematically underpredict \OVI\, column densities. Utilizing two suites run with different simulation codes, the \MM\, simulations (Marvelous Massive Dwarfs and Marvel-ous Dwarfs) and the publicly available FIRE-2 simulations, we explore the role of subgrid models and the resulting CGM phase in shaping \OVI\, production. By comparing observationally derived \OVI\, masses to the mass of oxygen produced over the galaxies' star formation history, we find evidence for an underproduction of oxygen for low-mass simulated galaxies. Despite clear differences in feedback implementation, CGM structure, and metal mixing, we find that \OVI\, in both suites generally self-selects cool/warm ($\rm log\, T\,/K \sim 4.5$), diffuse ($\rm log\,n_{gas}\,/cm^{-3} \sim -5.0 $), and moderately metal-enriched ($\rm log\, Z/Z_{\odot}  \sim -1 $) material at large radii from the galaxy. We show that neither the choice of ultraviolet background nor plausible variations in CGM thermal structure can close the gap with observations. Taken together, our results point to a possible underproduction of oxygen in low-mass galaxies.  Feedback prescriptions contribute via insufficient metal transport to large radii. Hence, the \OVI\, deficit may motivate an investigation of current modeling choices for supernova yields, star formation, and feedback in low-metallicity environments.

\end{abstract}


\section{Introduction}\label{sec:intro}

Feedback processes are fundamental mediators of galaxy evolution across the galaxy mass regime. For dwarf galaxies, stellar feedback from supernovae (SNe) is expected to regulate star formation \citep[e.g.][]{Dekel_Silk86,Somerville_Dave15,Keller_16} and drive metal and energy transport from within the galaxy into the surrounding environment \citep[e.g.,][]{McQuinn_15,TumCGMreview, FaucherGigure_Oh23}. The expelled material is subsequently distributed throughout the diffuse, multiphase circumgalactic medium (CGM) or beyond the halo entirely. Therefore, accurately mapping and simulating the CGM of dwarf galaxies is critical for understanding the impact of feedback in the low-mass regime and placing strong constraints on galaxy evolution models.

Over the last decade, there has been a growing number of observational works aiming to detect the CGM of dwarf galaxies in UV/optical absorption. Although low metal ion detections (\CII, \CIV, \SiII, \SiIII, and \SiIV) have been constrained to only the inner CGM \citep[$\sim 0.5 R_{vir}$,][]{Bordoloi_14, LiangChen_14, Burchett_16, Johnson_17, QuBregman_22, Zheng_24,Mishra_24}, the higher ion \OVI\, has been commonly detected at projected distances up to twice the virial radius of the halo \citep{Johnson_17, Tchernyshyov_22, Qu_24, Mishra_24, Dutta_25_col}. Given the dearth of low metal ion detections and the abundance and extent of \OVI\, detections, \OVI\, is likely the best ion to fully map the CGM of dwarf galaxies. 

Unlike for higher mass galaxies, the low virial masses of dwarf galaxies are not expected to significantly retain a hot, $T\sim10^{5.5}\rm\, K$, gas phase where collisional ionization most effectively ionizes oxygen into \OVI\, (i.e., the temperature where ionization equilibrium fractions peak). Observational \citep[e.g.,][]{Tchernyshyov_22, Dutta_25_col, Johnson26} and theoretical \citep[e.g.,][]{Gutcke2017, RocaFabrega_19, Cook_24} works have demonstrated that, for dwarf galaxies, a significant portion of \OVI\, likely comes from a cooler, diffuse phase where photoionization from the ultraviolet background (UVB) is the dominant ionization mechanism. Line of sight kinematic analyses of \OVI\, detections have demonstrated that this phase is typically kinematically bound to the halo \citep[e.g.,][]{Mishra_24, Dutta_25_kine, Johnson26}. Under the simplifying assumption that hot gas would likely be strongly outflowing for these halo masses, the kinematically bound \OVI\, detections further disfavor a hot phase origin for \OVI. Yet, since the thermal broadening of \OVI\, is comparable to that of observed \OVI\, absorption line widths, a hot phase origin for \OVI\, cannot be fully ruled out from these observations alone.

On the theoretical side, simulations tend to agree that low-mass halos do not contain a significant hot CGM phase \citep{RocaFabrega_19, Li_21, Cook_24, Tung_25, Piacitelli_25} and find that the CGM is a significant mass and metal reservoir for dwarf galaxies \citep{Christensen_18, Hafen_19, Piacitelli_25}. However, up to now, no simulation has reproduced the observed \OVI\, column densities around dwarf galaxies, often underpredicting columns by $>1\rm\, dex$ \citep{Suresh_17,Li_21,Cook_24, Piacitelli_25}. This underprediction of \OVI\, column densities has also been frequently found in $L_*$ galaxies as well \citep[e.g.,][]{Hummels_13,Cook_24,Lucchini26}

It is not readily clear what is driving this under-prediction of \OVI\, in simulations. Previous work has demonstrated that feedback strength and efficiency influence the production and distribution of \OVI\, \citep[e.g.,][]{ Oppenheimer_16, Rahmati_16, Liang_16, Mina_21, 2025arXiv250819396B}. \citet{RocaFabrega_19} utilizes two simulations, VELA and NIHAO, which implement different codes and models for stellar feedback to compare CGM and \OVI\, properties from dwarf to Milky Way-mass (MW) galaxies. They find that, while the general behavior in mass and redshift is consistent across suites, the radial distribution and maximal fraction of collisionally ionized \OVI\, depend strongly on feedback recipes. More recently, \citet{Rey_25} and \citet{2026arXiv260213394R} use three cosmological zoom-in simulations of the same galaxy and varying feedback models to find distinct CGM differences and column densities across models. These results motivate a systematic comparison of simulations with distinct feedback models to better understand how subgrid physics impacts \OVI\, production. 

Theoretical work has shown that black hole feedback can also strongly shape the \OVI\, content in the CGM of high-mass galaxies and enhance \OVI\, column densities by pushing more metal-enriched material out to large radii \cite[e.g.][]{Nelson2018, Sanchez2019, Sanchez_24}. A similar argument may hold for dwarf galaxies in that stronger SNe or BH feedback may lead to higher \OVI\, columns. 
The thermodynamic state of the CGM is also likely important: in the Marvel-ous Dwarfs and Marvelous Massive Dwarfs simulations (\MM), \citet{Piacitelli_25} finds that at $z=0$ a substantial portion of oxygen mass resides in other oxygen ionization states. Taken together, these findings suggest that the simulated underproduction of \OVI\, may arise from either insufficient metal enrichment of the outer CGM or an incorrect distribution of gas across temperature phases. In this work, we aim to identify the dominant driver of this discrepancy in current simulations, a necessary step toward reconciling simulations with observations.

This paper is organized as follows: Section \ref{sec:simulations} provides details on the simulation suites used in this work, FIRE-2 and \MM, highlighting key similarities and differences that may impact our results.  Section \ref{sec:OVIobservations} compares both suites with current \OVI\, column densities, and derived \OVI\, masses in the CGM, demonstrating that both sets are discrepant with observational results. Section \ref{sec:CGMproperties} compares the two suites in terms of CGM mass/metal content and distribution, as well as CGM physical conditions (temperature, density, metallicity) across radii. In Section \ref{sec:OVIproperties}, we compare the \OVI\, mass distribution and its associated physical conditions. Finally, Section \ref{sec:discussion} quantifies the gap between simulated and observed CGM \OVI\, reservoirs, evaluates the impact of the ultraviolet background and CGM thermodynamic structure, and explores potential pathways to reconcile simulations with observations. We summarize our conclusions in Section \ref{sec:summary}.

\section{Simulations}\label{sec:simulations}
As explained in Section \ref{sec:intro}, previous works have found a relationship between CGM properties, \OVI, and feedback recipes in simulations. To explore this connection, we use two suites of cosmological zoom-in simulations run with two different hydrodynamic codes: \changa\, \citep{CHANGA} and GIZMO \citep{Hopkins2015_GIZMO}. In addition to differing subgrid physics, these suites provide a large number of low-mass galaxies at modern resolutions, which enables our work to better identify systematic differences in the CGM of dwarf galaxies by combining a larger sample of well-resolved galaxies. In this section, we summarize the modeling for both simulation suites, organized by the subgrid models (Section \ref{subsec:physics}), and present the selection criteria and final sample of galaxies analyzed in this work (Section \ref{subsec:finalsample}).

\subsection{Subgrid Models}\label{subsec:physics}
\subsubsection{Hydrodynamics}
The first suite is the \MM\, simulations which are run with \changa, an N-Body, smoothed particle hydrodynamics code \citep{CHANGA}. The \MM\, sample is comprised of two suites: the Marvel-ous Dwarfs (Marvel) and the Marvelous Massive Dwarfs (MMD). These simulations and the subgrid physics implemented have been shown to produce galaxies with dark matter cores and reproduce observed scaling relations for stellar mass with halo mass, metallicity, luminosity, gas content, size, and specific star formation rates \citep{Munshi21,Azartash-Namin24,Riggs_24, Ruan2025, Cruz_25, Piacitelli_25, 2025arXiv251026875G} and have been used to measure the shapes of stellar and dark matter in dwarfs \citep{Keith2025} and explore the effects of self-interacting dark matter \citep{2026arXiv260123264E}. The \MM\, suite ranges in force resolution ($60 /87 \rm\, pc$), gas particle mass resolution ($1410/3300 \rm\, M_{\odot}$), initial star particle mass resolution ($420/994 \rm\, M_{\odot}$), and dark matter particle mass resolution ($6650/17900 \rm\, M_{\odot}$); where the higher resolution values correspond to Marvel.

The second suite is the publicly released FIRE-2 simulations \citep{FIRE_publicdata}. FIRE-2 is run using the multi-method gravity and hydrodynamics code \gizmo\, \citep{Hopkins2015_GIZMO}, which models hydrodynamics using the mesh-free finite-mass Godunov method. Further, the volumes range in baryonic mass resolution ($880-7100\,\rm M_{\odot}$) and dark matter mass resolution ($4400-39000\,\rm M_{\odot}$). Importantly, initial gas and star particle masses are equivalent in FIRE-2.

Neither simulation suite includes magnetohydrodynamics or cosmic ray (CR) energy transport. However, \citet{Su_17} finds that magnetic fields do not strongly affect dwarf galaxy evolution, which implies that the absence of magnetohydrodynamics in this work is unlikely to have a major impact on our conclusions. In Section \ref{subsec:quantgap}, we discuss the potential effects that CR physics may have on setting the \OVI\, production in dwarfs.

\subsubsection{Cooling}
Metal line cooling in \MM\, is modeled across temperatures $10-10^{9}\rm\,K$ using cooling and heating \cloudy\, \citep{Cloudy_Ferland17} tables calculated before runtime and assuming the \citet{HM2012} redshift-dependent UV photoionizing background \citep{Shen2010}. In addition to metal cooling, the total radiative cooling in \MM\, also incorporates cooling/heating from Compton and free-free processes and cooling from primordial species (\HI, H$_2$, \HII, \HeI, \HeII, \HeIII) as well as prescriptions for H$_2$ self-shielding and dust shielding of \HI. 

Gas heating and cooling in FIRE-2 is modeled across temperatures ranging from $10 - 10^{10}\rm \, K$  using \cloudy\, tables that assume the \citet{FaucherGiguère2009} ultraviolet background (UVB; see Section \ref{subsec:disc_vary} and \citet{Taira_25} for more detailed comparisons of the two UVB models) as well as gas self-shielding \citep{Hopkins2014}. The upper temperature bound of $T= 10^{10}\rm \, K$ for gas cooling is a factor of 10 higher than \MM\, however, since $<0.001\%$ of gas mass in either simulation exists above $\rm log\, T\,/K>8$, this is unlikely to strongly affect our conclusions. Cooling mechanisms include metallicity-dependent fine-structure, molecular, and metal-line cooling. 

\subsubsection{Star formation}
Star formation in \MM\, is only permitted in gas particles that are sufficiently cold ($T<1000\,\rm{K}$) and dense ($n >0.1\, \rm{m}_{\rm{H}}\, \rm{cm}^{-3}$) \citep{Christensen2012}. For particles that satisfy these criteria, star formation then occurs probabilistically, depending on the local free-fall time and the fraction of molecular hydrogen present (see \citet{Christensen2012} for more details). Additionally, a fixed star formation efficiency parameter, $c_{0}^{*}=0.1$, is adopted.  We note that despite the density threshold for star formation being relatively low/diffuse, star formation usually occurs only in gas particles of $n > 100\, \rm{m}_{\rm{H}}\, \rm{cm}^{-3}$ due to the dependence on molecular hydrogen. Once star formation has been triggered, a star particle is created with $30\%$ of the parent gas particle's mass and with an initial stellar mass distribution according to the \citet{Kroupa2001_IMF} IMF for stars of individual masses across 0.1 - 100 $M_{\odot}$ (identical to FIRE-2). 

Star formation in FIRE-2 occurs in gas that is dense ($n > 1000\,\rm{cm}^{-3}$), molecular, self-gravitating \citep[according to][]{Hopkins_13}, Jeans unstable, and self-shielding \citep[according to][]{KrumholzGnedin_11}. Together, these conditions result in star formation occurring at a rate of $\dot{\rho}_* = \rho_{mol}/t_{ff}$, where $\rho_{mol}$ and $t_{ff}$ are the density of molecular gas and the local gas free-fall time, respectively \citep{Hopkins2014}. Once star formation has been triggered, a star particle is created with $100\%$ of the parent gas mass and with an initial stellar mass distribution according to the \citet{Kroupa2001_IMF} IMF for stars of individual masses across 0.1 - 100 $M_{\odot}$ (identical to \MM).  

\subsubsection{Feedback}
Feedback mechanisms in \MM\, include stellar and black hole (BH) processes. BH formation and feedback follow prescriptions presented in \citet{Tremmel2015,Tremmel2017}. Mass and metal return from SNe and stellar winds, as well as thermal energy from SNe, are deposited into the nearest neighboring gas particle. Stellar feedback from SNe II follows the ``superbubble'' feedback model \citep{Keller14}. SNe II feedback energy ($E_{SN} = 1.0\times10^{51}\,\rm erg$) diffuses into the surrounding gas particles according to thermal conduction and evaporation subgrid models \citep{CowieMcKee_77, Keller14}.  To prevent numerical overcooling, feedback-heated particles are treated as multiphase fluid elements or ``two-phase particles,'' if the mass of the particles is not fully heated. These two phases maintain pressure equilibrium with one another and have distinct masses, densities, and temperatures. Mass exchange between phases is calculated via the subgrid thermal evaporation model. Once the cold-phase mass has been fully evaporated or if the hot phase cools below $10^5~\rm{K}$, the two-phase particle returns to the single-phase state. For analysis in this work, we split the two phases into their individual components of temperature, density, and mass. An additional channel of stellar feedback in \MM\, is Lyman-Werner radiation from young stars, which can heat and photodissociate H$_2$ and suppress star formation \citep{Christensen2012}.

Feedback mechanisms in FIRE-2 include only stellar processes but simulate a variety of different stellar feedback channels. 1) Feedback from SNe (types Ia and II), which occur stochastically, deposit the ejecta energy, momentum, mass, and metals into the surrounding gas directed radially from the star. Rates for SNe Ia are based on \citet{Mannucci_06} and rates for SNe II are based on STARBURST99 \citep{Starburst99} and SNe II deposit $E = 1.0\times10^{51}\,\rm erg$. At each timestep, when a SN occurs, gas neighbors are selected if the gas is within the kernel radius of the star particle or the star resides within the kernel radius of the gas. SN ejecta is then distributed to each selected neighbor such that the ejecta is distributed isotropically, conserves mass, momentum, and energy, and is weighted based on the ``effective face'' of the gas seen by the star (see \citet{Hopkins_18SNe} for more details). 2) Stellar winds are modeled like SNe but as continuous sources of feedback rather than discrete and follow rates from STARBURST99 \citep{Starburst99}. The contribution of feedback from winds is dominated by O, B, and asymptotic giant branch stars. 3) Local and long-range momentum flux from UV and optical radiation acts as radiation pressure, and 4) photoionization and photoheating, which treat individual star particles as sources producing UV flux and account for absorption from intervening material \citep{Hopkins2014}. Energy input from stellar winds and UV/optical radiation can disrupt/destroy the stellar birth cloud, thereby making the subsequent SNe feedback more effective at distributing material. We emphasize that this is not explicitly modeled in \MM\, and is likely an important modeling difference across suites for this work.

\subsubsection{Metal Production}
The \MM\, suite tracks the evolution of hydrogen, helium, oxygen, iron, and total metallicity.  SN Ia events eject $1.4 M_{\odot}$ of mass, including $0.63 M_{\odot}$ of iron, and $0.13 M_{\odot}$ of oxygen \citep[following yields presented in ][]{Thielemann86}, which is deposited into the surrounding gas particles. Type II SNe, or core-collapse SNe (ccSNe), are modeled for stars within the stellar mass range of $M=8-40 M_{\odot}$ and the ejected oxygen and iron mass from a given ccSNe event scales as a power-law with the mass of the progenitor star according to \citet{WoosleyWeaver_95, Raiteri96}. Mass and metal return via winds is modeled following \citet{Weidemann_87} and is included for stars within the mass range of $1-8 M_{\odot}$. Mass and metals are then imparted to gas particles in the vicinity of the star particle.

FIRE-2 independently tracks the evolution of hydrogen, helium, carbon, nitrogen, oxygen, neon, magnesium, silicon, sulfur, calcium, and iron, and total metallicity. SN Ia events eject metals based on yields from \citet{Iwamoto_99}, which results in an ejection of $\sim0.73 M_{\odot}$ of iron, and $\sim0.14 M_{\odot}$ of oxygen. Both SNe Ia yields are slightly higher than \MM\, and also include a dependence on the progenitor metallicity. SNe II events are modeled for $M>8 M_{\odot}$ stars and eject metals according to the yields from \citet{Nomoto_06}. Winds are modeled across star masses but are dominated by O, B, and asymptotic giant branch stars and eject metals based on a compilation of yields from \citet{Hoek_97}, \citet{Marigo_01}, and \citet{Izzard_04}.

\subsubsection{Metal Diffusion}
Subgrid metal diffusion across both suites follows the \citet{Smagorinsky63} model for the atmospheric boundary layer, which depends on the relative shearing motions of particles and gas metallicities. The scheme for metal diffusion in \MM\, is presented in \citet{Shen2010} and adopts a coefficient value of $C=0.03$, and the models for FIRE-2 are presented in  \citet{Hopkins2017, Su_17,Escala2018} and adopt a coefficient value of $C=0.003$. Further, to mitigate numerical issues with cooling, FIRE-2 gas is initialized at a metallicity floor of $Z_i/Z_{\odot} = 10^{-4}$; no metallicity floor is initialized for \MM.

\subsection{Galaxy Selection \& Definitions}\label{subsec:finalsample}
\begin{figure*}
    \centering
    \includegraphics[width=0.85\textwidth]{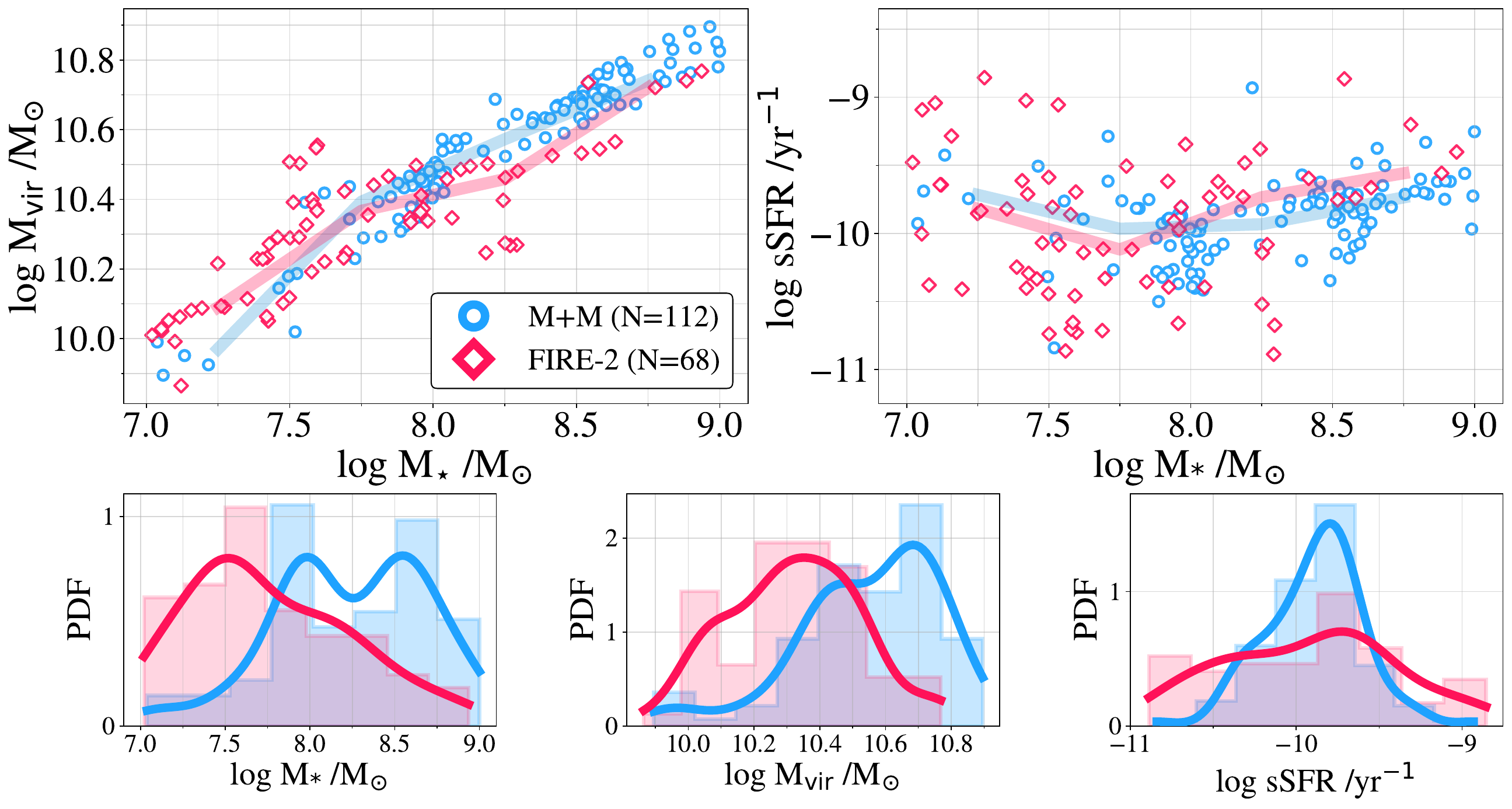}
    \caption{Sample of simulated galaxies selected using criteria detailed in Section \ref{subsec:finalsample}. Galaxies selected from FIRE-2 are shown as red diamonds, while \MM\, galaxies are shown as blue circles. \textit{Upper Left}: Halo Mass (\Mvir) versus Stellar Mass (\Mstar) relation for both suites. \textit{Upper Right}: Specific star formation rates (sSFR) across $M_*$. Star formation is measured over a timespan of 100 Myr for both suites. Transparent red and blue lines correspond to the median of either suite calculated in 4 equally sized bins and plotted at the center of each bin. \textit{Bottom Row}: Probability density functions (PDF) of \Mstar, \Mvir, and sSFR for both suites following the same color scheme as above. Transparent PDFs represent simulation data binned in 9 equally sized bins, while solid lines correspond to Gaussian kernel-density estimated PDFs. Although we see slight differences across suites, such as biases to lower or higher masses, we find adequate agreement between \MM\, and FIRE-2, motivating the comparison of CGM properties between the two suites.  }
    \label{fig:Sample}
\end{figure*}

Across simulations, galaxies are selected based on identical criteria. Simulation snapshots are selected in the redshift range $0.07 \leq z \leq0.7$  to match current \OVI\, observations \citep[e.g.,][]{Mishra_24} and spaced by $1\rm\,Gyr$. Within these snapshots, galaxies are selected based on stellar mass (\Mstar) to satisfy: $7\leq {\rm log}(M_*/M_{\odot}) \leq 9$, and an isolated environment criterion. In this work, we define ``isolated'' as $>1\, \rm Mpc$ away from MW-mass galaxy (${\rm log}(M_*/M_{\odot}) > 10$) and $>350\,\rm kpc$ from a dwarf galaxy (${\rm log}(M_*/M_{\odot}) > 6$). Only halos containing less than 2\% of their dark matter mass in low-resolution (high-mass) dark matter particles are selected. Finally, to match observational definitions for star-forming galaxies, galaxies were also selected to be currently star-forming (${\rm log}(\rm{sSFR}/{\rm yr}^{-1}) > -11$) in the 100 Myr before a given snapshot.

These requirements are used to select similar galaxies in terms of mass, cosmic age, environment, and activity. We remove four halos from FIRE-2 \textsf{m12r} due to the contamination of overly metal-enriched surroundings (i.e., they are extreme outliers in metallicity) likely from a neighboring galaxy. Although we searched all \textsf{m10}, \textsf{m11}, and \textsf{m12} volumes of the FIRE-2 public data, only some volumes yielded halos fitting our selection criteria. The volumes included in this work are as follows: \textsf{m11b}, \textsf{m11i}, \textsf{m11q}, \textsf{m12b}, \textsf{m12c}, \textsf{m12f}, \textsf{m12m}, \textsf{m12r}, \textsf{m12w}, and \textsf{m12z}.

This sample selection yields a total of 112 galaxies from \MM\, and 68 from FIRE-2. Figure \ref{fig:Sample} compares the final samples from either suite. For both FIRE-2 (red diamonds) and \MM\, (blue circles) galaxies, the stellar mass to halo mass relation and specific star formation rates (sSFR) are shown in the top left and right panels, respectively. The bottom row shows probability density histograms for stellar mass, virial mass, and specific star formation rate (sSFR).

After applying the above selection criteria, the suites yield galaxies that reside in similar halo masses and are undergoing similar star formation rates, for a given stellar mass. However, the histograms in Figure \ref{fig:Sample} demonstrate that the FIRE-2 sample has more galaxies at lower stellar and virial masses than \MM. The effects of this mass bias on our results are addressed in the following sections.  Furthermore, \MM\, galaxies are typically undergoing stellar activity of $\rm{log\, sSFR /yr^{-1}} \sim -10$ whereas the FIRE-2 sample tends to span a wider range in sSFR while still producing a similar median to \MM. This difference is in line with results presented in \citet{Iyer_20}, which found that FIRE-2 galaxies undergo more bursty star formation histories than Marvel galaxies. 

Further, although we impose an isolation criterion on our galaxy selection, this does not guarantee the absence of metal enrichment from other galaxies. This statement is particularly relevant for the \textsf{m12} volumes in the FIRE-2 suite, which include a massive galaxy in addition to the selected dwarfs. Nonetheless, we are able to identify robust trends across suites related to subgrid physics, and we verify that our results hold for both the \textsf{m12} and \textsf{m11} volumes, indicating they are not solely driven by the presence of a MW–mass galaxy.

In addition to selecting galaxies of similar environments, masses, and activity, we also adopt the \citet{HM2012} UVB table when modeling the \OVI\, ion fractions for both simulations to eliminate the effects of differing modeling decisions. However, we explore the sensitivity to the assumed UVB in Section \ref{subsec:disc_vary}.

For the analysis in this paper, when averaging over multiple galaxies, we break the simulated galaxies in either suite into a low-mass ($M_*<10^8 M _{\odot}$) and high-mass ($M_*>10^8 M _{\odot}$) sample to minimize mass trends imprinting on our results. Additionally, we often normalize values by the galaxy \Mstar, halo virial mass (\Mvir), or halo virial radius ($R_{vir}$). For both suites, \Mvir\, and $R_{vir}$ are defined by enclosing a mean density 200 times the critical density of the Universe. Across simulations, we treat metallicity ($Z$) as the mass fraction of a given species relative to the particle/cell mass. In this work, we focus our investigation on oxygen. We further normalize our metallicity values with respect to the solar abundance ($Z_{\odot}=0.0134$) reported in \citet{Asplund2009}.

\begin{figure*}[!t]
    \centering
    \includegraphics[width=0.8\textwidth]{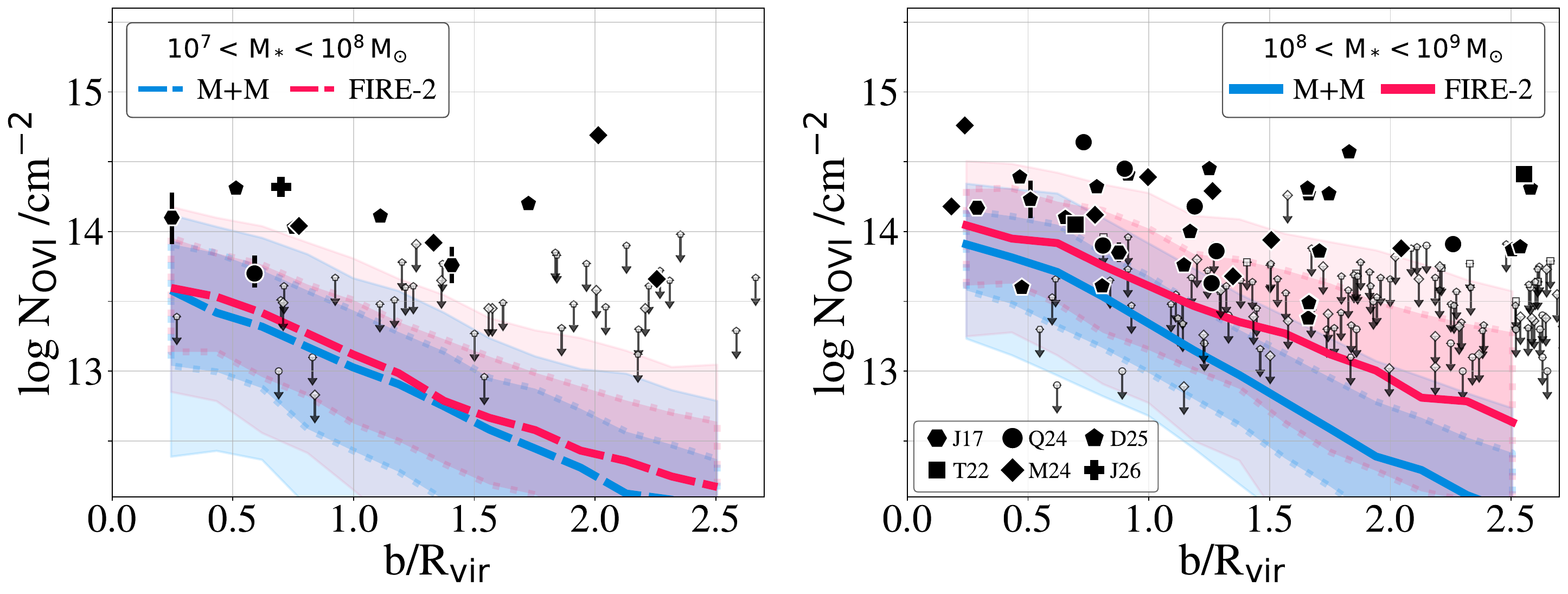}

    \caption{\OVI\, column density (\NOVI) profiles for both suites (colors as in Figure \ref{fig:Sample}) split into low-mass (\textit{left}, $M_*<10^8 M_{\odot}$) and high-mass (\textit{right}, $M_*>10^8 M_{\odot}$). Median \NOVI\, values and 16th-84th percentile ranges are shown as solid lines and shaded bands, and are calculated within $0.2 b/R_{vir}$ wide bins. Observational data from \citet{Johnson_17} (J17), \citet{Tchernyshyov_22} (T22), \citet{Qu_24} (Q24), \citet{Mishra_24} (M24), \citet{Dutta_25_col} (D25), \citet{Johnson26} (J26) are shown as hexagons, squares, circles, diamonds, pentagons, and crosses, respectively. Observationally derived detections of \OVI\, are shown as solid markers, while non-detections are shown as open-faced markers with a downward-pointing arrow. The FIRE-2 suite produces greater \NOVI\, at all impact parameters compared to \MM. However, both suites generally underproduce \NOVI\, at all impact parameters compared to observations, especially at large radii.}
    \label{fig:CompObs_NOVI}
\end{figure*}

\section{Comparison to OVI Observations}\label{sec:OVIobservations}

To begin our investigation of \OVI, we first assess the underprediction of \OVI\, in simulations by comparing to observed column densities (\NOVI) and observationally derived \OVI\, masses (\MOVI). Previous work has demonstrated that both the FIRE-2 \citep{Li_21} and \MM\, \citep{Piacitelli_25} simulations produce significantly lower \NOVI\, values than those observed \citep{Johnson_17,Qu_24, Mishra_24,Johnson26}. 

\subsection{OVI Column Densities}

We present the column densities for the simulated galaxies studied in this work in Figure \ref{fig:CompObs_NOVI}.  Similar to \citet{Li_21} and \citet{Piacitelli_25}, we utilize \textsf{TRIDENT} \citep{Trident} to generate synthetic sightlines and derive column densities. We adopt the \citet{HM2012} UVB table for both suites to eliminate the effects of differing UVB and explore the effects from the assumed UVB in Section \ref{subsec:disc_vary}. For each galaxy, we model 300 sightlines that are randomly oriented with respect to the disk, centered at random impact parameters, $b$, within the range $0.15 \leq b/R_{vir} \leq 2.5$. Each sightline is $1.25\,\rm Mpc$ in length to avoid intersecting another galaxy's CGM while still probing the full environment of the dwarf galaxy. We then combine the \NOVI\, values for all galaxies in two stellar mass bins, $10^{7}\leq M_* \leq 10^{8} M_{\odot}$ and $10^{8}\leq M_* \leq 10^{9} M_{\odot}$, and find the median values and 16th-84th percentiles across $b/R_{vir}$ and plot them in Figure \ref{fig:CompObs_NOVI}. We compare these simulated values to a subset of observational data from \citet{Johnson_17} (J17), \citet{Tchernyshyov_22} (T22), \citet{Qu_24} (Q24), \citet{Mishra_24} (M24), \citet{Dutta_25_col} (D25), and \citet{Johnson26} (J26). From these observational works, we select only galaxies that fall within our mass range, $10^{7}\leq M_* \leq 10^{9} M_{\odot}$ and plot them in their respective panels in Figure \ref{fig:CompObs_NOVI}.

As shown in previous works, both suites underpredict \NOVI\, compared to observations. For both low-mass ($M_*<10^8 M _{\odot}$, left panel) and high-mass galaxies ($M_*>10^8 M _{\odot}$, right panel), the median \NOVI\, profiles for both suites agree within $1\sigma$. However, the median values for FIRE-2 are systematically higher than \MM, in particular at larger impact parameters (i.e., $b/R_{vir}>1.5$) for high-mass galaxies. Specifically, at $b/R_{vir}\sim2$ the median \NOVI\, profiles of high-mass FIRE-2 galaxies are $\sim4\times$ higher than \MM\, (subsequent sections will investigate the source of this difference).

Compared to observations, simulated low-mass galaxies tend to underpredict observed \OVI\, detections by $\sim0.5-1.5\rm \, dex$ across $b/R_{vir}$ values. High-mass galaxies show greater \NOVI\, values and are in closer agreement with observations, with FIRE-2 reproducing a higher fraction of observed detections within $1\sigma$ for this mass regime. However, neither simulation reproduces the highest observed detections, and the medians of both lie below almost all observed detections. Further, both simulations show a stronger decline in typical \NOVI\, values with galaxy stellar mass than the dependence seen in observations \citep[e.g.,][]{Tchernyshyov_22, Qu_24}. 
\begin{figure*}[!t]
    \centering
    \includegraphics[width=0.65\textwidth]{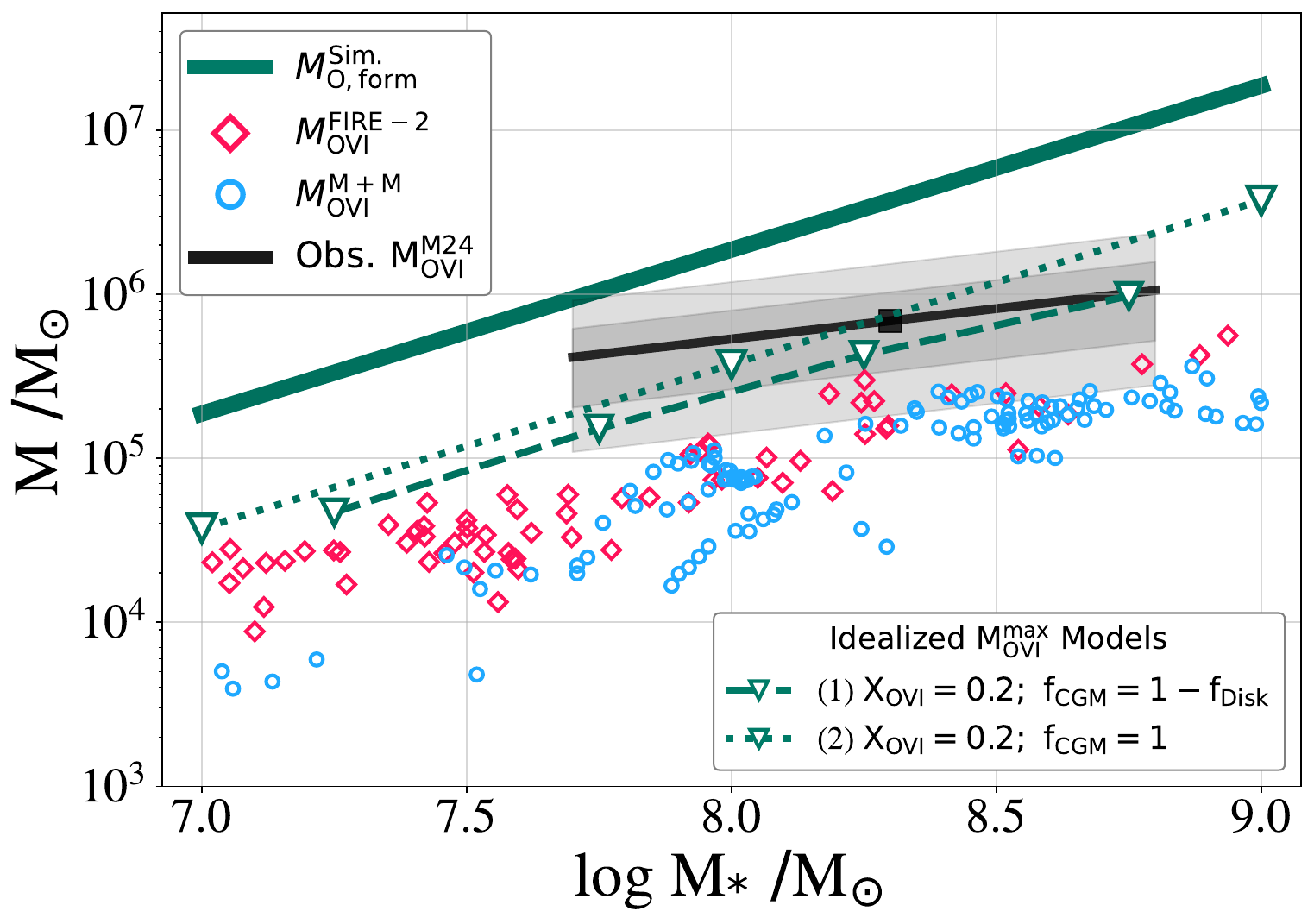}
    \caption{Current and maximum possible \OVI\, masses (\MOVI) in simulations. Current \MOVI\, within $0.15 \leq r/R_{\rm vir} \leq 2.5$ are shown for FIRE-2 ($M^{\rm FIRE\text{-}2}_{\rm OVI}$; red diamonds) and \MM\ ($M^{\rm M+M}_{\rm OVI}$; blue circles) galaxies, compared to observational estimates from \citet{Mishra_24} (\MishraMOVI; black line and shaded region). To estimate the maximum possible \MOVI, we use the total oxygen mass formed over the galaxy lifetime (\MOform; green solid line), computed from empirical yields following \citet{Hafen_19}. We then define $M_{\rm OVI}^{\rm max} = X_{\rm OVI} \cdot f_{\rm CGM} \cdot \text{\MOform}$, where $X_{\rm OVI}$ is the fraction of oxygen in the \OVI\, ionization state and $f_{\rm CGM}$ is the fraction of produced oxygen residing in the CGM. We consider two limiting cases (both adopting $X_{\rm OVI} = 0.2$): (1) a model assuming near-perfect, spatially uniform ionization while retaining the simulated disk oxygen fractions ($f_{\rm CGM} = 1- f_{\rm disk}$; dashed line), and (2) an extreme model in which all produced oxygen resides in the CGM ($f_{\rm CGM} = 1$; dotted line). While neither scenario is physically plausible, they provide upper bounds on the attainable \MOVI. Even in the case of perfect ionization (case 1), simulations fail to match the observed \OVI\ masses, implying insufficient CGM oxygen. In the more extreme limit (case 2), where all oxygen is placed in the CGM and optimally ionized, the maximum \MOVI\, for galaxies with $7.7 \leq \log(M_*/M_{\odot}) \leq 8.3$ remains below the median \MishraMOVI. This suggests that dwarf galaxies in this mass range have underproduced oxygen over their star formation histories, although this deficient oxygen production likely extends to both higher and lower masses.}
    \label{fig:CompObs_Mass}
\end{figure*}

\begin{figure}
    \centering
    \includegraphics[width=0.8\linewidth]{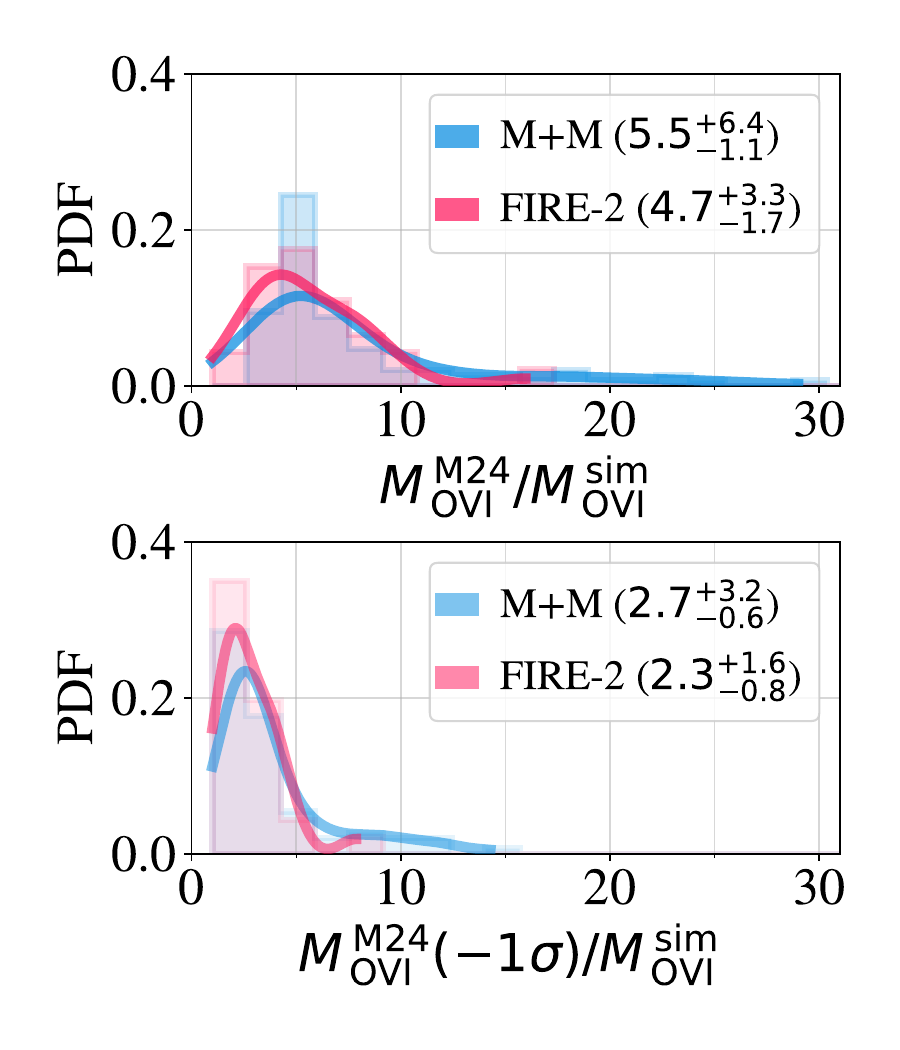}
    \caption{Gap between current simulated \OVI\, masses ($M_{\rm OVI}^{\rm sim}$) and the observationally derived \MishraMOVI\, for $7.7 \leq \log(M_*/M_{\odot}) \leq 8.8$ galaxies. \textit{Top panel:} The ratio between the median \MishraMOVI\, (black line in Figure \ref{fig:CompObs_Mass}) and $M_{\rm OVI}^{\rm sim}$. \textit{Bottom panel:} The ratio between the $1\sigma$ lower limit for \MishraMOVI\, (inner shaded region in Figure \ref{fig:CompObs_Mass}) and $M_{\rm OVI}^{\rm sim}$. 
    The ratios are computed for each simulated galaxy in both suites and plotted as histograms in 30 equally sized bins (transparent histograms) following the same color scheme as previous plots. Solid lines correspond to kernel-density estimated PDFs, plotted over the range of the simulation data. The median values and 16th-84th percentile ranges for both suites are provided in the legend.  To match the median \MishraMOVI, both suites require a $\sim5\times$ increase in their \OVI masses, while to be within $1\sigma$ of \MishraMOVI, a more modest increase of $2-3\times$ is required. See Appendix \ref{apx:IncreasedNOVI} for the resulting \NOVI\, values given a $3\times$ uniform increase in \OVI.
    }
    \label{fig:CompObs_MassRatios}
\end{figure}

\subsection{Total CGM OVI Mass}
In this section, we compare observationally derived \MOVI\, from \citet[][hereafter, \citetalias{Mishra_24}]{Mishra_24} to simulated values. While other studies also estimate \MOVI\, \citep[e.g.,][]{Johnson_17, Dutta_25_col, Tchernyshyov_22}, these measurements are typically limited to projected distances within $R_{vir}$. In contrast, \citetalias{Mishra_24} provides a sample with strong isolation criteria and mass estimates extending to $3R_{vir}$. From the observed column densities, \citetalias{Mishra_24} estimates the total \OVI\, mass within a radial annulus between $R_{\rm inner}$ and $R_{\rm outer}$ using the mean column density measured in that annulus, $\langle N_{\rm OVI} \rangle$. The mass is computed as, 
\begin{equation}\label{eqn:m24movi}
    M_{\rm OVI}^{M24} \approx \pi (R^2_{\rm outer}-R^2_{\rm inner}) ~m_{\rm OVI}~ \kappa_{\rm OVI} ~\langle N_{\rm OVI} \rangle,
\end{equation} 
where $m_{\rm OVI}$ is the mass of an \OVI\, ion and $\kappa_{\rm OVI}$ is the covering fraction of detections. This calculation is performed separately for strong ($N_{\rm OVI} > 10^{14}\,\mathrm{cm^{-2}}$) and weak ($10^{13.5} < N_{\rm OVI} < 10^{14}\,\mathrm{cm^{-2}}$) detections, with non-detections accounted for via bootstrapping, and the resulting masses are summed. The resulting total \OVI\ mass within $0 \leq b/R_{\rm vir} \leq 2$ is $\text{\MishraMOVI} = 6.9^{+3.4}_{-3.5} \times 10^5 M_\odot$, where the quoted uncertainties reflect the $1\sigma$ range from bootstrapping. 

It is important to highlight the uncertainties and assumptions built into the observations we compare to in this work. The first assumption we confront is potential dependencies on galaxy stellar mass that are not inherently accounted for when converting column density observations into \MOVI. For instance, \citetalias{Mishra_24} uses Equation \ref{eqn:m24movi} and combines detections across multiple galaxies to derive a single median \MOVI. Built into this equation is an assumption about the virial radius of the halo and, therefore, the stellar-to-halo mass relation. We account for this dependence by including a $\text{\Mstar} \propto R_{vir}$ scaling in \MishraMOVI\, (derived through private communication with the authors, and more details are provided in Appendix \ref{apx:mstardep}). There may also be an additional $\text{\NOVI} \propto \text{\Mstar}$ dependence that could alter the slope of the \MishraMOVI\, estimate; however, we do not account for it in this work. We assume a negligible $\text{\NOVI} \propto \text{\Mstar}$ dependence based on previous work \citep[e.g.,][]{Tchernyshyov_22, Dutta_25_col} that has analyzed \NOVI\, as a function of \Mstar\, and only found a weak dependence within this galaxy mass range. An increased observational sample may discover a stronger $\text{\NOVI} \propto \text{\Mstar}$ trend, but at the current state of the field, this is the most robust assumption.

Second, \citetalias{Mishra_24} notes that the few high \NOVI\, detections can dominate the signal in determining $\langle N_{\rm OVI} \rangle$ and result in elevated \MOVI\, values. To address this, the authors provide $1\sigma$ and $2\sigma$ error ranges based on bootstrapping. While this approach is the most reasonable for the current dataset, it assumes that the existing detections and non-detections adequately sample the true \NOVI\, distribution. Despite this limitation, the resulting mass estimate remains the most robust to date. Additionally, \NOVI\, detections across multiple works (i.e., \citet{Qu_24} and \citet{Johnson26}, also shown in Figure \ref{fig:CompObs_NOVI}) are consistent with those reported by \citetalias{Mishra_24}. An updated \MOVI\, estimate could reduce the error ranges. However, the mean \NOVI\, from \citetalias{Mishra_24} is consistent with the mean across the total observational sample in Figure \ref{fig:CompObs_NOVI}, suggesting that the inferred \MOVI\, is unlikely to decrease significantly.

Figure \ref{fig:CompObs_Mass} shows \MishraMOVI\, as a solid black line and shaded regions. The plotted observations span the 16th–84th percentile range of the observed stellar mass distribution ($7.7 \leq {\rm log}(M_*/M_{\odot}) \leq 8.8$), with the shaded regions indicating the reported $1\sigma$ and $2\sigma$ uncertainties in \MishraMOVI. Simulated \MOVI\, values within $0.15\leq r/R_{vir}\leq 2.5$ (i.e., the CGM) are shown as red diamonds (FIRE-2) and blue circles (\MM). Green lines are explained in Section \ref{subsec:SOS}. Although some simulated $M_* \geq 10^{8.2} M_{\odot}$ galaxies lie within the $2\sigma$ ranges, both suites systematically lie below the median and $1\sigma$ ranges.

Within the galaxy mass range of $7.7 \leq {\rm log}(M_*/M_{\odot}) \leq 8.8$, we present the \OVI\, mass deficit for each galaxy in Figure \ref{fig:CompObs_MassRatios}. Specifically, the top panel shows the ratio between the median \MishraMOVI\, value and the simulated \MOVI, while the bottom panel shows the ratio between the \MishraMOVI\, $1\sigma$ lower limit and the simulated \MOVI. To match the median \MishraMOVI, both suites would require a $\sim 5\times$ increase in their CGM \OVI\, reservoirs. However, due to the significant scatter in simulated \MOVI, some galaxies would require larger boosts of $>10\times$ to reach the observed median. Although, as seen in Figure \ref{fig:CompObs_NOVI} and Figure \ref{fig:CompObs_Mass}, the scatter in both \MOVI\, and \NOVI\, can be $0.5-1~\rm dex$ at a given galaxy mass or impact parameter. Thus, in order to reproduce the observed column densities, every simulated galaxy likely does not need to exactly match the median \MishraMOVI. The right panel of Figure \ref{fig:CompObs_MassRatios} demonstrates that in order to lie within $1\sigma$ of \MishraMOVI\, both suites require a smaller increase of $2-3\times$ in their CGM \OVI\, reservoirs. Indeed, we find that globally increasing the simulated \MOVI\, values by a factor of $3$ provides a reasonable fit to the existing column densities observations (see Appendix \ref{apx:IncreasedNOVI} for more details).

\subsection{Identifying the Simulated Oxygen Shortage}\label{subsec:SOS}

Given that FIRE-2 and \MM\, both require similar total enhancements in their \MOVI\, values to reproduce observations, we combine the two suites to quantify the gap between simulations and observations in this section. We first determine the effect of large-scale metal loss beyond $2.5R_{vir}$ in the \OVI\, deficit by computing the fraction of retained oxygen within the CGM and disk. The retained oxygen fraction is found using the current total oxygen gas mass within  $0 \leq r/R_{vir}\leq2.5$ for each galaxy ($M_{\rm O}^{<2.5R_{vir}}$, not shown) and the total oxygen mass formed over the galaxy's lifetime (\MOform, green line in Figure \ref{fig:CompObs_Mass}). To determine \MOform, the same analysis as presented in \citet{Hafen_19} is used to derive an empirical oxygen yield ($y_{\rm O,\, box}$) for each simulation box following, 
\begin{equation}
    y_{\rm O,\, box} = (M_{\rm O,\,box} - Z_{floor}M_{\rm b,\, box})/ M_{\rm *,\,box},
\end{equation}
 where $M_{\rm O, box}$, $M_{\rm b,\, box}$, and $M_{\rm *,\, box}$ are the total oxygen mass (gas and stars), baryonic mass (gas and stars), and stellar mass in the full simulation volume, and $Z_{floor}$ is the implemented oxygen metallicity floor for the suite\footnote{For FIRE-2, $Z_{O, floor} \approx 8.65\times10^{-7}$, but for \MM\, $Z_{O, floor}=0$ as the suite does not implement a metallicity floor.}. \MOform\, is found with the current stellar mass of the galaxy via $\text{\MOform} =  y_{\rm O,\, box}\, M_{*}$. From these calculations, we find strong agreement between the mean empirical yields for the simulations: $y_{\rm O,\, box}=0.0187\pm0.0008$ for FIRE-2 and $y_{\rm O,\, box}=0.0190\pm0.0004$ for \MM. We find little variation within a given suite in $y_{\rm O,\, box}$ and the two suites have $y_{\rm O,\, box}$ values within $1\sigma$ of each other. \footnote{We validated this approach for \MM\ by computing \MOform\ directly from SNIa and SNII rates and metal deposition for winds following \citet{Piacitelli_25}, finding agreement with $y_{\rm O,\, box}\, M_{*}$ to within a factor of $1.00\pm0.03$. The empirical yield framework is adopted to determine \MOform\, for both suites in the interest of consistency across suites and with previous work in FIRE-2 \citep[e.g.,][]{Hafen_19}.}

\begin{figure*}[!t]
    \centering
    \includegraphics[width=0.9\textwidth]{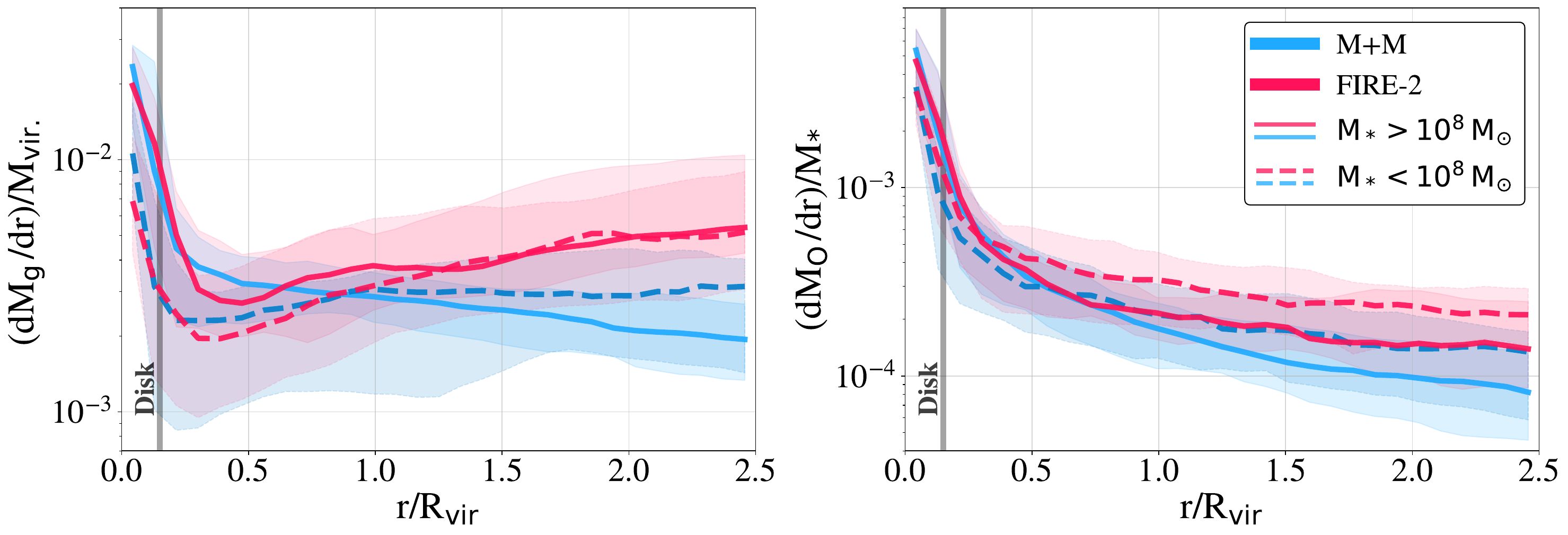}
    \caption{CGM mass comparisons for the \MM\, (blue) and FIRE-2 (red) samples. The left and right panels show results for gas mass ($\rm M_{g}$) and oxygen mass ($\rm M_{O}$), respectively. Medians for low-mass and high-mass samples are shown as dashed and solid lines, respectively. The gas mass profile ($\rm dM_{g}/dr$) is normalized by virial mass (\Mvir) while the oxygen mass profile ($\rm dM_{Z}/dr$) is normalized by \Mstar. The lines and shaded region represent the median and 16th-84th percentile within 30 equally sized bins between $0 \leq r/R_{vir} \leq 2.5$. We find that the total CGM gas and oxygen masses in both suites agree well with each other (not shown here). However, the radial distribution of CGM material differs between the simulation suites.} 
    \label{fig:CGMprop_mass}
\end{figure*}
Taking the median and standard deviation of the full simulation sample, we find that $77\pm17\%$ of oxygen produced is retained within $2.5R_{\rm vir}$, with higher mass galaxies retaining typically more of their metals ($80\pm 17\%$) than lower mass galaxies ($74 \pm 17\%$).  Therefore, although some oxygen is lost beyond $2.5R_{vir}$, metal loss alone cannot account for the \OVI\ deficit.

We further explore the capabilities of current simulations in achieving a $3\times$ increase in \MOVI. We do this using  
two test cases in Figure \ref{fig:CompObs_Mass}, which place upper bounds on the \MOVI\, achievable in simulations under highly idealized assumptions for \OVI. Both test cases follow the equation: 
\begin{equation}
    M_{\rm OVI}^{\rm max} = X_{\rm OVI} \cdot f_{\rm CGM} \cdot  \text{\MOform},
\end{equation}
where $X_{\rm OVI}$ is the fraction of oxygen that is ionized into \OVI\, and  $f_{\rm CGM}$ is the fraction of oxygen formed that will remain in the CGM. Both test cases assume near-perfect ionization into \OVI\, by adopting $X_{\rm OVI}=0.2$. Currently, both suites have median $X_{\rm OVI}$ values of $0.11$ and $0.9$ for low- and high-mass galaxies, respectively (see Section \ref{sec:OVIproperties} for more details); therefore, the choice of $X_{\rm OVI}=0.2$ results in twice the ionization efficiency relative to the fiducial suites.

In addition to $X_{\rm OVI}=0.2$, the first test case assumes no oxygen loss beyond $2.5R_{vir}$, while maintaining the current disk retention ($f_{\rm disk}$) in gas and stars. Under these assumptions, the retained CGM fraction is simply $f_{\rm CGM}=1-f_{\rm disk}$, where $f_{\rm disk}$ is computed by summing the total oxygen mass in all gas particles/cells and star particles within $r<0.15R_{vir}$. Taking the median $f_{\rm CGM}$ of the full simulated sample gives $f_{\rm CGM} = [0.73,0.74,0.67,0.47]$ for $\log(M_*/M_{\odot}) \sim [7.25,7.75,8.25,8.75]$ galaxies. This scenario, therefore, represents the maximum \MOVI\ achievable under idealized ionization and no metal loss to large scales (dashed green line in Figure \ref{fig:CompObs_Mass}). Even in this limit, only $\log(M_*/M_{\odot}) > 8.3$ galaxies show $M_{\rm OVI}^{\rm max}$ values within the $1\sigma$ bounds from \MishraMOVI. This test case demonstrates that preserving the interstellar medium (ISM) metal content, increasing the ionization of \OVI, and retaining all metals within $2.5R_{vir}$ could provide $\log(M_*/M_{\odot}) > 8.3$ galaxies with the necessary increase in \MOVI\, to match \MishraMOVI. On the other hand, the $M_{\rm OVI}^{\rm max}$ values for $\log(M_*/M_{\odot}) \leq 8.0$ galaxies lie below the $1\sigma$ bounds, indicating that these galaxies would require additional transfer of metals from the ISM into the CGM.

The second test case adopts $f_{\rm O,\, CGM} = 1$ and is shown as a green dotted line. This test case represents idealized ionization, no metal loss to large scales, and the CGM containing $100\%$ of oxygen formed, and therefore $0\%$ retention within the disk's gas and stars. The second test demonstrates that only ${\rm log}(M_*/M_{\odot}) \geq 8.3$ galaxies have formed enough oxygen to possibly allow agreement with the median \MishraMOVI. For ${\rm log}(M_*/M_{\odot}) \leq 8.3$ galaxies, the $M_{\rm OVI}^{\rm max}$ values lie within the $1\sigma$ across the stellar masses observed. Compared to the previous test case, this case increases the $M_{\rm OVI}^{\rm max}$ by $\sim50\%$ for these lower-mass galaxies due to the already small fraction of metals retained in their ISM.   

We emphasize that neither test case is physically plausible: both require the CGM to occupy a narrow phase space, and the second additionally depletes the disk of oxygen, violating the mass–metallicity relation \citep[e.g.][]{Tremonti_04}. These scenarios, therefore, represent strict upper limits on the \MOVI\, achievable in current simulations. Nonetheless, even under idealized assumptions of ionization efficiency and negligible large-scale metal loss, $7.7 \leq {\rm log}(M_*/M_{\odot}) \leq 8.3$ galaxies would require evacuating their ISM of metals to reside within the $1\sigma$ bounds. Our findings, therefore, suggest that these galaxies may have underproduced oxygen over their star formation histories: a tension we term the Simulated Oxygen Shortage (SOS).

Figures \ref{fig:CompObs_NOVI} and \ref{fig:CompObs_Mass} illustrate that the \OVI\, deficit in simulations is an issue of both feedback and oxygen production. As oxygen production and distribution are primarily governed by ccSNe, simulations likely need to revisit their models for one or more of either oxygen yields, star formation, adopted IMF, and SNe rates to address the SOS, and metal diffusion and feedback to address the even lower columns at high $b/R_{vir}$. The need for more feedback or more efficient feedback is especially relevant for \MM\, given that the  \NOVI\, values at large impact parameters are much lower than FIRE-2.

In Section \ref{subsec:quantgap}, we estimate the additional oxygen mass and number of ccSNe required to close this gap between observed and simulation \OVI\, column densities. The remainder of this paper examines how differences in subgrid physics affect the CGM \OVI\, properties. By better understanding the relationship between subgrid physics, the CGM, and \OVI, we aim to inform future work that will endeavor to close this gap between current simulations and observations. 

\section{CGM Properties across Suites}\label{sec:CGMproperties}
%
%
As described in Section \ref{sec:simulations}, the two suites adopt differing physical prescriptions. Here, we seek to understand how these differences influence CGM structure in the FIRE-2 and \MM\, simulations.

For the isolated galaxies we study here, feedback processes are the key drivers of metal transport from the ISM and into the CGM or intergalactic medium (IGM). In addition to distributing metals, feedback-heated gas can also distribute energy into the halo \citep[e.g.,][]{Pandya2021}, thereby affecting the thermal state of the CGM gas. Both the mass of oxygen and the physical conditions of a given gas parcel will set the predicted mass in \OVI\, from the simulations, and therefore differences in feedback prescriptions may play an important role in setting the total \OVI\, mass produced. 

Although not shown, we find little systematic variation across suites in terms of total CGM mass and oxygen content. However, the radial dependence of these properties does differ by suite. Figure \ref{fig:CGMprop_mass} compares the two suites in terms of their typical gas mass (left panel) and oxygen mass (right panel) radial distribution. 

\begin{figure*}[!t]
    \centering

    \includegraphics[width=0.95\textwidth]{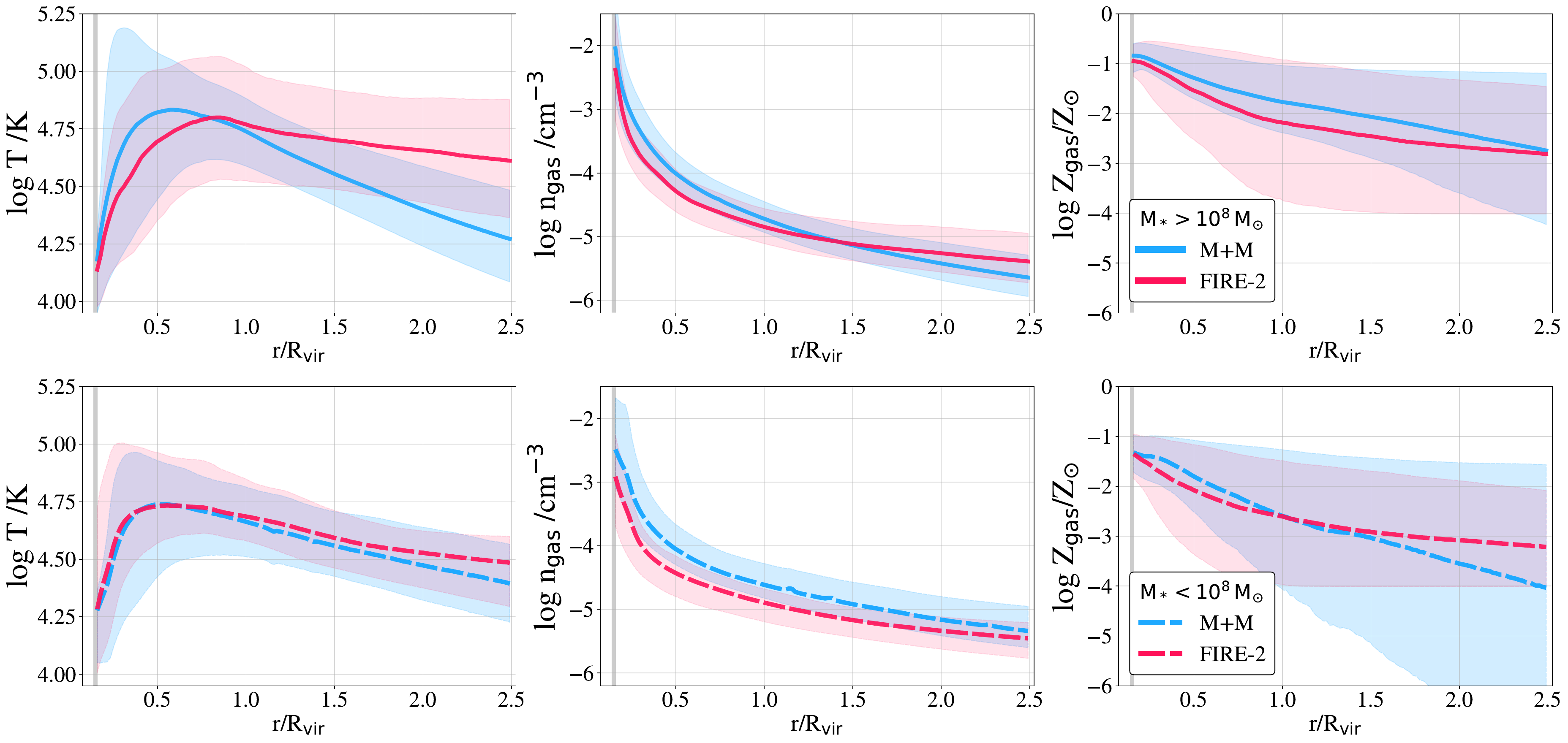}
    \caption{CGM condition comparisons for the \MM\, (blue) and FIRE-2 (red) samples separated into high (top row) and low (bottom row) stellar mass bins. Plotted is the CGM gas temperature (\textit{left}), gas density (\textit{center}), and gas-phase metallicity relative to solar (\textit{right}). Data is stacked for all galaxies in a given suite and mass bin, the mass-weighted average (lines) and associated 16th-84th percentiles (shaded regions) are plotted. For all three properties presented here, we find systematic differences across suites that likely imprint on the production of \OVI.}
    
    \label{fig:CGMprop_cond}
\end{figure*}

These radial distributions are made by summing the total simulated mass in gas or oxygen within 30 equally sized bins between $0 \leq r/R_{vir} \leq 2.5$. For convenience, we denote the boundary at $r/R_{vir}=0.15$ between the disk and CGM as a gray line. We separate the suites into low-mass (dashed lines) and high-mass (solid lines) samples and provide 16th-84th percentile ranges for each radial bin (shaded regions). Further, we normalize the gas mass distribution by the galaxy virial mass (\Mvir) and the oxygen mass distribution by \Mstar\, to better isolate the effects due to modeling rather than simply galaxy mass.

Within the disk and CGM regions, both suites have similar total gas masses and oxygen masses. However, at $r/R_{vir}\sim0.5$, FIRE-2 galaxies tend to have slightly less total gas mass for a given \Mvir\, than \MM. With increasing distance from the galaxy, FIRE-2 galaxies have a positively increasing gas mass profile, whereas \MM\, galaxies have a decreasing or relatively constant mass profile for the high- and low-mass samples, respectively. Finally, in the outer halo ($r/R_{vir} > 1$), FIRE-2 galaxies have preferentially more gas mass than both their own inner halos and \MM\, galaxies at the same distance. In particular, at $r/R_{vir} = 2$, the median gas mass profiles for FIRE-2 galaxies are $\sim 1.5-2\times$ greater than \MM.

The differences seen in the gas mass profiles of both suites are likely to be tied to differences in feedback efficiency. Specifically, \citet{Pandya2021} showed that FIRE-2 mass, momentum, and metal outflow rates increase with increasing distance from a dwarf galaxy while energy loading factors remain relatively constant. This result indicates that outflows sweep up material as they travel through the CGM in FIRE-2, which aligns with the findings in Figure \ref{fig:CGMprop_mass}, where we can see more gas mass is pushed to or preserved at larger radii in FIRE-2 than \MM\, (see also de la Cruz et al. (2026, \textit{in prep}).

The right panel of Figure \ref{fig:CGMprop_mass} shows the oxygen mass radial distribution. For both suites, oxygen mass decreases with increasing radii. Although both suites tend to be within $1\sigma$ of each other, we find systematic differences across suites. Comparing the median profile for high-mass galaxies, FIRE-2 and \MM\, have similar oxygen masses within $R_{vir}$, but FIRE-2 tends to have $1.2-1.7\times$ more oxygen within $1R_{vir}-2.5R_{vir}$. For low-mass galaxies, both the inner and outer CGM in FIRE-2 have between $1.3-1.7\times$ more oxygen than \MM. By repeating this calculation, removing the gas at the metallicity floor in FIRE-2, we find the same results. Therefore, the oxygen difference across $r/R_{vir}$ is tied to FIRE-2 feedback being more efficient than \MM\, at enriching the outer CGM in metals.

Figure \ref{fig:CGMprop_cond} summarizes the CGM temperature, density, and gas-phase metallicity \citep[oxygen mass fraction relative to solar oxygen abundance according to ][]{Asplund2009} as a function of radius. Beginning with the temperature profile (left column), the outer halo for both high- and low-mass galaxies is $1.1-1.8\times$ warmer in FIRE-2 than in \MM. This behavior can likely be attributed to preventative feedback in FIRE-2, which distributes energy to large radii and heats CGM gas, preventing accretion \citep{Pandya2020,Pandya2021}.  de la Cruz et al. (2026, \textit{in prep}) finds that \MM\, galaxies exhibit energy loadings an order of magnitude lower than FIRE-2, demonstrating that outflows in \MM\, do not drive CGM heating as effectively as they do in FIRE-2.

In terms of gas density,  the inner CGM ($r/R_{vir}\sim 0.5$) in \MM\, is $2-2.5\times$ denser than FIRE-2. However, there is a transition at larger radii, with FIRE-2 becoming denser than \MM. For high-mass galaxies, this occurs at $r/R_{vir}\sim 1.5$, while for low-mass galaxies it occurs at a larger distance, not shown in Figure \ref{fig:CGMprop_cond}, of $r/R_{vir}\sim 3.0$. 

Metallicities also differ by suites. In the high-mass bin, \MM\, galaxies have a more metal-enriched CGM by a factor of $1.2-2.6$ at all radii. In the low-mass bin, \MM\, galaxies have $1-2\times$ higher metallicities within $r/R_{vir}<1$; however, at larger radii, FIRE-2 exhibits higher metallicities by a factor of $\sim 1-6$. Importantly, we note the presence of the metallicity floor of $Z/Z_{\odot} = 10^{-4}$ implemented in FIRE-2, which is not present in \MM, but can be seen in the FIRE-2 results at larger radii. The lower average metallicities in FIRE-2 are consistent with Figure \ref{fig:CGMprop_mass}, which shows higher oxygen and gas masses at most radii. The increased gas mass dilutes the oxygen content, resulting in lower average metallicities than in \MM.

Overall, within $1R_{vir}$, high-mass FIRE-2 galaxies host a CGM that is cooler, less dense, and less metal-enriched compared to \MM;  beyond $1R_{vir}$, this generally reverses, where FIRE-2 becomes hotter and denser, while still being less metal-enriched. In the low-mass sample, both suites show similar CGM temperatures across radii, but \MM\, is systematically denser at all radii. Within $1R_{vir}$, low-mass galaxies in \MM\, have a more metal-enriched CGM than FIRE-2 galaxies of the same mass. At large radii, the CGM in FIRE-2 then transitions to being more metal-enriched than \MM. Together, these radially dependent differences underscore how differences in subgrid physics and feedback modeling imprint on CGM conditions.

\section{OVI Properties across Suites}\label{sec:OVIproperties}

Within our sample of galaxies, we can see systematic differences between simulations in terms of the CGM mass and metal distributions, as well as physical conditions. These differences are a product of modeling choices and prescriptions, and we now extend this investigation to \OVI\, and discern how these differences in modeling impact \OVI\, production in both suites. 
\begin{figure*}
    \centering
    \includegraphics[width=0.95\textwidth]{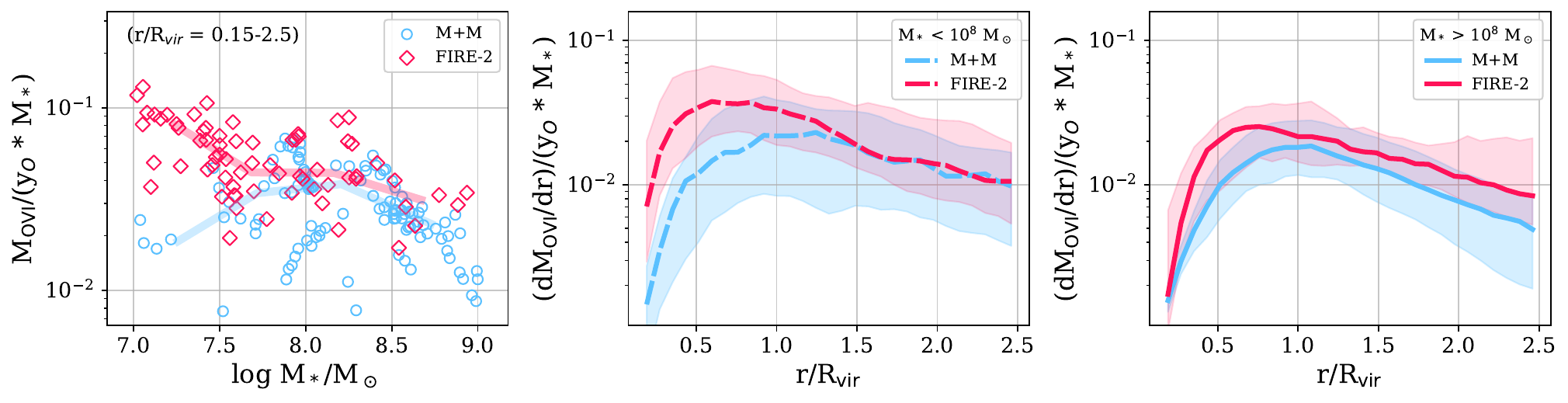}

    \caption{Mass of \OVI\, within $0.15 \leq r /R_{vir} \leq 2.5$ of each galaxy relative to the mass of oxygen produced by the galaxy ($y_OM_*$, from Figure \ref{fig:CompObs_Mass}). The left panel shows the summed \OVI\, mass divided by $y_OM_*$ (i.e., the fraction of $y_OM_*$ residing in \MOVI) versus stellar mass for \MM\, (blue circles) and FIRE-2 (red diamonds). Median values for both suites are plotted as in Figure \ref{fig:Sample}. The center and right panels show median radial profiles of ${dM_{OVI}}/dr$ normalized by $y_OM_*$ for low- and high-mass samples, respectively, with shaded regions indicating the 16th-84th percentile range. Although there is significant scatter in $\text{\MOVI}/y_OM_*$, the median fraction in FIRE-2 is $1.3\times$ greater than in \MM---except at the lowest mass bin where FIRE-2 is $4\times$ greater. This higher \OVI\, content in FIRE-2 persists across most radii, peaking at $\sim 0.75\, R_{vir}$ compared to $\sim 1.0\, R_{vir}$ in \MM, indicating that each suite produces \OVI\, at systematically different radii due to differences in CGM structure and conditions (Figures \ref{fig:CGMprop_mass} and \ref{fig:CGMprop_cond}). }
    \label{fig:OVIprop_MOVI}
\end{figure*}

\subsection{OVI Mass Distribution}\label{subsec:NOVI_MOVI}

To anchor our investigation of \OVI\, across suites, Figure \ref{fig:OVIprop_MOVI} presents a comparison of total \OVI\, mass and its radial distribution between FIRE-2 and \MM. The left panel shows the fraction of oxygen mass formed ($y_OM_*$, see also Figure \ref{fig:CompObs_Mass}) that exists in \OVI\, around each galaxy (measured within $0.15 \leq r/R_{vir} \leq 2.5$) as a function of \Mstar. The center and right panels show how this \OVI\, mass is distributed radially by plotting $dM_{OVI}/dr$ profiles, normalized by $y_OM_*$. The \OVI\ mass profiles show the median profile for both suites, and the shaded region shows the 16th-84th percentile. 

Overall, we find that on average \OVI\, typically comprises $2-8\%$ of oxygen mass formed for both suites. \MM\, galaxies typically exhibit lower fractions, $<1\%$ in some cases, while FIRE-2 galaxies tend to exhibit slightly higher fractions, especially for the lowest mass galaxies. For low-mass galaxies, the \OVI\, mass profiles show that FIRE-2 has higher \OVI\, production until $1.5R_{vir}$, then both suites correlate well from $1.5R_{vir}$ to $2.5R_{vir}$. For high-mass galaxies, we see FIRE-2 consistently has higher \OVI\, masses from $0.15 \leq r/R_{vir} \leq 2.5$, with a peak at around $0.75R_{vir}$. For \MM\, we see a peak in production slightly further out at around $1.0R_{vir}$.  These differences in radial profiles reflect differences in the underlying CGM mass distribution and thermodynamic conditions. The following section will disentangle the effects of these two in setting the \OVI\, production. 

The higher \OVI\, mass at large radii in FIRE-2 compared to \MM\, explains FIRE-2's shallower decline in \NOVI\, as a function of impact parameter in Figure \ref{fig:CompObs_NOVI}. In the future, \MM\, likely needs to produce more \OVI\, at large radii in order to produce a shallower \NOVI\, profile.

\subsection{OVI Conditions}\label{subsec:OVIconditions}
\begin{figure*}
    \centering
    \includegraphics[width=0.95\textwidth]{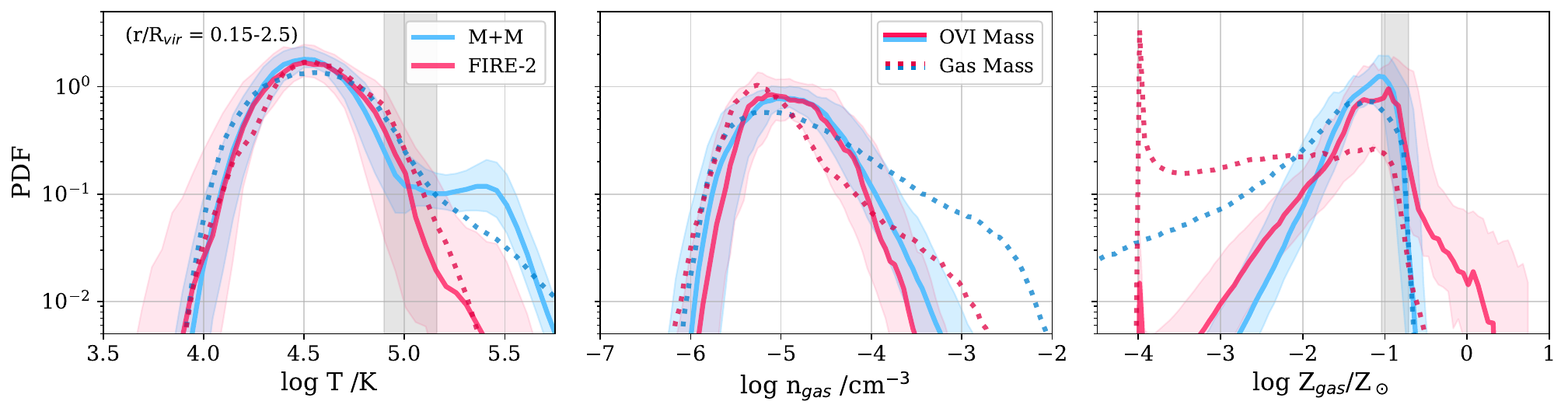}

    \caption{Probability density functions for gas temperature (left), density (center), and metallicity (right) of gas within $0.15 \leq r/R_{vir} \leq 2.5$. Each panel shows median \OVI\, mass-weighted PDFs (solid lines) and gas mass-weighted PDFs (dotted lines) for \MM\, (blue) and FIRE-2 (red) galaxies. Shaded regions represent the 16th-84th percentile for the \OVI\, mass-weighted distributions. The horizontal axes for each PDF display log T, log $n_{gas}$, and log $Z_{gas}/Z_{\odot}$, respectively. For reference, we include the $25-75\rm \,th$ percentile ranges for the virial temperatures and mean ISM ($r/R_{vir}\leq 0.1$) metallicities are shown as gray shaded regions in the left and right panels, respectively. \MM\, shows a secondary peak at $\rm log \,T \,/\rm{K} =  5.3-5.5$ in the temperature distribution and extends to higher densities in the gas mass-weighted PDF, indicating a more multiphase CGM compared to FIRE-2. Although the \OVI\, mass-weighted distributions similarly peak at $\rm log Z/Z_{\odot}  \sim -1 $, the gas mass-weighted distributions reveal that the CGM in \MM\, is more well-mixed in metals than in FIRE-2. }
    \label{fig:OVIprop_PDFcond}
\end{figure*}

Section \ref{sec:CGMproperties} demonstrated clear differences in the CGM in FIRE-2 and \MM, in particular, the abundance of oxygen mass as a function of radius, as well as the CGM physical conditions. In this section, we aim to disentangle whether the greater mass of \OVI\, in FIRE-2 seen in Section \ref{subsec:NOVI_MOVI}, is driven primarily by the larger oxygen mass (Figure \ref{fig:CGMprop_mass}) or the difference in physical conditions (Figure \ref{fig:CGMprop_cond}). Further, we present an analysis of how these physical conditions imprint on the \OVI\, ionization mechanisms.

Figure \ref{fig:OVIprop_PDFcond} presents the probability density functions (PDFs) for temperature (left), density (center), and metallicity (right) for gas within $0.15 \leq r/R_{vir} \leq 2.5$ for each suite. The left panel shows the temperature ($\rm log~ T$) PDF, the center panel shows the density (log $n_{gas}$) PDF, and the right panel shows the metallicity (log $Z_{gas}$) PDF relative to solar ($Z_{\odot}$). The solid lines represent the \OVI\, mass-weighted PDFs, and dotted lines represent the gas mass-weighted PDFs, where blue and red lines correspond to \MM\, and FIRE-2, respectively. The shaded regions show the 16th-84th percentile for the \OVI\, mass-weighted distributions.

Across the three properties presented in Figure \ref{fig:OVIprop_PDFcond}, we see that \OVI\, production is largely insensitive to the simulation suite. When comparing the peaks of the \OVI\,mass-weighted PDFs, both FIRE-2 and \MM\, primarily produce \OVI\, in cool/warm ($\rm log\, T\,/K \sim 4.5$), diffuse, ($\rm log\,n_{gas}\,/cm^{-3} \sim -5.0 $), and moderately metal-enriched ($\rm log Z/Z_{\odot}  \sim -1 $) material. This similarity in \OVI\ conditions across suites indicates that differences in CGM properties and subgrid physics have a limited impact on the characteristic physical conditions governing bulk \OVI\, production. However, there are clear differences in the wings of these distributions and the CGM gas mass-weighted PDFs. 

In the temperature PDF (left panel of Figure \ref{fig:OVIprop_PDFcond}), \OVI\, mass in \MM\, shows a secondary peak at $\rm log T\,/K \sim 5.3-5.5$, where \OVI\, ion fractions reach a maximum, which is not present in FIRE-2. The gas mass PDFs reveal that this bimodality in \MM\, is due to galaxies retaining a greater portion of gas mass in this temperature phase. In \MM, this phase of gas represents $0.1-2\%$ of CGM mass and is comprised of SNe-heated and ejected material that is actively cooling down to virial temperatures. This phase is also typically only present within $r/R_{vir}\leq 0.75$ \citep[see also][]{Piacitelli_25} since at larger distances this gas will have cooled out of this temperature phase and into the virial component of the CGM. In FIRE-2, gas in this temperature phase typically only comprises $0-0.01\%$ of CGM mass. However, this gas phase can exist out to $r/R_{vir}\leq 2.5$ for FIRE-2 galaxies, likely due to preventative feedback in FIRE-2 being more efficient at heating gas at large radii.

In the center panel of Figure \ref{fig:OVIprop_PDFcond}, the gas mass-weighted density PDFs for both suites show a broader bimodal distribution, with reservoirs of diffuse ($\rm log\, n_{gas} \sim -5$) and dense ($\rm log\, n_{gas} \sim -3$) gas. FIRE-2 galaxies tend to have a larger fraction of mass in the diffuse component compared to \MM, while \MM\, galaxies have a larger fraction of mass in their dense phase compared to FIRE-2. These findings are similar to those found in Figure \ref{fig:CGMprop_cond}, where the CGM in \MM\, tends to be denser within $r/R_{vir}\leq 1$.

The \OVI\, mass-weighted metallicity PDF (right panel of Figure \ref{fig:OVIprop_PDFcond}) shows that FIRE-2 spans a broader range in metallicities compared to \MM, extending to higher and lower metallicities. In particular, a fraction of \OVI\, in FIRE-2 is produced in near solar metallicity ($\rm log Z/Z_{\odot} \sim 0 $) material. We find that this high metallicity material is present at all radii out to $r/R_{vir} =2.5$. The preservation of this high metallicity material is related to both feedback and metal diffusion modeling. First, for this material to be present at all radii, feedback in FIRE-2 must be effective at ejecting ISM material to large distances. Second, this material also implies that ISM outflows---although it sweeps up CGM material as found in Section \ref{sec:CGMproperties}---largely retains its original ISM metal content rather than diffusing and mixing into the CGM, a finding presented in \citet{Pandya2021}.

When considering the gas mass-weighted PDF, we can see distinct differences in the CGM metal substructure across suites. FIRE-2 shows a flat distribution across $-3.9 \leq \rm log Z/Z_{\odot} \leq -1 $ and a strong peak of pristine IGM gas at $\rm log Z/Z_{\odot} \sim -4 $, corresponding to the implemented metallicity floor. The gas mass-weighted PDFs in \MM\, have a strong peak at $\rm log Z/Z_{\odot} \sim -1.5 $ and a steady drop-off to lower metallicities. These differences imply that the metal content of the CGM in \MM\, is more well-mixed than FIRE-2, where the majority of the CGM exists at a common metallicity. A direct comparison of metal diffusion coefficients between SPH and MFM simulations is not straightforward; however, these results are broadly consistent with FIRE-2 adopting a smaller coefficient in its metal diffusion subgrid model than \MM\, (Section \ref{sec:simulations}).

We stress that although these suites differ in how effectively metals mix into the CGM, this parameter space is largely unconstrained, and so we cannot make any argument on which modeling may be more physical. Rather, our work aims to understand how \OVI\, properties may trace CGM structure and modeling choices.

\subsection{OVI Ionization}\label{subsec:OVIIonization}

\begin{figure}
    \centering
    \includegraphics[width=0.4\textwidth]{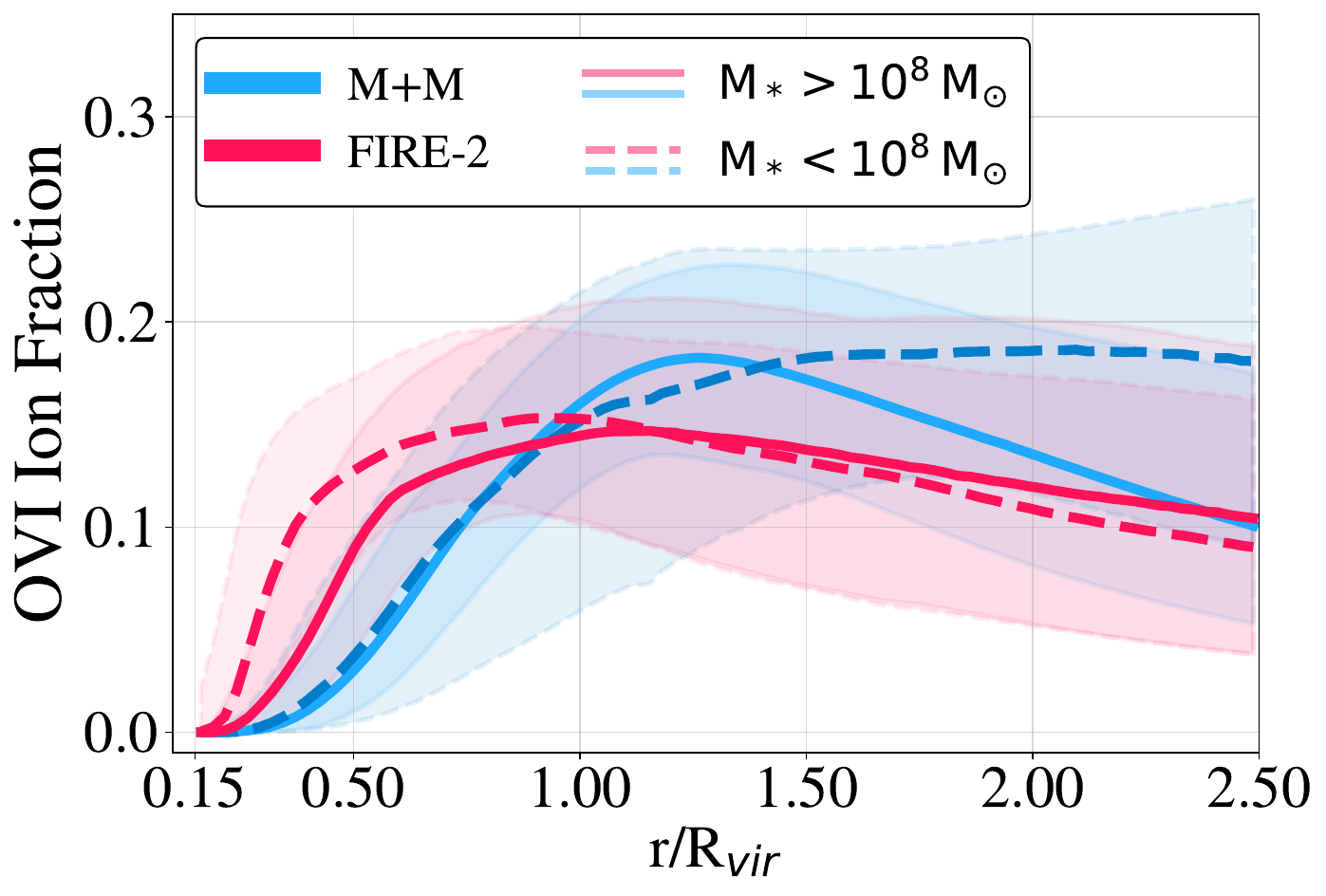}
    
    \caption{\OVI\, ion fractions as a function of radius. This shows the median fraction of oxygen that is in the \OVI\, state within a 100 equally sized radial bins within $0.15\leq r/R_{vir}\leq2.5$ and the 16th-84th percentile ranges as shaded regions.  Medians are calculated from the high-mass (solid lines) and low-mass (dashed lines) galaxies of either suite. These \OVI\, ion fractions are modeled by \cloudy\, and are based solely on the physical conditions of the gas. Differences in \OVI\, ion fractions across simulation suites reflect variations in CGM physical conditions but are small compared to the differences in CGM oxygen mass, indicating that the larger \OVI\, mass in FIRE-2 (Figure \ref{fig:OVIprop_MOVI}) is primarily driven by its larger oxygen reservoir (Figure \ref{fig:CGMprop_mass}).}
    \label{fig:OVIprop_ionfrac}
\end{figure}

The differences in CGM and \OVI\, conditions impact the ion fractions by setting how effectively \OVI\, will be ionized. Figure \ref{fig:OVIprop_ionfrac} shows the radial profile of \OVI\, ion fractions for either suite (colors as in previous figures) and split into low-mass (dashed lines) and high-mass samples (solid lines). Specifically, this plot shows, based on the physical conditions of the gas within a radial shell, the median fraction of oxygen mass that will exist in the \OVI\, state. These ion fractions allow us to discern whether the increased FIRE-2 \OVI\, mass is primarily due to the physical conditions of the gas or the mass of oxygen in the CGM. Both suites show small \OVI\, ion fractions within $0.15 R_{vir}$ and generally increasing \OVI\, ion fractions with increasing distance from the galaxy, driven by the decline in gas density. Across suites, the ideal \OVI\, conditions (where ion fractions are greatest) are achieved near or beyond  $r/R_{vir}\sim 1$. 

However, at $r/R_{vir}\sim0.5$, FIRE-2 galaxies tend to have \OVI\, ion fractions $2-3\times$ higher than \MM, because the inner CGM in FIRE-2 tends to be cooler and more diffuse than \MM\, (Figure \ref{fig:CGMprop_cond}). The \OVI\, ion fractions in FIRE-2 peaking at smaller radii explain the \OVI\, mass profile in Figure \ref{fig:OVIprop_MOVI}, where FIRE-2 galaxies produce more \OVI\, in the inner CGM. For $r/R_{vir}>1$, \MM\, galaxies have \OVI\, ion fractions $1-1.3\times$ higher than FIRE-2. Although CGM physical conditions vary across suites and give rise to differences in \OVI\, ion fractions, these differences are on the order of $\sim0.1$ and are therefore small in comparison to the differences in oxygen mass in the CGM (Figure \ref{fig:CGMprop_mass}). Thus, the differences in \OVI\, production across suites are primarily driven by the mass of oxygen in the CGM. 

\begin{figure}
    \centering
    \includegraphics[width=0.9\columnwidth]{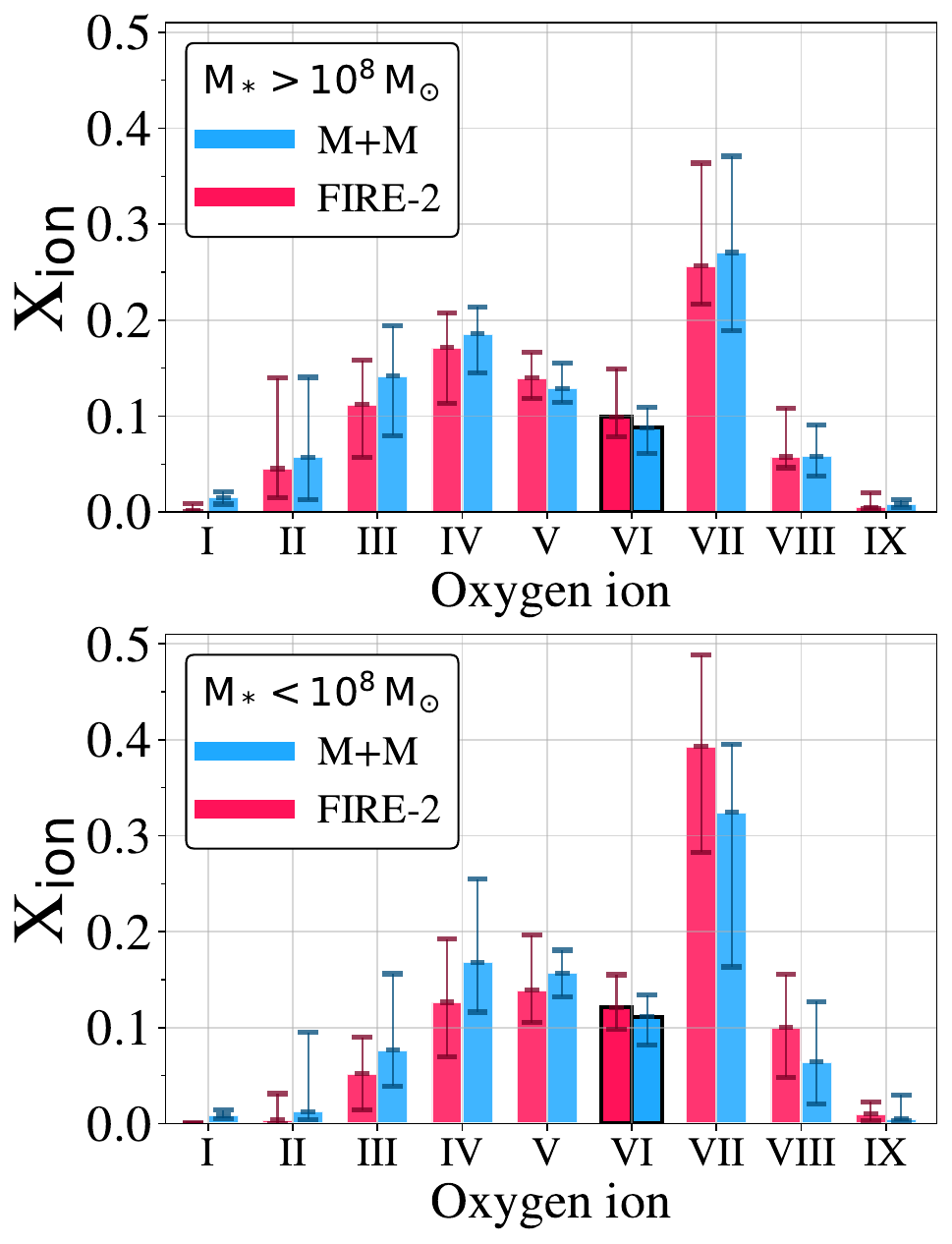}
    \caption{Total oxygen mass fractions ($X_{ion}$) distributed across ionization states within the CGM ($r/R_{\rm vir} = 0.15-2.5$) shown for both suites separated into high- and low-mass samples. Each bar represents the median mass fraction in a given ion, and errorbars represent the 16th-84th percentiles for each ion. Across suites, we find that a small fraction of oxygen resides in the \OVI\, state ($\sim10\%$) while a substantial fraction of material resides in \OVII\, as well as \OIV\, and \OV. }
    \label{fig:OVIprop_states}
\end{figure}

Differences in the physical conditions of the CGM will also affect the distribution of oxygen mass across its various ionization states. Figure \ref{fig:OVIprop_states} shows the fraction of oxygen mass, $f_{O}$, in each of its ionization states in the suites. For this analysis, all gas within $r/R_{vir} = 0.15-2.5$ of each galaxy is selected. For both suites, the scatter in $f_{O}$ (characterized by the error range on each bar in Figure \ref{fig:OVIprop_states}) for a given ion tends to encompass the median of the other suite, suggesting the two suites generally agree. However, we find slight differences in the values of $f_{O}$ across suites and differences in whether, in sum, higher or lower ionization states dominate the oxygen budget.

Across suites, \OVI\, tends to comprise a consistent $\sim10\%$ of the oxygen budget for all dwarf galaxies in FIRE-2 and \MM---notably a lower fraction than the ionization fractions presented in Figure \ref{fig:OVIprop_ionfrac} due to the large portion of oxygen mass existing at small radii (Figure \ref{fig:CGMprop_mass}). Ions like \OI, \OII, \OVIII, and \OIX\, each comprise $<10\%$ of the oxygen budget. Both suites agree well with one another that the vast majority of their oxygen budget is not in \OVI. In particular, $25-40\%$ of the oxygen mass resides in \OVII, $15-20\%$ in each of \OIV\, and \OV, and $\sim10\%$ in \OIII. 

\begin{figure}
    \centering
    \includegraphics[width=1\columnwidth]{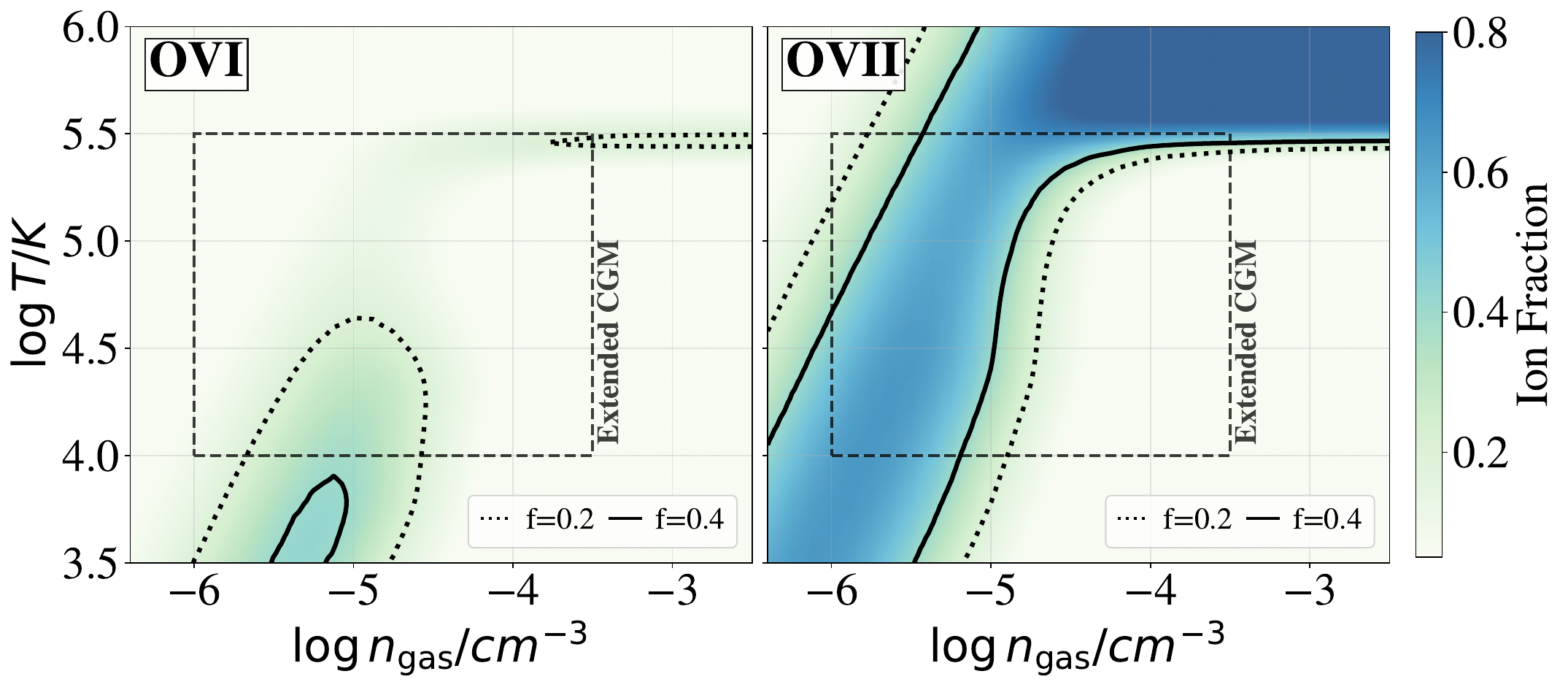}
    \caption{Ion fractions of \OVI\, and \OVII\, as a function of density and temperature taken from \textsf{TRIDENT} \citep{Trident} assuming the \citet{HM2012} UVB at $z=0.2$. Contours are added where the ion fractions reach $0.2$ (dotted) and $0.4$ (solid), and a rectangle is drawn to show the approximate phase space that the extended CGM of dwarf galaxies exists. \OVI\, ion fractions generally reach a maximum of $0.3$ for diffuse, cool gas. For the conditions where \OVI\, ion fractions are greatest, \OVII\, is greater, illustrating the large mass fractions in \OVII\, seen in Figure \ref{fig:OVIprop_states}.}
    \label{fig:FOVIMap}
\end{figure}

Ultimately, in both FIRE-2 and \MM\, the majority of oxygen is residing in \OVII, due to the stability of the \OVII\, electron configuration ($\rm1s^2$) and correspondingly low \OVII$\rightarrow$\OVI\, recombination rates \citep[e.g.,][]{Nahar_99}. This greater \OVII\, production can also be seen in Figure \ref{fig:FOVIMap}, which shows the ion fractions as modeled by \textsf{TRIDENT} for both \OVI\, and \OVII\, as a function of gas density and temperature. In the same phase of gas where \OVI\, ion fractions are highest, \OVII\, ion fractions are generally $2-4\times$ greater.

Significant fractions of oxygen also exist in \OIV\, and \OV. Low-mass FIRE-2 galaxies tend to have fractionally more oxygen in ionization states higher than \OVI\, compared to \MM, while the reverse is true for states lower than \OVI\, for high- and low-mass \MM\, galaxies. The dominance of higher ions in FIRE-2 is driven by the hotter outer CGM in FIRE-2.  The dominance of low ions in \MM\, is driven by the denser inner CGM (Figure \ref{fig:CGMprop_cond}).


Unlike lower ionization states, \OVI\, can be produced by both photoionization (PI) and collisional ionization (CI), depending on the phase of the gas. CI dominates at higher temperatures, almost independent of density, where the average velocity of particles is great enough to sufficiently ionize. The CI peak for \OVI\, can be seen in Figure \ref{fig:FOVIMap} where the dotted contours show \OVI\, ion fractions reach a local maximum of 0.2 at $T\approx10^{5.5}\rm K$ and $n>10^{-4}\rm cm^{-3}$.  PI from the UVB dominates in cooler and diffuse gas, seen in Figure \ref{fig:FOVIMap} where the \OVI\, ion fractions reach a second maximum for $T<10^4\,\rm K$ and $n<10^{-4.5}\rm cm^{-3}$: notably reaching higher ion fractions than the CI phase. Importantly, there is no well-defined phase of gas where the dominant mechanism transitions from one to the other. Rather, at intermediate temperatures, both CI and PI play a role in placing oxygen into the \OVI\, state. 

To determine the fraction of \OVI\, mass produced by CI and PI, we utilize data from \citet{SD93} assembled into tabular form \citep{Smith_17} which provides the ion fractions of various elements based solely on collisional ionization. Since this modeling excludes photoionization, the ion fractions have no redshift dependence. To simplify calculations, the model only depends on temperature and not density, as the temperature is the first-order driver of collisional ionization. 

Using these tables, we determine the mass of collisionally ionized \OVI\, on a particle-by-particle basis. To then determine the mass of photoionized \OVI, we subtract the mass of collisionally ionized \OVI\, from the total \OVI\, mass given by the CI+PI \cloudy\, tables. By default, \cloudy\, will model both of these ionization processes. We have redone these calculations using the density-dependent temperature threshold for CI adopted in \citet{RocaFabrega_19}, which classifies the mass of \OVI\, on a particle-by-particle basis as either fully CI or PI. Although the same general findings persist with both methods, we adopt our method using \citet{SD93} data as it better captures the gradation of ionization mechanisms in temperature space. 

\begin{figure}
    \centering
    \includegraphics[width=0.4\textwidth]{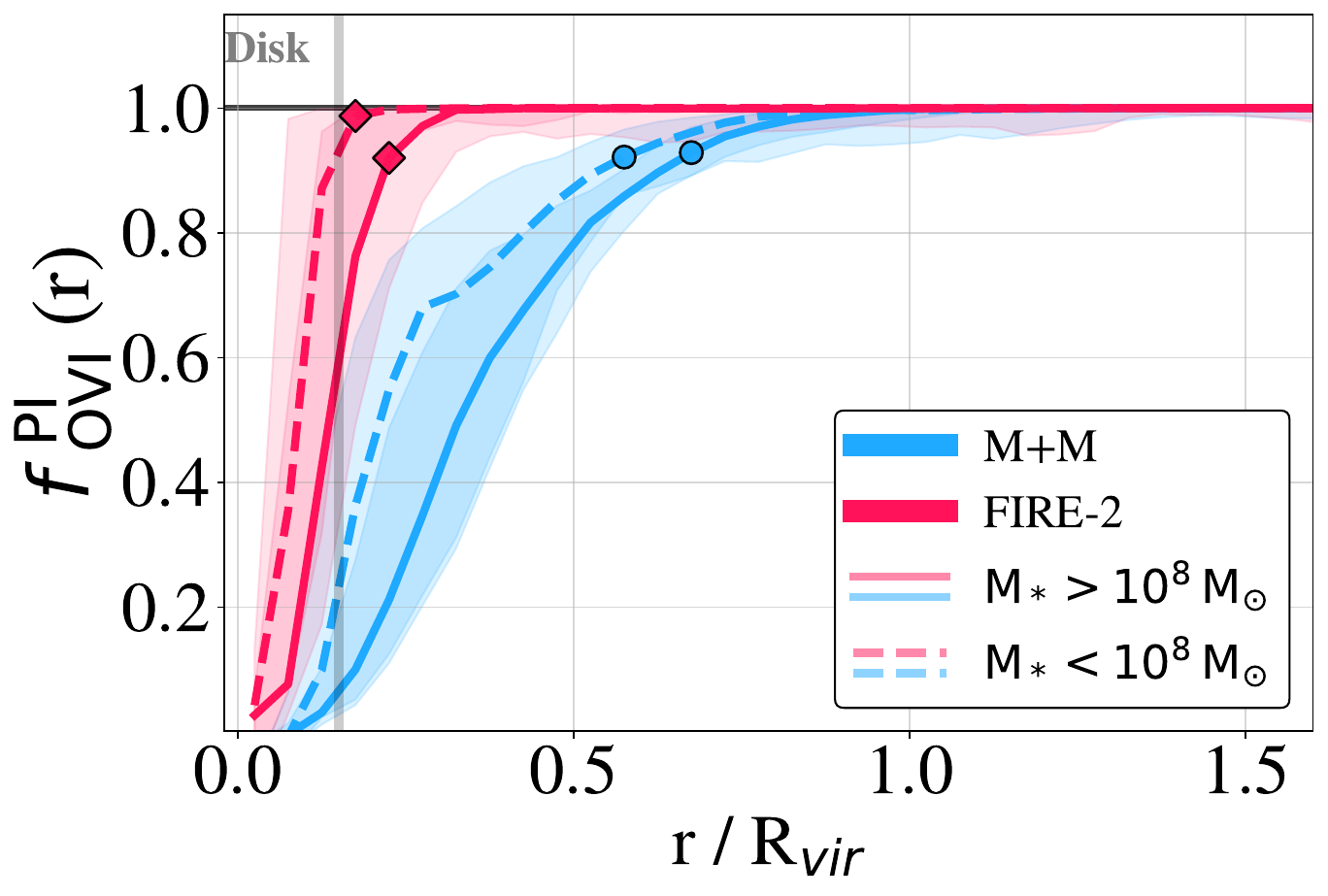}
    \caption{\OVI\, ionization mechanism profiles for the \MM\, (blue, circle) and FIRE-2 (red, diamond) samples.  Across normalized radial distance, we show the fraction of \OVI\, produced by PI ($f^{\rm PI}_{\rm OVI}(r)$) for the high-mass (solid) and low-mass (dashed) galaxies. $f^{\rm PI}_{\rm OVI}(r)$ describes the fraction of \OVI\, mass within a given shell ($dr = 0.05\,R_{vir}$) that is produced via PI.  The solid line denotes the median value for a given suite, and the shaded region shows the 16th-84th percentiles. For $r/R_{vir}<0.5$, \MM\, galaxies produce less \OVI\, via PI (and therefore more CI \OVI) than FIRE-2 in the inner CGM, which is caused by feedback driving differences in CGM phase (Figures \ref{fig:CGMprop_cond} and \ref{fig:OVIprop_PDFcond}). However, globally, both suites produce the vast majority of \OVI\, via PI, especially at large radii.}
    \label{fig:OVIprop_ionmech}
\end{figure}

Given the lack of a significant hot ($T \sim 10^{5.5} ~\mathrm{K}$) CGM reservoir---where CI dominates \OVI\, ionization---$\gtrsim90\%$ of the \OVI\ mass in both suites is produced via PI. However, as noted in \citet{RocaFabrega_19}, the radial dependence of the \OVI\, ionization mechanism differs by suite. Figure \ref{fig:OVIprop_ionmech} shows the fraction of PI \OVI\, ($f^{\rm PI}_{\rm OVI}$) as a function of radius (i.e., the fraction is relative to the amount of \OVI\, in that shell) with the high- and low-mass samples being shown as solid or dashed lines. Markers for each line denote the radius at which $f^{\rm PI}_{\rm OVI}=0.9$. Broadly, both suites show an increasing $f^{\rm PI}_{\rm OVI}$ profile with increasing $r$ and find $f^{\rm PI}_{\rm OVI}=1$ for $r/R_{vir}>1$, indicating that the majority of \OVI\, is produced via PI. 

Differences between suites arise in the inner CGM. \MM\, tends to have a lower $f^{\rm PI}_{\rm OVI}$ value than FIRE-2 within $r/R_{vir}\sim 0.75$ for both high- and low-mass galaxies. The lower $f^{\rm PI}_{\rm OVI}$ values indicate \MM\, produces more \OVI\, via CI than FIRE-2. The larger CI \OVI\, phase in \MM\, is due to a significant portion of \OVI\, mass in the inner CGM existing in hotter gas, $T\sim10^{5.5}\,\rm K$, seen in Figure \ref{fig:OVIprop_PDFcond}.

This radial dependence has been previously noted in MW-mass halos in the VELA simulations. \citet{Strawn_21} finds that within the inner CGM of MW-mass halos, \OVI\, is primarily produced in hot outflows (via CI), but there is a radial dependence where at higher radii a cool PI medium is the primary source of \OVI. \MM\, galaxies tend to be consistent with this picture; however, due to the dearth of $\rm log\, T\,/K \sim 5.5$ gas, FIRE-2 has a larger PI \OVI\, reservoir at all radii.

\section{Discussion}\label{sec:discussion}

As shown in Section \ref{sec:OVIobservations}, the underprediction of \OVI\, columns in both suites appears to be primarily driven by a lack of oxygen mass in the simulated CGM. Although a redistribution and enhanced ionization of oxygen can achieve observationally consistent \MOVI\, values for higher mass dwarfs, lower-mass dwarfs would require evacuating their ISM in order to produce \MOVI\, within $1\sigma$ of the observations. 
Since these galaxies would require $0\%$ disk metal retention to possibly replicate observed \OVI\, masses, we argue that these galaxies show signs of a Simulated Oxygen Shortage (SOS) such that they have likely underproduced oxygen over their star formation histories. Sections \ref{sec:CGMproperties} and \ref{sec:OVIproperties} reinforce that distinct differences in CGM properties and structure have little effect on the global production of \OVI. Rather, the enhanced \MOVI\, in FIRE-2 at all radii are the dominant drivers behind the greater \NOVI\, values compared to \MM\, in Figure \ref{fig:CompObs_NOVI}.

In Section \ref{subsec:disc_vary}, we explore how UVB modeling and global CGM gas phase impacts the total mass of \OVI\, produced. Section \ref{subsec:quantgap} quantifies the mass and ccSNe needed to resolve the underprediction of \OVI\, columns under the assumption that star formation models and ccSNe yields are at the core of this issue. Section \ref{subsec:quantgap} then discusses constraints/complications with this increase in the feedback and oxygen budget and identifies possible solutions to the SOS.

\subsection{OVI Dependence on UVB \& CGM Conditions}\label{subsec:disc_vary}

\subsubsection{Varying the UVB}
In this section, we briefly explore the impact that adopting the \citet[][hereafter, \citetalias{HM2012}]{HM2012} UVB may have on \OVI\, production across suites. We compare the effects of switching from \citetalias{HM2012} to the UVB adopted by FIRE-2 during runtime: \citet[][hereafter, \citetalias{FaucherGiguère2009}]{FaucherGiguère2009}. 

\begin{figure}[!t]
    \centering
    \includegraphics[width=0.4
    \textwidth]{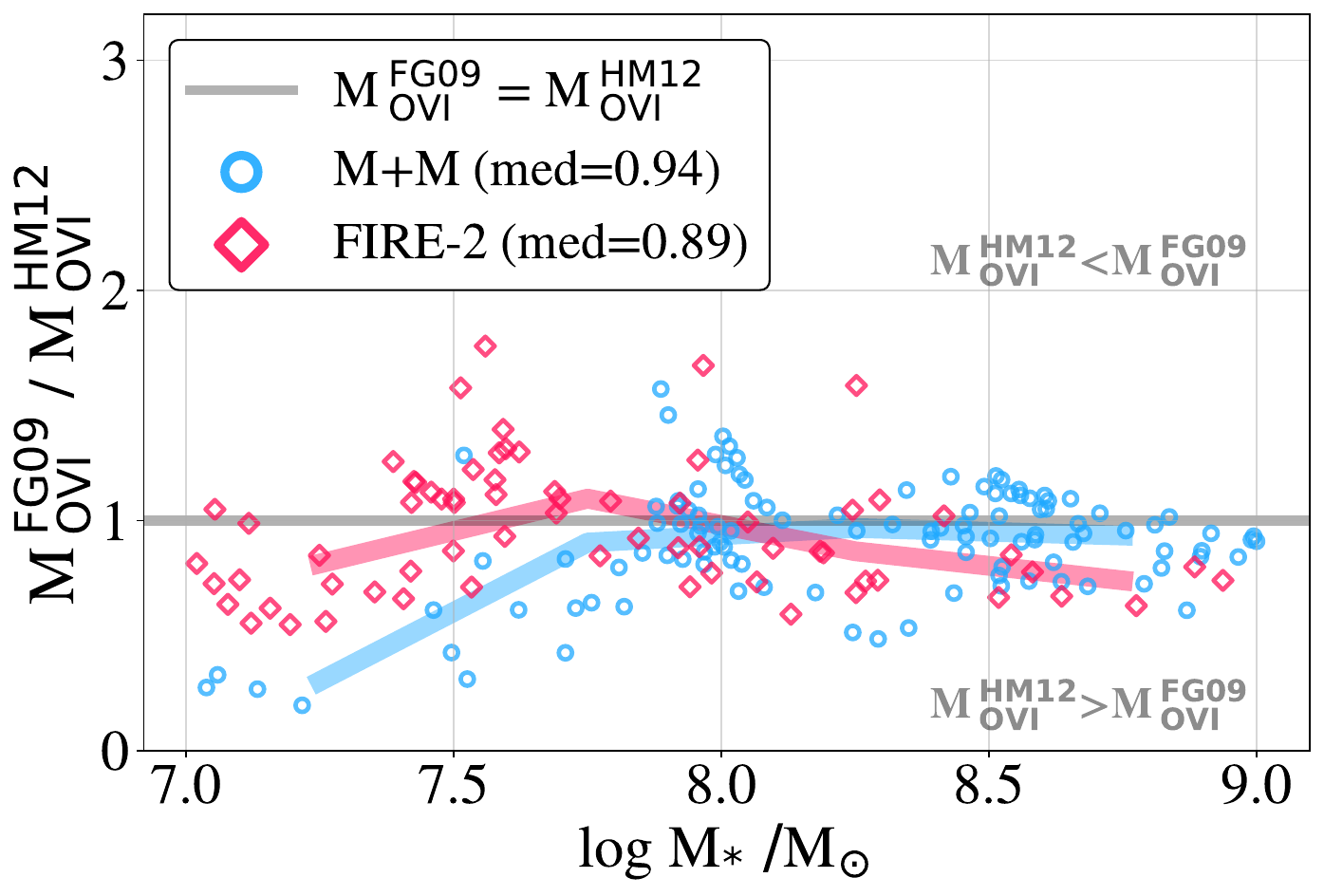}

    \caption{Effects of UVB choice on total \MOVI. Individual points represent the ratio of total \OVI\, mass as computed with \citetalias{HM2012} ($\rm M\,_{\rm OVI}^{\rm HM12}$, the fiducial UVB adopted in this work) versus \citetalias{FaucherGiguère2009} ($\rm M\,_{\rm OVI}^{\rm FG09}$) for a single CGM. Solid lines represent the median values for both suites in 4 equally sized stellar mass bins, while the median of the full suite is provided in the legend. We find the choice of \citetalias{HM2012} over \citetalias{FaucherGiguère2009} generally increases \MOVI, but this difference is typically small with \citetalias{FaucherGiguère2009} reducing \MOVI\, by a factor of 0.9.}
    
    \label{fig:Disc_UVB}
\end{figure}

Figure \ref{fig:Disc_UVB} shows the change in total \OVI\, mass when switching the assumed UVB when modeling \cloudy\, ion fractions between \citetalias{HM2012} and \citetalias{FaucherGiguère2009}. We find individual galaxies can experience a boost or reduction in \MOVI\, when switching between models, but both suites generally produce more \OVI\, with \citetalias{HM2012} over \citetalias{FaucherGiguère2009}. In particular, for high-mass galaxies, \citetalias{HM2012} yields $1.03 - 1.07\times$ and $1.15 - 1.35\times$ more \OVI\, for \MM\, and FIRE-2, respectively. At the lowest mass end, \OVI\, production strongly prefers \citetalias{HM2012} with some \MM\, galaxies experiencing a $3.34\times$ increase in $M_{\rm OVI}$, while FIRE-2 galaxies are enhanced to a lesser degree ($1.23\times$). These changes in \MOVI\, with UVB are due to the differences in the ion fractions as a function of CGM phase. FG09 \OVI\, ion fractions peak at slightly lower densities but comparable temperatures to \citetalias{HM2012}. This difference results in the \OVI\, mass produced by \citetalias{FaucherGiguère2009} being on average $\sim1.5\times$ more diffuse than \citetalias{HM2012}. These findings are in line with \citet{Taira_25}, a recent work from the FOGGIE collaboration, which found that the \citetalias{HM2012} UVB (and another UVB from \citet{Puchwein19}) enhances \NOVI\, compared to \citetalias{FaucherGiguère2009}. The authors explain that the larger \NOVI\, values are due to the UVB model having a greater flux at the ionization energy of \OVI.

Although the choice of UVB between \citetalias{HM2012} and \citetalias{FaucherGiguère2009} does change the total \MOVI\, in both suites, whether this change is an enhancement or reduction varies from galaxy to galaxy. Furthermore, the magnitude of the enhancement is typically less than a factor of $2$, with \citetalias{HM2012} (the UVB adopted throughout this work) producing the highest \MOVI\, values. This brief investigation demonstrates that the choice of \citetalias{HM2012} or \citetalias{FaucherGiguère2009} UVBs has a relatively minor effect on the total \MOVI, although we discuss how further modifications to the UVB modeling may lead to larger enhancements in Section \ref{subsec:quantgap}.

\subsubsection{Varying the CGM Phase}\label{subsubsec:phasevary}
\begin{figure}[!t]
    \centering
    \includegraphics[width=0.47\textwidth]{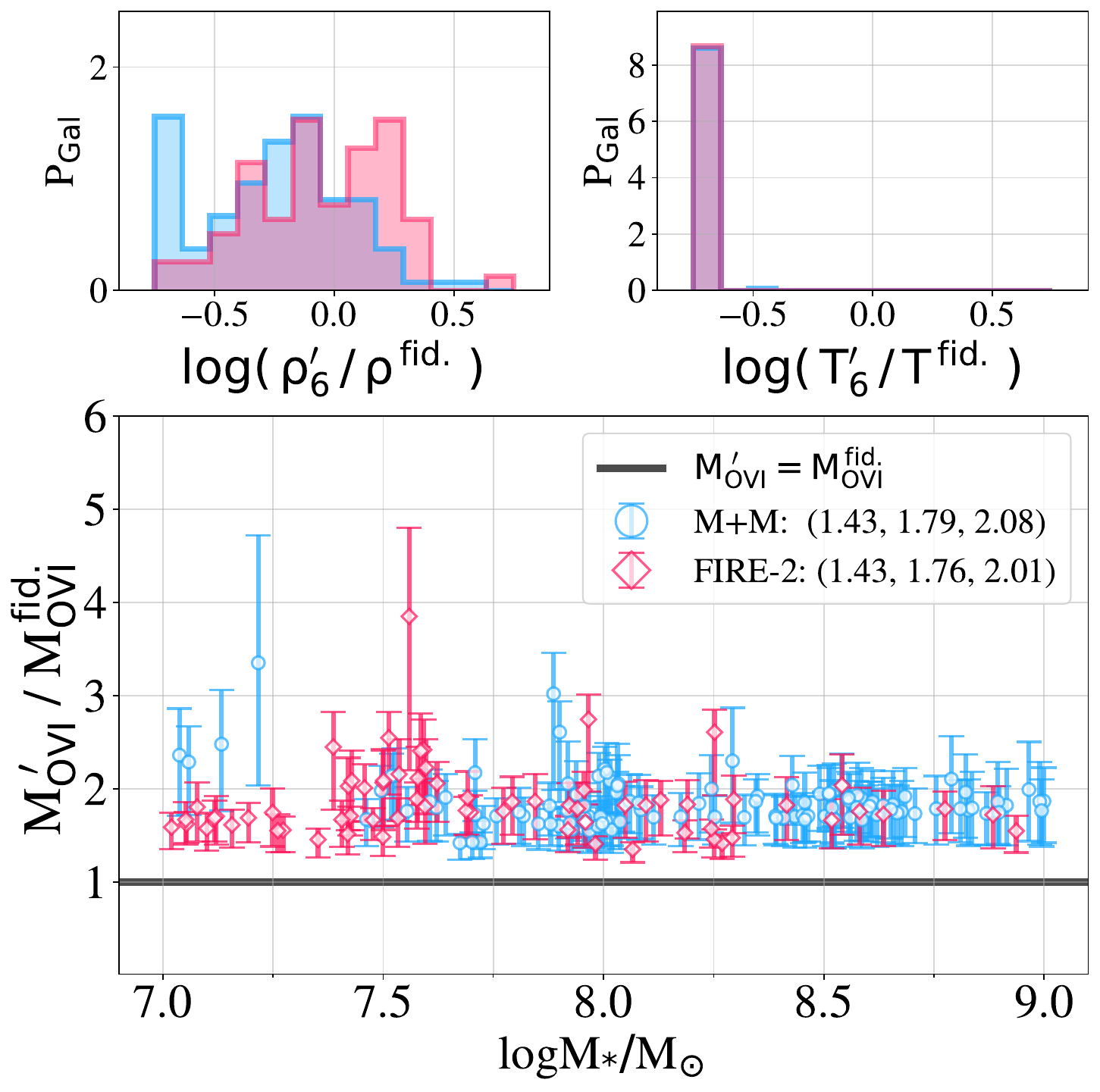}
    \caption{Effects of CGM phase permutations and their impact on \MOVI. Diffuse CGM gas ($n < 6 \times 10^{-4}\ \mathrm{cm}^{-3}$, $0.15 \leq r/R_{\rm vir} \leq 2.5$) is rescaled in density and temperature by factors of up to $2$, $3$, and $6$ to assess how global phase changes affect the total \MOVI. \textit{Top row:} Distributions of the density (\textit{left}) and temperature (\textit{right}) scaling factors that maximize \MOVI\, for the full factor-of-$6$ variation.\textit{ Bottom row:} Maximum \MOVI, obtained across all density–temperature permutations. Markers indicate the maximum values for factor-of-$3$ variations, while error bars span the range from factor-of-$2$ to factor-of-$6$. Median values for each variation are listed in the legend. Simulations can enhance \MOVI\, by $<2\times$ under these permutations; however, it would require a $6\times$ global cooling of the CGM. }
    \label{fig:Disc_CGMtoggle}
\end{figure}

The thermal state of CGM gas will set the ion fractions; thus, it is important to understand how varying the CGM physical conditions may change \MOVI. In this section, we provide a brief investigation of how altering the global CGM physical state may enhance \MOVI.

Here, we select all diffuse gas ($n<6\times 10^{-4}\, \rm cm^{-3}$) within $0.15 \leq r/R_{vir} \leq 2.5$ of each galaxy and permute the density and temperature of the gas to see how the total \MOVI\, changes. This selection criteria removes cool/dense gas from this analysis and selects the diffuse/virialized component of the CGM, as this is where the bulk of \OVI\, arises (Figure \ref{fig:OVIprop_PDFcond}).

For each CGM, we adjust the gas temperature and density by uniform factors to map the effects of CGM phase on \MOVI\, production. We consider three ranges of variation, allowing the CGM properties to vary within factors of $2$, $3$, and $6$ about their fiducial values. For each range, we sample temperature with 15 logarithmically spaced factors and density with 7 logarithmically spaced factors. Of all permutations, we select the conditions that produce the greatest \MOVI, and plot these ideal, fine-tuned factors in Figure \ref{fig:Disc_CGMtoggle}.

The upper left and upper right panels of Figure \ref{fig:Disc_CGMtoggle} show histograms of the ratios of ``the most ideal'' \OVI\, CGM densities and temperatures relative to the fiducial values for the factor-of-$6$ variation. Although neither suite uniformly favors an increase or decrease in density, both suites overwhelmingly favor a reduction in CGM temperature at the extreme of the varied range (reduced by a factor of $10^{0.75}$). This bias towards a CGM temperature reduction is due to the already large fraction of \MOVI\, which is produced via PI (see Section \ref{subsec:OVIIonization} and Figure \ref{fig:FOVIMap}) being enhanced as the CGM temperature is decreased. Physically, this cooling of the CGM would likely require either increased metal cooling rates or decreased energy loading into the CGM. However, cooling the CGM to this degree would be a departure from virialization of the halo, as we are cooling the virialized CGM component to sub-virial temperatures.

The resulting total \MOVI\, for the factor-of-$2$, factor-of-$3$, and factor-of-$6$ variation is shown in the bottom panel of Figure \ref{fig:Disc_CGMtoggle}. For each idealized \MOVI\, value, we divide this by the fiducial \MOVI. We find that even with a factor-of-$6$ variation, simulations can typically only double their \MOVI\, reservoirs. Although this brief investigation proves the phase of CGM gas can enhance the total $M_{\rm OVI}$ somewhat, we emphasize that only under a rather extreme cooling of the CGM do we find noticeable enhancement in the total \MOVI.

\subsection{Toward Resolving the OVI Deficit}\label{subsec:quantgap}

In this section, we move beyond identifying the \OVI\, deficit to exploring possible avenues for resolving it. We emphasize that none of the solutions discussed here are straightforward or well-understood at the current point in the field. Each avenue carries significant potential complications that would likely impact other well-established aspects of galaxy evolution. We first briefly discuss unmodeled physical processes that may support \OVI\, production. We then turn to what we consider to be the fundamental issue: the Simulated Oxygen Shortage (SOS). Under the assumption that oxygen underproduction is the primary driver of the \OVI\, deficit, we estimate the total oxygen mass needed to exactly reproduce \MishraMOVI, and then convert this into an approximate additional rate of ccSNe. We show that while such increases are motivated by recent observational work, they would likely introduce serious tensions with galaxy scaling relations and the broader baryon cycle. The preliminary discussion presented here is aimed at quantifying the scale of the problem and motivating future work that can more thoroughly explore this parameter space. 

\subsubsection{Possible Effects from Unmodeled Physics}\label{subsec:stepsforward}

Based on previous works, we begin by discussing the potential effect that increased numerical resolution and the inclusion of cosmic rays (CR) may have on the \OVI\, production.


\textit{Enhancing Resolution:} Currently, there is not a consensus on the impact of resolution on \OVI\, production. Although we do not see trends with \MOVI\, and simulation resolution, the resolution for both suites may be too coarse to resolve the true physical origins of \OVI. \citet{Rey2024} found that increased spatial resolution of outflowing gas (down to $18\,\rm pc$) in the inner CGM of a $M_* = 10^8 M_{\odot}$ led to a $50\%$ increase in the covering fraction of sightlines with $N_{\rm OVI} \geq 10^{13} \rm cm^{-2}$. This increase is driven by the increased resolution, enabling the outflowing gas to reach higher temperatures and stay hotter for longer. Despite the increase in \NOVI, the simulations did not produce $N_{\rm OVI}\sim 10^{14} ~\rm cm^{-2}$ and thus are still underpredicting \NOVI. However, other works have found that increased resolution at large radii leads to slightly lower \NOVI\, values \citep[e.g.,][]{Hummels19} or relatively the same \NOVI\, values \citep[e.g.,][]{Lucchini26}. These results imply that increased resolution in the current simulations may increase the fraction of CI \OVI. However, if the bulk of \OVI\, arises from large-scale virialized gas and is photoionized, then the resolutions of FIRE-2 and \MM\, are likely sufficient to resolve this phase and are not the main drivers of the gap with observations.

\textit{Cosmic Rays:} Current uncertainties in CR modeling have yielded differing conclusions regarding their role in dwarf galaxies. Previous work from IllustrisTNG \citep{Ramesh_24} has demonstrated that cosmic rays are unlikely to substantially modify the CGM of dwarf galaxies. On the other hand, using the SURGE simulations, \citet{Bieri_26} finds that for $M_{vir}<10^{12}M_{\odot}$ halos, CRs are especially important regulators of both star formation and CGM properties, driving stronger outflows at $R_{vir}$. Results from FIRE-2 \citep{Hopkins20,Ji_20, Li_21} find that CRs are generally negligible for $M_{vir}<10^{11}M_{\odot}$ halos and minimally alter galaxy properties. Yet, for $M_{vir}> 10^{11}M_{\odot}$ halos, CRs can support up to $3\times$ higher \OVI\, column densities due to CR pressure changing the CGM phase non-linearly \citep{Ji_20}. Further, \citet{Li_21} finds CRs produce slightly higher \OVI\, equivalent widths for a $M_{vir}=10^{11.2}~M_{\odot}$ halo. These differences in the role of CRs in dwarf galaxies are likely driven by differing implementations of CR physics and assumptions about CR diffusion into the CGM. In fact, previous work has demonstrated that the effects on the CGM are sensitive to differing CR diffusion coefficients and the inclusion/exclusion of different processes (i.e., Alfvén cooling) \citep[e.g.,][]{Hopkins20, Buck_20, Bieri_26}. Given the uncertainties in modeling, CR physics may play an important role in enhancing \OVI\, production in the CGM via altering the CGM phase or increasing the metal content. 

\textit{Radiative Transfer:} Given the majority of the simulated \OVI\, content is produced via PI (Section \ref{sec:OVIproperties}), an enhanced UVB is likely to boost the \OVI\, content. Section \ref{subsec:disc_vary} demonstrates that the total \MOVI\, changes relatively little between \citetalias{HM2012} and \citetalias{FaucherGiguère2009}. However, if these models underpredict the flux of ionizing photons, both may underproduce \OVI; although notably \citetalias{HM2012} already adopts the highest flux at the ionization energy of \OVI\, ($138\rm\, eV$) compared to \citetalias{FaucherGiguère2009} and the other models included in \citet{Taira_25}.  In this direction, explicitly modeling the radiative transfer (RT) of ionizing flux from the galaxy may boost \OVI. \citet{2025arXiv250819396B} finds minimal enhancements to \OVI\ column densities when modeling RT for both UVB and stellar radiation, which may reflect the limited stellar flux at $138\rm eV$ \citep[e.g.,][]{Baumschlager_24}. This could be enhanced due to changes in the assumed stellar binary population or if RT from AGN is also explicitly modeled. Observational work has sought a link between the presence of a supermassive black hole (SMBH) and \CIV\, CGM column densities for dwarf galaxies, although no identifiable trend has been found \citep{Garza_24}. Nonetheless, \OVI\, may be more sensitive to the local AGN ionizing radiation than \CIV\, and the inclusion of this feedback channel could boost ion fractions. This, however, is reliant on the SMBH actively accreting, and the ionizing flux reaching out to large distances where \OVI\, resides.
 
\subsubsection{Increasing Oxygen Budget \& Caveats}
\begin{figure}
    \centering
    \includegraphics[width=0.95\linewidth]{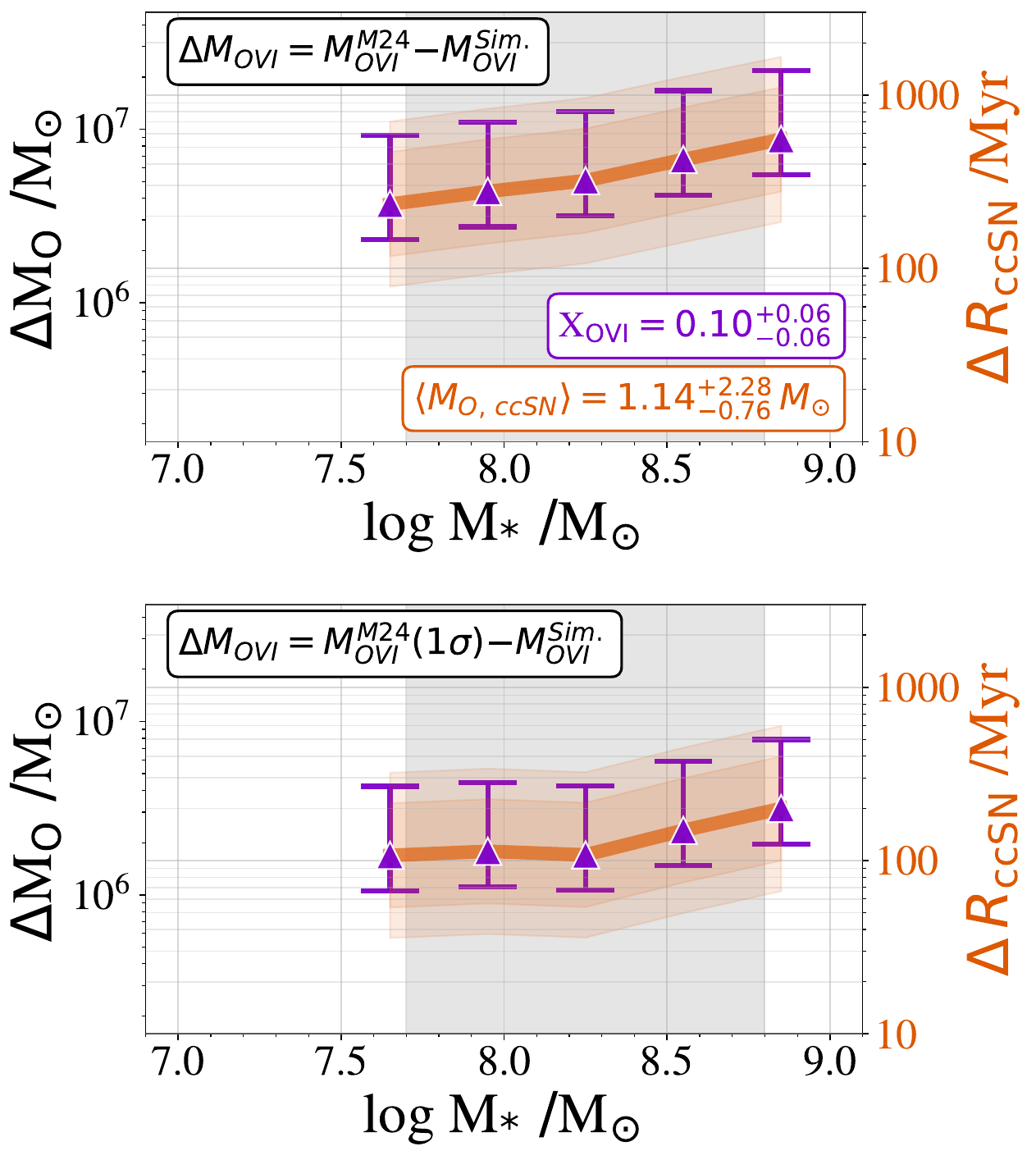}
    \caption{Additional mass of oxygen ($\Delta M_{\rm O}$, left axis, purple points) and rate of ccSNe per Myr ($\Delta R_{\rm ccSN}$, right axis, orange lines/regions) needed to match the median \MishraMOVI\,(top panel) and the $1\sigma$ \MishraMOVI\, lower limit (bottom panel). These values are derived following the steps in Section \ref{subsec:quantgap} and assuming $\Delta\text{\MOVI\,}=10^{5.5-5.9}M_{\odot}$ and $\Delta\text{\MOVI\,}=10^{5.2-5.5}M_{\odot}$ is needed to match the median and $1\sigma$ lower limit of \MishraMOVI, respectively. $\Delta M_{\rm O}$ adopts a value of $0.10$ for the fraction of the CGM oxygen budget residing in \OVI\, ($X_{\rm OVI}$) and varies it between $0.04-0.16$, shown as errorbars (see also Figure \ref{fig:OVIprop_states}). $\Delta R_{\rm ccSN}$ assumes the average mass of oxygen produced in a ccSN ($\langle M_{\rm O, ccSN}\rangle$) is $1.14 ~M_{\odot}$ (solid orange line) and varies this value by $2\times$ and $3\times$ (orange shaded regions).}
    \label{fig:Disc_MassNeeded}
\end{figure}

As shown in Figure \ref{fig:CompObs_Mass}, galaxies in both suites require an increase in their \MOVI\, to match \MishraMOVI. In terms of total \OVI\, mass needed ($\Delta\text{\MOVI\,}$), to match the median \MishraMOVI\,, simulated $7.5 \leq {\rm log}(M_*/M_{\odot}) \leq 9.0$ galaxies would need  $\Delta\text{\MOVI\,}=10^{5.6-5.9}M_{\odot}$. To place the simulated \MOVI\, values within the $1\sigma$ of \MishraMOVI, an additional $\Delta\text{\MOVI\,}=10^{5.2-5.5}M_{\odot}$ would be needed. As we compare with observational results from \citetalias{Mishra_24}, we focus this investigation on $7.5 \leq {\rm log}(M_*/M_{\odot}) \leq 9.0$ galaxies. We compare a slightly broader mass range than considered in Section \ref{sec:OVIobservations} under the assumption that the underproduction of oxygen likely extends beyond the range highlighted in Figure \ref{fig:CompObs_Mass}.

Assuming \OVI\, comprises some fraction, $X_{\rm OVI}$, of the total oxygen budget, the additional oxygen mass needed in the CGM is found using $\Delta M_{\rm O} = \Delta\text{\MOVI\,} /X_{\rm OVI}$. Figure \ref{fig:Disc_MassNeeded} presents the estimated $\Delta M_{\rm O} (\rm M_{\odot})$ (left axis, purple points). A value of $X_{\rm OVI}=0.10$ is adopted to match $X_{\rm OVI}$ in the current simulations (Figure \ref{fig:OVIprop_states}), and a range of $X_{\rm OVI}=0.04 - 0.16$ is shown as errorbars.  Assuming $X_{\rm OVI}=0.10$, $\rm M_* \sim 10^{8.3} M_{\odot}$ galaxies likely require $\Delta M_{\rm O}\approx 3\times 10^6 M_{\odot}$ or $\Delta M_{\rm O}\approx 1\times 10^6 M_{\odot}$ to match the median or lie within $1\sigma$ of \MishraMOVI, respectively. Increasing $X_{\rm OVI}$ by 0.06 reduces this mass slightly ($\Delta M_{\rm O}\approx 2\times 10^6 M_{\odot}$ to reach the median) while reducing $X_{\rm OVI}$ by 0.06 increases this mass significantly to $\Delta M_{\rm O}> 10^7 M_{\odot}$ to reach the median. As simulations aim to close this gap with \OVI\, observations, it will likely be critical to preserve similar or achieve higher $X_{\rm OVI}$ values in the CGM while also increasing $ M_{\rm O}$.

It is critical to address the potential ramifications of increasing the CGM oxygen content (and therefore the CGM metallicity) to this degree. First, enhanced metallicity will also enhance CGM gas cooling times, which could also: a) reshape the phase of the CGM by supporting more efficient cooling to lower temperatures, b) enhance inflow processes feeding the galaxy through time through faster accretion, and c) reduce outflows, potentially enhancing recycling and reducing the large-scale metal distribution. Together, in addition to changing the baryon cycle of the galaxy, these changes would also likely change CGM column densities across various ions, which are currently in broad agreement with low ion observations for the \MM\, suite. \citep{Piacitelli_25}

Further complications arise when considering that both suites tend to reasonably reproduce the stellar mass-metallicity relation (MZR) (\citealp[{\rm FIRE:}][]{Ma_16, Escala2018}; \citealp[{\rm M+M:}][]{Piacitelli_25}) and gas-phase MZR (\citealp[{\rm FIRE:}][]{Ma_16, Porter22}; \citealp[{\rm M+M:}][]{Christensen2016, Azartash-Namin24}). Because these relations probe distinct baryonic components and are calibrated to different elements (e.g., oxygen versus iron), their agreement with observations has been interpreted as evidence that the simulated disks retain a realistic fraction of metals. Although the observed MZR scatter permits a range of disk metallicities, increasing the oxygen production would likely still require enhanced inflows and outflows to both match \MishraMOVI\, and maintain a similar disk metal retention.

\subsubsection{Increasing Core-Collapse SNe Rates \& Caveats}

As oxygen is understood to be primarily produced/distributed by ccSNe, the SOS may likely imply missing high-mass stars or an underproduction of oxygen for a given ccSN, or a combination of both. Further, simply matching the \MishraMOVI\, alone is not sufficient; \NOVI\, columns must agree across observations and simulations. Observations detect \OVI\, out to large distances (Figure \ref{fig:CompObs_NOVI}), thus more oxygen is likely required to be produced \textit{and} distributed out to large radii. 

Metal transport driven by feedback is necessary to distribute metals to large radii and reproduce the observed \NOVI\, across impact parameters, while metal production provides the necessary oxygen to distribute. To first order, increased oxygen production would require more ccSNe than are being simulated currently or an increase in the metal yield per ccSN. In the SNe community, \citet{Pessi25} find that the rate of ccSNe increases with decreasing stellar mass in the dwarf regime. In particular, \citet{Pessi25} finds that $M_{*} = 10^{7.8-8.4}M_{\odot}$ galaxies experience approximately $2-6\times$ more ccSNe per unit stellar mass than a $M_{*} = 10^{9.7}M_{\odot}$ galaxy  \citep[e.g., M33;][]{Corbelli_14}. This inverse trend implies that the stellar IMF may be dependent on host galaxy/local ISM properties like metallicity. Evidence for a metallicity-dependent IMF has also been presented in previous findings \citep[e.g.,][]{Geha13,Li_23,Pessi23}. However, we emphasize that a non-universal IMF remains a topic of ongoing debate, with no clear consensus in the community.

Nonetheless, implementing a metallicity-dependent IMF in simulations would likely result in larger oxygen gas reservoirs being produced, and the increased number of ccSNe would supply additional feedback energy to distribute these metals into the outer CGM; yet, issues regarding the increased budget are non-trivial (see below for further discussion). Indeed, metallicity-dependent SNe rates have already been explored in FIRE-2. \citet{Gandhi_22} investigated models in which the SN Ia rate increases with decreasing metallicity, resulting in better agreement with stellar mass-metallicity relations in low-mass galaxies with minimal changes in stellar masses. Further work applying a similar framework to ccSNe rates may hold promise in closing this gap with \OVI\, observations. 

In this direction, we convert the $\Delta M_{\rm O}$ calculated in the previous section directly into the additional rate of ccSNe per Myr. This exercise is useful in quantifying the additional SNe that the SOS in Figure \ref{fig:CompObs_Mass} would suggest, assuming the SOS is born from issues in stellar modeling. $\Delta M_{\rm O}$  can be converted into an approximate additional rate of ccSNe per Myr ($\Delta R_{\rm ccSN} /\rm Myr^{-1}$; right axis, orange lines/regions) by assuming an average mass of oxygen produced in a ccSN ($\langle M_{\rm O, ccSN}\rangle$) and dividing by the Hubble time in Myr\footnote{For this work, we adopt a Hubble time to estimate an average $\Delta R_{\rm ccSN}$. However, in reality, these simulated dwarfs do not experience constant SFR across cosmic time, but we leave a more detailed investigation for future work.}. An average mass of $\langle M_{\rm O, ccSN}\rangle=1.14 \,M_{\odot}$ is chosen for $\Delta R_{\rm ccSN} (\rm Myr^{-1})$ and is shown as a solid orange line. Shaded regions reflect varying $\langle M_{\rm O, ccSN}\rangle=1.14 \,M_{\odot}$ by a factor of $2$ and $3$. We find that  $\rm M_* \sim 10^{8.3} M_{\odot}$ galaxies would require an additional $300-400\, \rm{ccSNe/Myr}$ throughout a Hubble time to match \MishraMOVI---which constitutes roughly a $3\times$ increase from current SFR (assuming the same IMF modeling). Increasing the yield per ccSNe can soften this requirement to $100-200\, \rm{ccSNe/Myr}$ (a $\sim1.5\times$ increase). However, simulations would also need to increase the yield per ccSN for all metals to preserve abundance ratios. To agree within $1\sigma$ of \MishraMOVI, the suites require fewer additional ccSNe ($\leq100, \rm{ccSNe/Myr}$), yet this would still constitute a $\sim2\times$ increase from current SFR with the same yields. These results suggest that both $\langle M_{\rm O, ccSN}\rangle$ and star formation models may need to be readdressed to match current observations.

This preliminary quantification is solely based on bringing the simulated \MOVI\, values into perfect alignment with the median and $1\sigma$ bounds of \MishraMOVI. In reality, due to the observed mixture of detections and non-detections, not all dwarf galaxies have a uniformly \OVI-enriched CGM (seen in the $1\sigma$ and $2\sigma$ ranges provided for \MishraMOVI). This variability in \MOVI\, will likely soften some of the requirements for simulations to match observations, but does not resolve the total deficit in oxygen production. Additionally, these calculations assume that the additional oxygen is produced and retained over a Hubble time, and that the dominant production channel is ccSNe under the same \citet{Kroupa2001_IMF} IMF. Deviations from these assumptions would modify the inferred requirements. 

Furthermore, contributions to the oxygen production from  SNe Ia should also be considered. Although ccSNe are the dominant production sites of oxygen ($\langle M_{\rm O, ccSN}\rangle = 1.14 \rm\, M_{\odot}$), SNe Ia also contribute approximately $10\%$ that of ccSNe ($\langle M_{\rm O, SN\,Ia}\rangle \sim 0.14 \rm\, M_{\odot}$) to the total oxygen reservoirs. Given this, modifications to the modeling of SNe Ia (such as the assumed progenitor systems and delay time distributions) will likely also increase the total oxygen production. We leave a detailed investigation exploring the parameter space that would place simulated galaxies within the full $1\sigma$ and $2\sigma$ ranges for future work.

However, increasing $R_{\rm ccSN}$ cannot be considered in isolation, as it is likely to introduce tensions with other well-established galaxy scaling relations and baryon cycle processes, in addition to the effects of increasing the CGM metal content noted previously. For instance, although increasing the number of ccSNe would supply both the increased oxygen production and metal transport, the increased feedback budget would likely also shape the evolution and growth of the galaxy. Additionally, increasing the number of ccSNe and therefore stellar feedback may alter the stellar masses and the resulting stellar-to-halo mass relation.  Since star formation in these simulations is self-regulating \citep[e.g.,][]{Christensen_14,Hopkins_14_SF,Benincasa_16}, the overall stellar masses may not change. However, if the increased feedback budget is too great, star formation may be over-suppressed, and the galaxies quenched, leading to stellar masses that fall below the stellar-to-halo mass relation. 

Since SNe Ia release less feedback energy than ccSNe, an increased rate of SNe Ia would pose a smaller linear increase in the feedback budget, potentially alleviating the concern of over-suppressing star formation. However, the smaller feedback energy input then poses concerns for these events effectively distributing the oxygen produced to large radii where it is needed most to support \NOVI\, values.


Given all of the factors discussed here, it is not clear if reproducing the observed \NOVI\, values requires a few simple modeling changes, or a complex rework of dwarf galaxy evolution. However, we have been unable to identify changes in the phase structure of the CGM that can close the gap between simulations and observations, suggesting that dwarf galaxies need an increase in their oxygen production and reservoirs. Doing so likely requires a reshaping of the star formation history in the galaxies, yet this may have deleterious impacts on the current agreement with galaxy scaling relations.  Additional work is needed to ascertain the scope of the required changes and their impacts.

\section{Summary} \label{sec:summary}
 
We present an investigation of \OVI\, in the CGM of simulated dwarf galaxies from the FIRE-2 and \MM\, (Marvelous Massive Dwarfs and Marvel-ous Dwarfs) suites. Neither simulation suite reproduces the observed \OVI\, columns in dwarf galaxy halos. We therefore investigate the origin of this discrepancy. Our main conclusions are:

\begin{enumerate}
    \item Although the CGM gas and oxygen content are broadly consistent across suites, differences in subgrid physics implementations lead to differences in CGM radial properties and physical conditions. For example:
    \begin{enumerate}
        \item Feedback in FIRE-2 preferentially pushes gas and oxygen mass to higher radii (Figure \ref{fig:CGMprop_mass}), whereas the CGM within the halos of \MM\, galaxies tends to be warmer and higher metallicity (Figure \ref{fig:CGMprop_cond}). 
        \item The simulations also differ in their metal enrichment pathways: outflows in FIRE-2 can reach large distances while retaining their original metallicity, whereas metals in \MM, galaxies mix more efficiently, producing a more homogeneous CGM metal distribution (Figure \ref{fig:OVIprop_PDFcond}).
    \end{enumerate}

    \item Despite differences in CGM properties, \OVI\, properties tend to be well-aligned across suites. FIRE-2 and \MM\, agree that \OVI\, is primarily produced via photoionization (Figure \ref{fig:OVIprop_ionmech}) and in cool/warm ($\rm log\, T\,/K \sim 4.5$), diffuse, ($\rm log\,n_{gas}\,/cm^{-3} \sim -5.0 $), and moderately metal-enriched ($\rm log Z/Z_{\odot}  \sim -1 $) material (Figure \ref{fig:OVIprop_PDFcond}). Both suites further agree that the most favorable conditions for \OVI\, lie beyond the halos of dwarf galaxies (Figure \ref{fig:OVIprop_ionfrac}).

    \item The insensitivity of \OVI\, conditions across simulation suites, despite CGM differences, suggests that CGM thermodynamic structure is not the primary driver of the simulated \OVI\, deficit. Although CGM conditions influence the dominant ionization mechanism (Figure \ref{fig:OVIprop_ionmech}), they cannot resolve the discrepancy with observations. In Section \ref{subsec:disc_vary}, we demonstrate that even strong global cooling of the CGM produces only modest increases in \MOVI\, (Figure \ref{fig:Disc_CGMtoggle}).

    \item To close the gap between simulated and observed \OVI\, reservoirs, simulations likely require:
    \begin{enumerate}
        \item Greater total oxygen production over the lifetime of the galaxy to address the Simulated Oxygen Shortage (SOS). In our sample, only ${\rm log}(M_*/M_{\odot}) \geq 8.3$ galaxies have produced sufficient oxygen to reproduce the median \MishraMOVI\, under idealized ionization assumptions (Figure \ref{fig:CompObs_Mass}). To match the median \MishraMOVI\,, ${\rm log}(M_*/M_{\odot}) \sim 8.3$ would require additional oxygen mass of $\Delta M_{\rm O}\approx 3\times 10^6 M_{\odot}$, while lying within $1\sigma$ requires a more modest $\Delta M_{\rm O}\approx 1\times 10^6 M_{\odot}$. If this additional oxygen mass is sourced entirely from an increased rate of ccSNe, this would constitute an additional $100-400$ ccSNe per Myr (Figure \ref{fig:Disc_MassNeeded}) or a $2-3\times$ increase from current models.  However, increasing the total yield per ccSN or the ionization efficiency of \OVI\, can soften these requirements.
        
        \item Efficient metal loading that transports oxygen to large radii. More total oxygen production is required to have the available material, but feedback must also be effective at dispersing the oxygen produced out to large radii ($r/R_{vir} \sim 2.5$) to increase simulated \NOVI\, values.
    \end{enumerate}
\end{enumerate}

Taken together, our results suggest that the \OVI\, deficit in simulations appears to be primarily an oxygen underproduction problem rather than a CGM phase structure problem. Addressing this issue may require readdressing current metal production/star formation models. Current findings in the SNe community suggest that in low-metallicity environments, the number of ccSNe may be underpredicted, potentially due to a metallicity-dependent IMF \citep[e.g.,][]{Pessi25}. If confirmed, a higher rate of ccSNe would naturally increase oxygen production and stellar feedback and could help reconcile simulations with observations. However, the resulting increase in feedback and CGM oxygen content could likely introduce complex, coupled changes to the baryon cycle that could potentially break agreement with key galaxy scaling relations.

Future work will investigate the effects of an environmentally dependent IMF on dwarf galaxy evolution and CGM column densities, as well as better constrain the parameter space that simultaneously addresses the SOS while preserving agreement with established galaxy scaling relations. In addition, kinematic analyses of synthetic \OVI\, absorption may help determine whether simulations reproduce the physical origins of the observed \OVI\, gas.

If the interpretation regarding an under-modeled rate of ccSNe is correct, the implications would extend beyond the \OVI\, underprediction in simulations. Observational models for metal production and star formation histories that do not account for this would also be impacted. Though these conclusions remain speculative, they warrant detailed future investigation.

\bigskip
\section*{Acknowledgments}
We thank Philip Hopkins, Yakov Faerman, and Sean Johnson for helpful conversations during this work. Resources supporting this work were provided by the NASA High-End Computing (HEC) Program through the NASA Advanced Supercomputing (NAS) Division at Ames Research Center. Some of the simulations were performed using resources made available by the Flatiron Institute. The Flatiron Institute is a division of the Simons Foundation. This work used Stampede2 at the Texas Advanced Computing Center (TACC) through allocation MCA94P018 from the Advanced Cyberinfrastructure Coordination Ecosystem: Services \& Support (ACCESS) program, which is supported by U.S. National Science Foundation grants \#2138259, \#2138286, \#2138307, \#2137603, and \#2138296. D.R.P and A.M.B are supported by NASA Grant 80NSSC24K0894. A.M.B acknowledges support by grant FI-CCA-Research-00011826 from the Simons Foundation. J.W. is supported by a grant from NSERC (National Science and Engineering Research Council) Canada. C.C. was supported by the NSF under CAREER grant AST-1848107, and this work was performed in part at Aspen Center for Physics, which is supported by NSF grant PHY-2210452. N.N.S. was supported by the National Science Foundation MPS-Ascend award ID 2212959.

\bibliography{sample701}{}

@ARTICLE{Hopkins2015_GIZMO,
       author = {{Hopkins}, Philip F.},
        title = "{A new class of accurate, mesh-free hydrodynamic simulation methods}",
      journal = {\mnras},
         year = 2015,
        month = jun,
       volume = {450},
       number = {1},
        pages = {53-110},
          doi = {10.1093/mnras/stv195},
archivePrefix = {arXiv},
       eprint = {1409.7395},
 primaryClass = {astro-ph.CO},
       adsurl = {https://ui.adsabs.harvard.edu/abs/2015MNRAS.450...53H}
}

@ARTICLE{HM2012,
       author = {{Haardt}, Francesco and {Madau}, Piero},
        title = "{Radiative Transfer in a Clumpy Universe. IV. New Synthesis Models of the Cosmic UV/X-Ray Background}",
      journal = {\apj},
         year = 2012,
        month = feb,
       volume = {746},
       number = {2},
          eid = {125},
        pages = {125},
          doi = {10.1088/0004-637X/746/2/125},
       adsurl = {https://ui.adsabs.harvard.edu/abs/2012ApJ...746..125H}
}

@ARTICLE{FaucherGiguère2009,
       author = {{Faucher-Gigu{\`e}re}, Claude-Andr{\'e} and {Lidz}, Adam and {Zaldarriaga}, Matias and {Hernquist}, Lars},
        title = "{A New Calculation of the Ionizing Background Spectrum and the Effects of He II Reionization}",
      journal = {\apj},
         year = 2009,
        month = oct,
       volume = {703},
       number = {2},
        pages = {1416-1443},
          doi = {10.1088/0004-637X/703/2/1416},
archivePrefix = {arXiv},
       eprint = {0901.4554},
 primaryClass = {astro-ph.CO},
       adsurl = {https://ui.adsabs.harvard.edu/abs/2009ApJ...703.1416F}
}

@ARTICLE{Kroupa2001_IMF,
       author = {{Kroupa}, Pavel},
        title = "{On the variation of the initial mass function}",
      journal = {\mnras},
         year = 2001,
        month = apr,
       volume = {322},
       number = {2},
        pages = {231-246},
          doi = {10.1046/j.1365-8711.2001.04022.x},
archivePrefix = {arXiv},
       eprint = {astro-ph/0009005},
 primaryClass = {astro-ph},
       adsurl = {https://ui.adsabs.harvard.edu/abs/2001MNRAS.322..231K}
}

@ARTICLE{CHANGA,
       author = {{Menon}, Harshitha and {Wesolowski}, Lukasz and {Zheng}, Gengbin and {Jetley}, Pritish and {Kale}, Laxmikant and {Quinn}, Thomas and {Governato}, Fabio},
        title = "{Adaptive techniques for clustered N-body cosmological simulations}",
      journal = {Computational Astrophysics and Cosmology},
         year = 2015,
        month = mar,
       volume = {2},
          eid = {1},
        pages = {1},
          doi = {10.1186/s40668-015-0007-9},
archivePrefix = {arXiv},
       eprint = {1409.1929},
 primaryClass = {astro-ph.IM},
       adsurl = {https://ui.adsabs.harvard.edu/abs/2015ComAC...2....1M}
}

@ARTICLE{Shen2010,
       author = {{Shen}, S. and {Wadsley}, J. and {Stinson}, G.},
        title = "{The enrichment of the intergalactic medium with adiabatic feedback - I. Metal cooling and metal diffusion}",
      journal = {\mnras},
         year = 2010,
        month = sep,
       volume = {407},
       number = {3},
        pages = {1581-1596},
          doi = {10.1111/j.1365-2966.2010.17047.x},
archivePrefix = {arXiv},
       eprint = {0910.5956},
 primaryClass = {astro-ph.CO},
       adsurl = {https://ui.adsabs.harvard.edu/abs/2010MNRAS.407.1581S}
}

@ARTICLE{Christensen2012,
       author = {{Christensen}, Charlotte and {Quinn}, Thomas and {Governato}, Fabio and {Stilp}, Adrienne and {Shen}, Sijing and {Wadsley}, James},
        title = "{Implementing molecular hydrogen in hydrodynamic simulations of galaxy formation}",
      journal = {\mnras},
         year = 2012,
        month = oct,
       volume = {425},
       number = {4},
        pages = {3058-3076},
          doi = {10.1111/j.1365-2966.2012.21628.x},
archivePrefix = {arXiv},
       eprint = {1205.5567},
 primaryClass = {astro-ph.CO},
       adsurl = {https://ui.adsabs.harvard.edu/abs/2012MNRAS.425.3058C}
}

@ARTICLE{Ruan2025,
       author = {{Ruan}, Dilys and {Brooks}, Alyson M. and {Cruz}, Akaxia and {Peter}, Annika H.~G. and {Keller}, Benjamin W. and {Quinn}, Thomas and {Wadsley}, James and {Adams}, Elizabeth A.~K.},
        title = "{Predictions for detecting a turndown in the baryonic Tully{\textendash}Fisher relation}",
      journal = {\mnras},
         year = 2025,
        month = aug,
       volume = {541},
       number = {3},
        pages = {2180-2196},
          doi = {10.1093/mnras/staf1099},
archivePrefix = {arXiv},
       eprint = {2503.16607},
 primaryClass = {astro-ph.GA},
       adsurl = {https://ui.adsabs.harvard.edu/abs/2025MNRAS.541.2180R}
}

@ARTICLE{Keith2025,
       author = {{Keith}, Blake and {Munshi}, Ferah and {Brooks}, Alyson M. and {Van Nest}, Jordan and {Engelhardt}, Anna and {Cruz}, Akaxia and {Keller}, Ben and {Quinn}, Thomas and {Wadsley}, James},
        title = "{A MARVEL-ous Study of How Well Galaxy Shapes Reflect Dark Matter Halo Shapes in Cold Dark Matter Simulations}",
      journal = {\apj},
         year = 2025,
        month = jun,
       volume = {986},
       number = {2},
          eid = {138},
        pages = {138},
          doi = {10.3847/1538-4357/add40d},
archivePrefix = {arXiv},
       eprint = {2501.16317},
 primaryClass = {astro-ph.GA},
       adsurl = {https://ui.adsabs.harvard.edu/abs/2025ApJ...986..138K}
}

@ARTICLE{Piacitelli_25,
       author = {{Piacitelli}, Daniel R. and {Brooks}, Alyson M. and {Christensen}, Charlotte and {Sanchez}, N. Nicole and {Faerman}, Yakov and {Shen}, Sijing and {Cruz}, Akaxia and {Keller}, Ben and {Quinn}, Thomas R. and {Wadsley}, James},
        title = "{Marvelous Metals: Surveying the Circumgalactic Medium of Simulated Dwarf Galaxies}",
      journal = {\apj},
         year = 2025,
        month = nov,
       volume = {993},
       number = {2},
          eid = {230},
        pages = {230},
          doi = {10.3847/1538-4357/ae06a0},
archivePrefix = {arXiv},
       eprint = {2505.08861},
 primaryClass = {astro-ph.GA},
       adsurl = {https://ui.adsabs.harvard.edu/abs/2025ApJ...993..230P}
}

@ARTICLE{FIRE_publicdata,
       author = {{Wetzel}, Andrew and {Hayward}, Christopher C. and {Sanderson}, Robyn E. and {Ma}, Xiangcheng and {Angl{\'e}s-Alc{\'a}zar}, Daniel and {Feldmann}, Robert and {Chan}, T.~K. and {El-Badry}, Kareem and {Wheeler}, Coral and {Garrison-Kimmel}, Shea and {Nikakhtar}, Farnik and {Panithanpaisal}, Nondh and {Arora}, Arpit and {Gurvich}, Alexander B. and {Samuel}, Jenna and {Sameie}, Omid and {Pandya}, Viraj and {Hafen}, Zachary and {Hummels}, Cameron and {Loebman}, Sarah and {Boylan-Kolchin}, Michael and {Bullock}, James S. and {Faucher-Gigu{\`e}re}, Claude-Andr{\'e} and {Kere{\v{s}}}, Du{\v{s}}an and {Quataert}, Eliot and {Hopkins}, Philip F.},
        title = "{Public Data Release of the FIRE-2 Cosmological Zoom-in Simulations of Galaxy Formation}",
      journal = {\apjs},
         year = 2023,
        month = apr,
       volume = {265},
       number = {2},
          eid = {44},
        pages = {44},
          doi = {10.3847/1538-4365/acb99a},
archivePrefix = {arXiv},
       eprint = {2202.06969},
 primaryClass = {astro-ph.GA},
       adsurl = {https://ui.adsabs.harvard.edu/abs/2023ApJS..265...44W}
}

@ARTICLE{Bordoloi_14,
       author = {{Bordoloi}, Rongmon and {Tumlinson}, Jason and {Werk}, Jessica K. and {Oppenheimer}, Benjamin D. and {Peeples}, Molly S. and {Prochaska}, J. Xavier and {Tripp}, Todd M. and {Katz}, Neal and {Dav{\'e}}, Romeel and {Fox}, Andrew J. and {Thom}, Christopher and {Ford}, Amanda Brady and {Weinberg}, David H. and {Burchett}, Joseph N. and {Kollmeier}, Juna A.},
        title = "{The COS-Dwarfs Survey: The Carbon Reservoir around Sub-L* Galaxies}",
      journal = {\apj},
         year = 2014,
        month = dec,
       volume = {796},
       number = {2},
          eid = {136},
        pages = {136},
          doi = {10.1088/0004-637X/796/2/136},
archivePrefix = {arXiv},
       eprint = {1406.0509},
 primaryClass = {astro-ph.GA},
       adsurl = {https://ui.adsabs.harvard.edu/abs/2014ApJ...796..136B}
}

@ARTICLE{LiangChen_14,
       author = {{Liang}, Cameron J. and {Chen}, Hsiao-Wen},
        title = "{Mining circumgalactic baryons in the low-redshift universe}",
      journal = {\mnras},
         year = 2014,
        month = dec,
       volume = {445},
       number = {2},
        pages = {2061-2081},
          doi = {10.1093/mnras/stu1901},
archivePrefix = {arXiv},
       eprint = {1402.3602},
 primaryClass = {astro-ph.CO},
       adsurl = {https://ui.adsabs.harvard.edu/abs/2014MNRAS.445.2061L}
}

@ARTICLE{Burchett_16,
       author = {{Burchett}, Joseph N. and {Tripp}, Todd M. and {Bordoloi}, Rongmon and {Werk}, Jessica K. and {Prochaska}, J. Xavier and {Tumlinson}, Jason and {Willmer}, C.~N.~A. and {O'Meara}, John and {Katz}, Neal},
        title = "{A Deep Search for Faint Galaxies Associated with Very Low Redshift C IV Absorbers. III. The Mass- and Environment-dependent Circumgalactic Medium}",
      journal = {\apj},
         year = 2016,
        month = dec,
       volume = {832},
       number = {2},
          eid = {124},
        pages = {124},
          doi = {10.3847/0004-637X/832/2/124},
archivePrefix = {arXiv},
       eprint = {1512.00853},
 primaryClass = {astro-ph.GA},
       adsurl = {https://ui.adsabs.harvard.edu/abs/2016ApJ...832..124B}
}

@ARTICLE{Johnson_17,
       author = {{Johnson}, Sean D. and {Chen}, Hsiao-Wen and {Mulchaey}, John S. and {Schaye}, Joop and {Straka}, Lorrie A.},
        title = "{The Extent of Chemically Enriched Gas around Star-forming Dwarf Galaxies}",
      journal = {\apjl},
         year = 2017,
        month = nov,
       volume = {850},
       number = {1},
          eid = {L10},
        pages = {L10},
          doi = {10.3847/2041-8213/aa9370},
archivePrefix = {arXiv},
       eprint = {1710.06441},
 primaryClass = {astro-ph.GA},
       adsurl = {https://ui.adsabs.harvard.edu/abs/2017ApJ...850L..10J}
}

@ARTICLE{QuBregman_22,
       author = {{Qu}, Zhijie and {Bregman}, Joel N.},
        title = "{Absorption Line Search through Three Local Group Dwarf Galaxy Halos}",
      journal = {\apj},
         year = 2022,
        month = mar,
       volume = {927},
       number = {2},
          eid = {228},
        pages = {228},
          doi = {10.3847/1538-4357/ac51df},
archivePrefix = {arXiv},
       eprint = {2203.08246},
 primaryClass = {astro-ph.GA},
       adsurl = {https://ui.adsabs.harvard.edu/abs/2022ApJ...927..228Q}
}

@ARTICLE{Zheng_24,
       author = {{Zheng}, Yong and {Faerman}, Yakov and {Oppenheimer}, Benjamin D. and {Putman}, Mary E. and {McQuinn}, Kristen B.~W. and {Kirby}, Evan N. and {Burchett}, Joseph N. and {Telford}, O. Grace and {Werk}, Jessica K. and {Kim}, Doyeon A.},
        title = "{A Comprehensive Investigation of Metals in the Circumgalactic Medium of Nearby Dwarf Galaxies}",
      journal = {\apj},
         year = 2024,
        month = jan,
       volume = {960},
       number = {1},
          eid = {55},
        pages = {55},
          doi = {10.3847/1538-4357/acfe6b},
archivePrefix = {arXiv},
       eprint = {2301.12233},
 primaryClass = {astro-ph.GA},
       adsurl = {https://ui.adsabs.harvard.edu/abs/2024ApJ...960...55Z}
}

@ARTICLE{Mishra_24,
       author = {{Mishra}, Nishant and {Johnson}, Sean D. and {Rudie}, Gwen C. and {Chen}, Hsiao-Wen and {Schaye}, Joop and {Qu}, Zhijie and {Zahedy}, Fakhri S. and {Boettcher}, Erin T. and {Cantalupo}, Sebastiano and {Chen}, Mandy C. and {Faucher-Gigu{\'e}re}, Claude-Andr{\'e} and {Greene}, Jenny E. and {Li}, Jennifer I. -Hsiu and {Liu}, Zhuoqi (Will) and {Lopez}, Sebastian and {Petitjean}, Patrick},
        title = "{The Cosmic Ultraviolet Baryon Survey (CUBS). IX. The Enriched Circumgalactic and Intergalactic Medium Around Star-forming Field Dwarf Galaxies Traced by O VI Absorption}",
      journal = {\apj},
         year = 2024,
        month = nov,
       volume = {976},
       number = {1},
          eid = {149},
        pages = {149},
          doi = {10.3847/1538-4357/ad7b0a},
archivePrefix = {arXiv},
       eprint = {2408.11151},
 primaryClass = {astro-ph.GA},
       adsurl = {https://ui.adsabs.harvard.edu/abs/2024ApJ...976..149M}
}

@ARTICLE{Dutta_25_col,
       author = {{Dutta}, Sayak and {Muzahid}, Sowgat and {Schaye}, Joop and {Bouch{\'e}}, Nicolas F. and {Cantalupo}, Sebastiano and {Chen}, Hsiao-Wen and {Johnson}, Sean},
        title = "{MUSEQuBES: The Column Density, Covering Fraction, and Mass of O VI-bearing Gas in and Around Low-redshift Galaxies}",
      journal = {\apj},
         year = 2025,
        month = may,
       volume = {985},
       number = {1},
          eid = {44},
        pages = {44},
          doi = {10.3847/1538-4357/adc922},
archivePrefix = {arXiv},
       eprint = {2409.15423},
 primaryClass = {astro-ph.GA},
       adsurl = {https://ui.adsabs.harvard.edu/abs/2025ApJ...985...44D}
}

@ARTICLE{Tchernyshyov_22,
       author = {{Tchernyshyov}, Kirill and {Werk}, Jessica K. and {Wilde}, Matthew C. and {Prochaska}, J. Xavier and {Tripp}, Todd M. and {Burchett}, Joseph N. and {Bordoloi}, Rongmon and {Howk}, J. Christopher and {Lehner}, Nicolas and {O'Meara}, John M. and {Tejos}, Nicolas and {Tumlinson}, Jason},
        title = "{The CGM$^{2}$ Survey: Circumgalactic O VI from Dwarf to Massive Star-forming Galaxies}",
      journal = {\apj},
         year = 2022,
        month = mar,
       volume = {927},
       number = {2},
          eid = {147},
        pages = {147},
          doi = {10.3847/1538-4357/ac450c},
archivePrefix = {arXiv},
       eprint = {2110.13167},
 primaryClass = {astro-ph.GA},
       adsurl = {https://ui.adsabs.harvard.edu/abs/2022ApJ...927..147T}
}

@ARTICLE{Qu_24,
       author = {{Qu}, Zhijie and {Chen}, Hsiao-Wen and {Johnson}, Sean D. and {Rudie}, Gwen C. and {Zahedy}, Fakhri S. and {DePalma}, David and {Schaye}, Joop and {Boettcher}, Erin T. and {Cantalupo}, Sebastiano and {Chen}, Mandy C. and {Faucher-Gigu{\`e}re}, Claude-Andr{\'e} and {Li}, Jennifer I. -Hsiu and {Mulchaey}, John S. and {Petitjean}, Patrick and {Rafelski}, Marc},
        title = "{The Cosmic Ultraviolet Baryon Survey (CUBS). VII. On the Warm-hot Circumgalactic Medium Probed by O VI and Ne VIII at 0.4 {\ensuremath{\lesssim}} z {\ensuremath{\lesssim}} 0.7}",
      journal = {\apj},
         year = 2024,
        month = jun,
       volume = {968},
       number = {1},
          eid = {8},
        pages = {8},
          doi = {10.3847/1538-4357/ad410b},
archivePrefix = {arXiv},
       eprint = {2402.08016},
 primaryClass = {astro-ph.GA},
       adsurl = {https://ui.adsabs.harvard.edu/abs/2024ApJ...968....8Q}
}

@ARTICLE{Li_21,
       author = {{Li}, Fei and {Rahman}, Mubdi and {Murray}, Norman and {Hafen}, Zachary and {Faucher-Gigu{\`e}re}, Claude-Andr{\'e} and {Stern}, Jonathan and {Hummels}, Cameron B. and {Hopkins}, Philip F. and {El-Badry}, Kareem and {Kere{\v{s}}}, Du{\v{s}}an},
        title = "{Probing the CGM of low-redshift dwarf galaxies using FIRE simulations}",
      journal = {\mnras},
         year = 2021,
        month = jan,
       volume = {500},
       number = {1},
        pages = {1038-1053},
          doi = {10.1093/mnras/staa3322},
archivePrefix = {arXiv},
       eprint = {2010.13606},
 primaryClass = {astro-ph.GA},
       adsurl = {https://ui.adsabs.harvard.edu/abs/2021MNRAS.500.1038L}
}

@ARTICLE{SD93,
       author = {{Sutherland}, Ralph S. and {Dopita}, M.~A.},
        title = "{Cooling Functions for Low-Density Astrophysical Plasmas}",
      journal = {\apjs},
         year = 1993,
        month = sep,
       volume = {88},
        pages = {253},
          doi = {10.1086/191823},
       adsurl = {https://ui.adsabs.harvard.edu/abs/1993ApJS...88..253S}
}

@ARTICLE{Munshi21,
       author = {{Munshi}, Ferah and {Brooks}, Alyson M. and {Applebaum}, Elaad and {Christensen}, Charlotte R. and {Quinn}, T. and {Sligh}, Serena},
        title = "{Quantifying Scatter in Galaxy Formation at the Lowest Masses}",
      journal = {\apj},
         year = 2021,
        month = dec,
       volume = {923},
       number = {1},
          eid = {35},
        pages = {35},
          doi = {10.3847/1538-4357/ac0db6},
archivePrefix = {arXiv},
       eprint = {2101.05822},
 primaryClass = {astro-ph.GA},
       adsurl = {https://ui.adsabs.harvard.edu/abs/2021ApJ...923...35M}
}

@ARTICLE{Azartash-Namin24,
       author = {{Azartash-Namin}, Bianca and {Engelhardt}, Anna and {Munshi}, Ferah and {Keller}, B.~W. and {Brooks}, Alyson M. and {Van Nest}, Jordan and {Christensen}, Charlotte R. and {Quinn}, Tom and {Wadsley}, James},
        title = "{Bursting with Feedback: The Relationship between Feedback Model and Bursty Star Formation Histories in Dwarf Galaxies}",
      journal = {\apj},
         year = 2024,
        month = jul,
       volume = {970},
       number = {1},
          eid = {40},
        pages = {40},
          doi = {10.3847/1538-4357/ad49a5},
archivePrefix = {arXiv},
       eprint = {2401.06041},
 primaryClass = {astro-ph.GA},
       adsurl = {https://ui.adsabs.harvard.edu/abs/2024ApJ...970...40A}
}

@ARTICLE{Thielemann86,
       author = {{Thielemann}, F. -K. and {Nomoto}, K. and {Yokoi}, K.},
        title = "{Explosive nucleosynthesis in carbon deflagration models of Type I supernovae}",
      journal = {\aap},
         year = 1986,
        month = apr,
       volume = {158},
       number = {1-2},
        pages = {17-33},
       adsurl = {https://ui.adsabs.harvard.edu/abs/1986A&A...158...17T}
}

@ARTICLE{Raiteri96,
       author = {{Raiteri}, C.~M. and {Villata}, M. and {Navarro}, J.~F.},
        title = "{Simulations of Galactic chemical evolution. I. O and Fe abundances in a simple collapse model.}",
      journal = {\aap},
         year = 1996,
        month = nov,
       volume = {315},
        pages = {105-115},
       adsurl = {https://ui.adsabs.harvard.edu/abs/1996A&A...315..105R}
}

@ARTICLE{Weidemann_87,
       author = {{Weidemann}, V.},
        title = "{The initial-final mass relation : galactic disk and Magellanic Clouds.}",
      journal = {\aap},
         year = 1987,
        month = dec,
       volume = {188},
        pages = {74-84},
       adsurl = {https://ui.adsabs.harvard.edu/abs/1987A&A...188...74W}}

@ARTICLE{Keller14,
       author = {{Keller}, B.~W. and {Wadsley}, J. and {Benincasa}, S.~M. and {Couchman}, H.~M.~P.},
        title = "{A superbubble feedback model for galaxy simulations}",
      journal = {\mnras},
         year = 2014,
        month = aug,
       volume = {442},
       number = {4},
        pages = {3013-3025},
          doi = {10.1093/mnras/stu1058},
archivePrefix = {arXiv},
       eprint = {1405.2625},
 primaryClass = {astro-ph.GA},
       adsurl = {https://ui.adsabs.harvard.edu/abs/2014MNRAS.442.3013K}
}

@ARTICLE{CowieMcKee_77,
       author = {{Cowie}, L.~L. and {McKee}, C.~F.},
        title = "{The evaporation of spherical clouds in a hot gas. I. Classical and saturated mass loss rates.}",
      journal = {\apj},
         year = 1977,
        month = jan,
       volume = {211},
        pages = {135-146},
          doi = {10.1086/154911},
       adsurl = {https://ui.adsabs.harvard.edu/abs/1977ApJ...211..135C}
}

@ARTICLE{KrumholzGnedin_11,
       author = {{Krumholz}, Mark R. and {Gnedin}, Nickolay Y.},
        title = "{A Comparison of Methods for Determining the Molecular Content of Model Galaxies}",
      journal = {\apj},
         year = 2011,
        month = mar,
       volume = {729},
       number = {1},
          eid = {36},
        pages = {36},
          doi = {10.1088/0004-637X/729/1/36},
archivePrefix = {arXiv},
       eprint = {1011.4065},
 primaryClass = {astro-ph.CO},
       adsurl = {https://ui.adsabs.harvard.edu/abs/2011ApJ...729...36K}
}

@ARTICLE{Hopkins_13,
       author = {{Hopkins}, Philip F. and {Narayanan}, Desika and {Murray}, Norman},
        title = "{The meaning and consequences of star formation criteria in galaxy models with resolved stellar feedback}",
      journal = {\mnras},
         year = 2013,
        month = jul,
       volume = {432},
       number = {4},
        pages = {2647-2653},
          doi = {10.1093/mnras/stt723},
archivePrefix = {arXiv},
       eprint = {1303.0285},
 primaryClass = {astro-ph.CO},
       adsurl = {https://ui.adsabs.harvard.edu/abs/2013MNRAS.432.2647H}
}

@ARTICLE{RocaFabrega_19,
       author = {{Roca-F{\`a}brega}, S. and {Dekel}, A. and {Faerman}, Y. and {Gnat}, O. and {Strawn}, C. and {Ceverino}, D. and {Primack}, J. and {Macci{\`o}}, A.~V. and {Dutton}, A.~A. and {Prochaska}, J.~X. and {Stern}, J.},
        title = "{CGM properties in VELA and NIHAO simulations; the OVI ionization mechanism: dependence on redshift, halo mass, and radius}",
      journal = {\mnras},
         year = 2019,
        month = apr,
       volume = {484},
       number = {3},
        pages = {3625-3645},
          doi = {10.1093/mnras/stz063},
archivePrefix = {arXiv},
       eprint = {1808.09973},
 primaryClass = {astro-ph.GA},
       adsurl = {https://ui.adsabs.harvard.edu/abs/2019MNRAS.484.3625R}
}

@ARTICLE{Gutcke2017, 
       author = {{Gutcke}, Thales A. and {Stinson}, Greg S. and {Macci{\`o}}, Andrea V. and {Wang}, Liang and {Dutton}, Aaron A.},
        title = "{NIHAO - VIII. Circum-galactic medium and outflows - The puzzles of H I and O VI gas distributions}",
      journal = {\mnras},
         year = 2017,
        month = jan,
       volume = {464},
       number = {3},
        pages = {2796-2815},
          doi = {10.1093/mnras/stw2539},
archivePrefix = {arXiv},
       eprint = {1602.06956},
 primaryClass = {astro-ph.GA},
       adsurl = {https://ui.adsabs.harvard.edu/abs/2017MNRAS.464.2796G}
}

@ARTICLE{Cook_24,
       author = {{Cook}, Andrew W.~S. and {van de Voort}, Freeke and {Pakmor}, R{\"u}diger and {Grand}, Robert J.~J.},
        title = "{The halo mass dependence of physical and observable properties in the circumgalactic medium at z = 0}",
      journal = {\mnras},
         year = 2025,
        month = oct,
       volume = {543},
       number = {2},
        pages = {1224-1238},
          doi = {10.1093/mnras/staf1537},
archivePrefix = {arXiv},
       eprint = {2409.05578},
 primaryClass = {astro-ph.GA},
       adsurl = {https://ui.adsabs.harvard.edu/abs/2025MNRAS.543.1224C}
}

@ARTICLE{Johnson26,
       author = {{Johnson}, Sean D. and {Mishra}, Nishant and {Muzahid}, Sowgat and {Rudie}, Gwen C. and {Zahedy}, Fakhri S. and {Qu}, Zhijie and {Faucher-Gigu{\`e}re}, Claude-Andr{\'e} and {Stern}, Jonathan and {Li}, Jennifer I.-Hsiu and {Fuller}, Elise and {Cantalupo}, Sebastiano and {Chen}, Hsiao-Wen and {Kadri}, Ahmad and {Kumar}, Suyash and {Liu}, Zhuoqi (Will) and {Walth}, Gregory},
        title = "{MUSEQuBES: Physical Conditions, Origins, and Multielement Abundances of the Circumgalactic Medium of an Isolated, Star-forming Dwarf Galaxy at z = 0.57}",
      journal = {\apjl},
         year = 2026,
        month = jan,
       volume = {996},
       number = {2},
          eid = {L30},
        pages = {L30},
          doi = {10.3847/2041-8213/ae2f5f},
archivePrefix = {arXiv},
       eprint = {2510.06310},
 primaryClass = {astro-ph.GA},
       adsurl = {https://ui.adsabs.harvard.edu/abs/2026ApJ...996L..30J}
}

@ARTICLE{Hopkins2014,
       author = {{Hopkins}, Philip F. and {Kere{\v{s}}}, Du{\v{s}}an and {O{\~n}orbe}, Jos{\'e} and {Faucher-Gigu{\`e}re}, Claude-Andr{\'e} and {Quataert}, Eliot and {Murray}, Norman and {Bullock}, James S.},
        title = "{Galaxies on FIRE (Feedback In Realistic Environments): stellar feedback explains cosmologically inefficient star formation}",
      journal = {\mnras},
         year = 2014,
        month = nov,
       volume = {445},
       number = {1},
        pages = {581-603},
          doi = {10.1093/mnras/stu1738},
archivePrefix = {arXiv},
       eprint = {1311.2073},
 primaryClass = {astro-ph.CO},
       adsurl = {https://ui.adsabs.harvard.edu/abs/2014MNRAS.445..581H}
}

@ARTICLE{Hopkins2017,
       author = {{Hopkins}, Philip F.},
        title = "{Anisotropic diffusion in mesh-free numerical magnetohydrodynamics}",
      journal = {\mnras},
         year = 2017,
        month = apr,
       volume = {466},
       number = {3},
        pages = {3387-3405},
          doi = {10.1093/mnras/stw3306},
archivePrefix = {arXiv},
       eprint = {1602.07703},
 primaryClass = {astro-ph.IM},
       adsurl = {https://ui.adsabs.harvard.edu/abs/2017MNRAS.466.3387H}
}

@ARTICLE{Nelson2018,
       author = {{Nelson}, Dylan and {Kauffmann}, Guinevere and {Pillepich}, Annalisa and {Genel}, Shy and {Springel}, Volker and {Pakmor}, R{\"u}diger and {Hernquist}, Lars and {Weinberger}, Rainer and {Torrey}, Paul and {Vogelsberger}, Mark and {Marinacci}, Federico},
        title = "{The abundance, distribution, and physical nature of highly ionized oxygen O VI, O VII, and O VIII in IllustrisTNG}",
      journal = {\mnras},
         year = 2018,
        month = jun,
       volume = {477},
       number = {1},
        pages = {450-479},
          doi = {10.1093/mnras/sty656},
archivePrefix = {arXiv},
       eprint = {1712.00016},
 primaryClass = {astro-ph.GA},
       adsurl = {https://ui.adsabs.harvard.edu/abs/2018MNRAS.477..450N}
}

@ARTICLE{Sanchez2019,
       author = {{Sanchez}, N. Nicole and {Werk}, Jessica K. and {Tremmel}, Michael and {Pontzen}, Andrew and {Christensen}, Charlotte and {Quinn}, Thomas and {Cruz}, Akaxia},
        title = "{Not So Heavy Metals: Black Hole Feedback Enriches the Circumgalactic Medium}",
      journal = {\apj},
         year = 2019,
        month = sep,
       volume = {882},
       number = {1},
          eid = {8},
        pages = {8},
          doi = {10.3847/1538-4357/ab3045},
archivePrefix = {arXiv},
       eprint = {1810.12319},
 primaryClass = {astro-ph.GA},
       adsurl = {https://ui.adsabs.harvard.edu/abs/2019ApJ...882....8S}
}

@ARTICLE{Pandya2021,
       author = {{Pandya}, Viraj and {Fielding}, Drummond B. and {Angl{\'e}s-Alc{\'a}zar}, Daniel and {Somerville}, Rachel S. and {Bryan}, Greg L. and {Hayward}, Christopher C. and {Stern}, Jonathan and {Kim}, Chang-Goo and {Quataert}, Eliot and {Forbes}, John C. and {Faucher-Gigu{\`e}re}, Claude-Andr{\'e} and {Feldmann}, Robert and {Hafen}, Zachary and {Hopkins}, Philip F. and {Kere{\v{s}}}, Du{\v{s}}an and {Murray}, Norman and {Wetzel}, Andrew},
        title = "{Characterizing mass, momentum, energy, and metal outflow rates of multiphase galactic winds in the FIRE-2 cosmological simulations}",
      journal = {\mnras},
         year = 2021,
        month = dec,
       volume = {508},
       number = {2},
        pages = {2979-3008},
          doi = {10.1093/mnras/stab2714},
archivePrefix = {arXiv},
       eprint = {2103.06891},
 primaryClass = {astro-ph.GA},
       adsurl = {https://ui.adsabs.harvard.edu/abs/2021MNRAS.508.2979P}
}

@ARTICLE{Pandya2020,
       author = {{Pandya}, Viraj and {Somerville}, Rachel S. and {Angl{\'e}s-Alc{\'a}zar}, Daniel and {Hayward}, Christopher C. and {Bryan}, Greg L. and {Fielding}, Drummond B. and {Forbes}, John C. and {Burkhart}, Blakesley and {Genel}, Shy and {Hernquist}, Lars and {Kim}, Chang-Goo and {Tonnesen}, Stephanie and {Starkenburg}, Tjitske},
        title = "{First Results from SMAUG: The Need for Preventative Stellar Feedback and Improved Baryon Cycling in Semianalytic Models of Galaxy Formation}",
      journal = {\apj},
         year = 2020,
        month = dec,
       volume = {905},
       number = {1},
          eid = {4},
        pages = {4},
          doi = {10.3847/1538-4357/abc3c1},
archivePrefix = {arXiv},
       eprint = {2006.16317},
 primaryClass = {astro-ph.GA},
       adsurl = {https://ui.adsabs.harvard.edu/abs/2020ApJ...905....4P}
}

@ARTICLE{Asplund2009,
       author = {{Asplund}, Martin and {Grevesse}, Nicolas and {Sauval}, A. Jacques and {Scott}, Pat},
        title = "{The Chemical Composition of the Sun}",
      journal = {\araa},
         year = 2009,
        month = sep,
       volume = {47},
       number = {1},
        pages = {481-522},
          doi = {10.1146/annurev.astro.46.060407.145222},
archivePrefix = {arXiv},
       eprint = {0909.0948},
 primaryClass = {astro-ph.SR},
       adsurl = {https://ui.adsabs.harvard.edu/abs/2009ARA&A..47..481A}
}

@ARTICLE{Trident,
       author = {{Hummels}, Cameron B. and {Smith}, Britton D. and {Silvia}, Devin W.},
        title = "{Trident: A Universal Tool for Generating Synthetic Absorption Spectra from Astrophysical Simulations}",
      journal = {\apj},
         year = 2017,
        month = sep,
       volume = {847},
       number = {1},
          eid = {59},
        pages = {59},
          doi = {10.3847/1538-4357/aa7e2d},
archivePrefix = {arXiv},
       eprint = {1612.03935},
 primaryClass = {astro-ph.IM},
       adsurl = {https://ui.adsabs.harvard.edu/abs/2017ApJ...847...59H}
}

@ARTICLE{Suresh_17,
       author = {{Suresh}, Joshua and {Rubin}, Kate H.~R. and {Kannan}, Rahul and {Werk}, Jessica K. and {Hernquist}, Lars and {Vogelsberger}, Mark},
        title = "{On the OVI abundance in the circumgalactic medium of low-redshift galaxies}",
      journal = {\mnras},
         year = 2017,
        month = mar,
       volume = {465},
       number = {3},
        pages = {2966-2982},
          doi = {10.1093/mnras/stw2499},
archivePrefix = {arXiv},
       eprint = {1511.00687},
 primaryClass = {astro-ph.GA},
       adsurl = {https://ui.adsabs.harvard.edu/abs/2017MNRAS.465.2966S}
}

@ARTICLE{Porter22,
       author = {{Porter}, Lori E. and {Orr}, Matthew E. and {Burkhart}, Blakesley and {Wetzel}, Andrew and {Ma}, Xiangcheng and {Hopkins}, Philip F. and {Emerick}, Andrew},
        title = "{Spatially resolved gas-phase metallicity in FIRE-2 dwarfs: late-time evolution of metallicity relations in simulations with feedback and mergers}",
      journal = {\mnras},
         year = 2022,
        month = sep,
       volume = {515},
       number = {3},
        pages = {3555-3576},
          doi = {10.1093/mnras/stac1958},
archivePrefix = {arXiv},
       eprint = {2204.06572},
 primaryClass = {astro-ph.GA},
       adsurl = {https://ui.adsabs.harvard.edu/abs/2022MNRAS.515.3555P}
}

@ARTICLE{Rey2024,
       author = {{Rey}, Martin P. and {Katz}, Harley B. and {Cameron}, Alex J. and {Devriendt}, Julien and {Slyz}, Adrianne},
        title = "{Boosting galactic outflows with enhanced resolution}",
      journal = {\mnras},
         year = 2024,
        month = mar,
       volume = {528},
       number = {3},
        pages = {5412-5431},
          doi = {10.1093/mnras/stae388},
archivePrefix = {arXiv},
       eprint = {2302.08521},
 primaryClass = {astro-ph.GA},
       adsurl = {https://ui.adsabs.harvard.edu/abs/2024MNRAS.528.5412R}
}

@ARTICLE{Christensen2016,
       author = {{Christensen}, Charlotte R. and {Dav{\'e}}, Romeel and {Governato}, Fabio and {Pontzen}, Andrew and {Brooks}, Alyson and {Munshi}, Ferah and {Quinn}, Thomas and {Wadsley}, James},
        title = "{In-N-Out: The Gas Cycle from Dwarfs to Spiral Galaxies}",
      journal = {\apj},
         year = 2016,
        month = jun,
       volume = {824},
       number = {1},
          eid = {57},
        pages = {57},
          doi = {10.3847/0004-637X/824/1/57},
archivePrefix = {arXiv},
       eprint = {1508.00007},
 primaryClass = {astro-ph.GA},
       adsurl = {https://ui.adsabs.harvard.edu/abs/2016ApJ...824...57C}
}

@ARTICLE{Escala2018,
       author = {{Escala}, Ivanna and {Wetzel}, Andrew and {Kirby}, Evan N. and {Hopkins}, Philip F. and {Ma}, Xiangcheng and {Wheeler}, Coral and {Kere{\v{s}}}, Du{\v{s}}an and {Faucher-Gigu{\`e}re}, Claude-Andr{\'e} and {Quataert}, Eliot},
        title = "{Modelling chemical abundance distributions for dwarf galaxies in the Local Group: the impact of turbulent metal diffusion}",
      journal = {\mnras},
         year = 2018,
        month = feb,
       volume = {474},
       number = {2},
        pages = {2194-2211},
          doi = {10.1093/mnras/stx2858},
archivePrefix = {arXiv},
       eprint = {1710.06533},
 primaryClass = {astro-ph.GA},
       adsurl = {https://ui.adsabs.harvard.edu/abs/2018MNRAS.474.2194E}
}

@ARTICLE{Su_17,
       author = {{Su}, Kung-Yi and {Hopkins}, Philip F. and {Hayward}, Christopher C. and {Faucher-Gigu{\`e}re}, Claude-Andr{\'e} and {Kere{\v{s}}}, Du{\v{s}}an and {Ma}, Xiangcheng and {Robles}, Victor H.},
        title = "{Feedback first: the surprisingly weak effects of magnetic fields, viscosity, conduction and metal diffusion on sub-L* galaxy formation}",
      journal = {\mnras},
         year = 2017,
        month = oct,
       volume = {471},
       number = {1},
        pages = {144-166},
          doi = {10.1093/mnras/stx1463},
archivePrefix = {arXiv},
       eprint = {1607.05274},
 primaryClass = {astro-ph.GA},
       adsurl = {https://ui.adsabs.harvard.edu/abs/2017MNRAS.471..144S}
}

@ARTICLE{Smagorinsky63,
       author = {{Smagorinsky}, J.},
        title = "{General Circulation Experiments with the Primitive Equations}",
      journal = {Monthly Weather Review},
         year = 1963,
        month = jan,
       volume = {91},
       number = {3},
        pages = {99},
          doi = {10.1175/1520-0493(1963)091<0099:GCEWTP>2.3.CO;2},
       adsurl = {https://ui.adsabs.harvard.edu/abs/1963MWRv...91...99S}
}

@ARTICLE{Pessi25,
       author = {{Pessi}, T. and {Desai}, D.~D. and {Prieto}, J.~L. and {Kochanek}, C.~S. and {Shappee}, B.~J. and {Anderson}, J.~P. and {Beacom}, J.~F. and {Dong}, S. and {Stanek}, K.~Z. and {Thompson}, T.~A.},
        title = "{Supernova rates and luminosity functions from ASAS-SN: II. 2014─2017 core-collapse supernovae and their subtypes}",
      journal = {\aap},
         year = 2025,
        month = nov,
       volume = {703},
          eid = {A34},
        pages = {A34},
          doi = {10.1051/0004-6361/202556799},
archivePrefix = {arXiv},
       eprint = {2508.10985},
 primaryClass = {astro-ph.HE},
       adsurl = {https://ui.adsabs.harvard.edu/abs/2025A&A...703A..34P}
}

@ARTICLE{Pessi23,
       author = {{Pessi}, Thallis and {Anderson}, Joseph P. and {Lyman}, Joseph D. and {Prieto}, Jose L. and {Galbany}, Llu{\'\i}s and {Kochanek}, Christopher S. and {S{\'a}nchez}, Sebastian F. and {Kuncarayakti}, Hanindyo},
        title = "{A Metallicity Dependence on the Occurrence of Core-collapse Supernovae}",
      journal = {\apjl},
         year = 2023,
        month = oct,
       volume = {955},
       number = {2},
          eid = {L29},
        pages = {L29},
          doi = {10.3847/2041-8213/acf7c6},
archivePrefix = {arXiv},
       eprint = {2306.11962},
 primaryClass = {astro-ph.SR},
       adsurl = {https://ui.adsabs.harvard.edu/abs/2023ApJ...955L..29P}
}

@ARTICLE{McQuinn_15,
       author = {{McQuinn}, Kristen B.~W. and {Skillman}, Evan D. and {Dolphin}, Andrew and {Cannon}, John M. and {Salzer}, John J. and {Rhode}, Katherine L. and {Adams}, Elizabeth A.~K. and {Berg}, Danielle and {Giovanelli}, Riccardo and {Haynes}, Martha P.},
        title = "{Leo P: How Many Metals Can a Very Low Mass, Isolated Galaxy Retain?}",
      journal = {\apjl},
         year = 2015,
        month = dec,
       volume = {815},
       number = {2},
          eid = {L17},
        pages = {L17},
          doi = {10.1088/2041-8205/815/2/L17},
archivePrefix = {arXiv},
       eprint = {1512.00459},
 primaryClass = {astro-ph.GA},
       adsurl = {https://ui.adsabs.harvard.edu/abs/2015ApJ...815L..17M}
}

@ARTICLE{Tung_25,
       author = {{Tung}, Pei-Cheng and {Chen}, Ke-Jung},
        title = "{Coevolution of Dwarf Galaxies and Their Circumgalactic Medium Across Cosmic Time}",
      journal = {\apj},
         year = 2025,
        month = jul,
       volume = {988},
       number = {1},
          eid = {127},
        pages = {127},
          doi = {10.3847/1538-4357/ade1d4},
archivePrefix = {arXiv},
       eprint = {2412.16440},
 primaryClass = {astro-ph.GA},
       adsurl = {https://ui.adsabs.harvard.edu/abs/2025ApJ...988..127T}
}

@ARTICLE{Keller_16,
       author = {{Keller}, B.~W. and {Wadsley}, J. and {Couchman}, H.~M.~P.},
        title = "{Cosmological galaxy evolution with superbubble feedback - II. The limits of supernovae}",
      journal = {\mnras},
         year = 2016,
        month = dec,
       volume = {463},
       number = {2},
        pages = {1431-1445},
          doi = {10.1093/mnras/stw2029},
archivePrefix = {arXiv},
       eprint = {1604.08244},
 primaryClass = {astro-ph.GA},
       adsurl = {https://ui.adsabs.harvard.edu/abs/2016MNRAS.463.1431K}
}

@ARTICLE{Dutta_25_kine,
       author = {{Dutta}, Sayak and {Muzahid}, Sowgat and {Schaye}, Joop and {Cantalupo}, Sebastiano and {Chen}, Hsiao-Wen and {Johnson}, Sean},
        title = "{MUSEQuBES: The Kinematics of O VI-bearing Gas in and around Low-redshift Galaxies}",
      journal = {\apj},
         year = 2025,
        month = feb,
       volume = {980},
       number = {2},
          eid = {264},
        pages = {264},
          doi = {10.3847/1538-4357/adabbd},
archivePrefix = {arXiv},
       eprint = {2409.15432},
 primaryClass = {astro-ph.GA},
       adsurl = {https://ui.adsabs.harvard.edu/abs/2025ApJ...980..264D}
}

@ARTICLE{Strawn_21,
       author = {{Strawn}, Clayton and {Roca-F{\`a}brega}, Santi and {Mandelker}, Nir and {Primack}, Joel and {Stern}, Jonathan and {Ceverino}, Daniel and {Dekel}, Avishai and {Wang}, Bryan and {Dange}, Rishi},
        title = "{O VI traces photoionized streams with collisionally ionized boundaries in cosmological simulations of z {\ensuremath{\sim}} 1 massive galaxies}",
      journal = {\mnras},
         year = 2021,
        month = mar,
       volume = {501},
       number = {4},
        pages = {4948-4967},
          doi = {10.1093/mnras/staa3972},
archivePrefix = {arXiv},
       eprint = {2008.11863},
 primaryClass = {astro-ph.GA},
       adsurl = {https://ui.adsabs.harvard.edu/abs/2021MNRAS.501.4948S}
}

@ARTICLE{Christensen_18,
       author = {{Christensen}, Charlotte R. and {Dav{\'e}}, Romeel and {Brooks}, Alyson and {Quinn}, Thomas and {Shen}, Sijing},
        title = "{Tracing Outflowing Metals in Simulations of Dwarf and Spiral Galaxies}",
      journal = {\apj},
         year = 2018,
        month = nov,
       volume = {867},
       number = {2},
          eid = {142},
        pages = {142},
          doi = {10.3847/1538-4357/aae374},
archivePrefix = {arXiv},
       eprint = {1808.07872},
 primaryClass = {astro-ph.GA},
       adsurl = {https://ui.adsabs.harvard.edu/abs/2018ApJ...867..142C}
}

@ARTICLE{Hafen_19,
       author = {{Hafen}, Zachary and {Faucher-Gigu{\`e}re}, Claude-Andr{\'e} and {Angl{\'e}s-Alc{\'a}zar}, Daniel and {Stern}, Jonathan and {Kere{\v{s}}}, Du{\v{s}}an and {Hummels}, Cameron and {Esmerian}, Clarke and {Garrison-Kimmel}, Shea and {El-Badry}, Kareem and {Wetzel}, Andrew and {Chan}, T.~K. and {Hopkins}, Philip F. and {Murray}, Norman},
        title = "{The origins of the circumgalactic medium in the FIRE simulations}",
      journal = {\mnras},
         year = 2019,
        month = sep,
       volume = {488},
       number = {1},
        pages = {1248-1272},
          doi = {10.1093/mnras/stz1773},
archivePrefix = {arXiv},
       eprint = {1811.11753},
 primaryClass = {astro-ph.GA},
       adsurl = {https://ui.adsabs.harvard.edu/abs/2019MNRAS.488.1248H}
}

@ARTICLE{Liang_16,
       author = {{Liang}, Cameron J. and {Kravtsov}, Andrey V. and {Agertz}, Oscar},
        title = "{Column density profiles of multiphase gaseous haloes}",
      journal = {\mnras},
         year = 2016,
        month = may,
       volume = {458},
       number = {2},
        pages = {1164-1187},
          doi = {10.1093/mnras/stw375},
archivePrefix = {arXiv},
       eprint = {1507.07002},
 primaryClass = {astro-ph.GA},
       adsurl = {https://ui.adsabs.harvard.edu/abs/2016MNRAS.458.1164L}
}

@ARTICLE{Rahmati_16,
       author = {{Rahmati}, Alireza and {Schaye}, Joop and {Crain}, Robert A. and {Oppenheimer}, Benjamin D. and {Schaller}, Matthieu and {Theuns}, Tom},
        title = "{Cosmic distribution of highly ionized metals and their physical conditions in the EAGLE simulations}",
      journal = {\mnras},
         year = 2016,
        month = jun,
       volume = {459},
       number = {1},
        pages = {310-332},
          doi = {10.1093/mnras/stw453},
archivePrefix = {arXiv},
       eprint = {1511.01094},
 primaryClass = {astro-ph.GA},
       adsurl = {https://ui.adsabs.harvard.edu/abs/2016MNRAS.459..310R}
}

@ARTICLE{Oppenheimer_16,
       author = {{Oppenheimer}, Benjamin D. and {Crain}, Robert A. and {Schaye}, Joop and {Rahmati}, Alireza and {Richings}, Alexander J. and {Trayford}, James W. and {Tumlinson}, Jason and {Bower}, Richard G. and {Schaller}, Matthieu and {Theuns}, Tom},
        title = "{Bimodality of low-redshift circumgalactic O VI in non-equilibrium EAGLE zoom simulations}",
      journal = {\mnras},
         year = 2016,
        month = aug,
       volume = {460},
       number = {2},
        pages = {2157-2179},
          doi = {10.1093/mnras/stw1066},
archivePrefix = {arXiv},
       eprint = {1603.05984},
 primaryClass = {astro-ph.GA},
       adsurl = {https://ui.adsabs.harvard.edu/abs/2016MNRAS.460.2157O}
}

@ARTICLE{Mina_21,
       author = {{Mina}, Mattia and {Shen}, Sijing and {Keller}, Benjamin Walter and {Mayer}, Lucio and {Madau}, Piero and {Wadsley}, James},
        title = "{The baryon cycle of Seven Dwarfs with superbubble feedback}",
      journal = {\aap},
         year = 2021,
        month = nov,
       volume = {655},
          eid = {A22},
        pages = {A22},
          doi = {10.1051/0004-6361/202039420},
archivePrefix = {arXiv},
       eprint = {2009.06646},
 primaryClass = {astro-ph.GA},
       adsurl = {https://ui.adsabs.harvard.edu/abs/2021A&A...655A..22M}
}

@ARTICLE{Ji_20,
       author = {{Ji}, Suoqing and {Chan}, T.~K. and {Hummels}, Cameron B. and {Hopkins}, Philip F. and {Stern}, Jonathan and {Kere{\v{s}}}, Du{\v{s}}an and {Quataert}, Eliot and {Faucher-Gigu{\`e}re}, Claude-Andr{\'e} and {Murray}, Norman},
        title = "{Properties of the circumgalactic medium in cosmic ray-dominated galaxy haloes}",
      journal = {\mnras},
         year = 2020,
        month = aug,
       volume = {496},
       number = {4},
        pages = {4221-4238},
          doi = {10.1093/mnras/staa1849},
archivePrefix = {arXiv},
       eprint = {1909.00003},
 primaryClass = {astro-ph.GA},
       adsurl = {https://ui.adsabs.harvard.edu/abs/2020MNRAS.496.4221J}
}

@ARTICLE{Ramesh_24,
       author = {{Ramesh}, Rahul and {Nelson}, Dylan and {Girichidis}, Philipp},
        title = "{IllustrisTNG + Cosmic Rays with a Simple Transport Model: From Dwarfs to L$^\star$ Galaxies}",
      journal = {arXiv e-prints},
         year = 2024,
        month = sep,
          eid = {arXiv:2409.18238},
        pages = {arXiv:2409.18238},
          doi = {10.48550/arXiv.2409.18238},
archivePrefix = {arXiv},
       eprint = {2409.18238},
 primaryClass = {astro-ph.GA},
       adsurl = {https://ui.adsabs.harvard.edu/abs/2024arXiv240918238R}
}

@ARTICLE{Cruz_25,
       author = {{Cruz}, Akaxia and {Brooks}, Alyson and {Lisanti}, Mariangela and {Peter}, Annika H.~G. and {Geda}, Robel and {Quinn}, Thomas and {Tremmel}, Michael and {Munshi}, Ferah and {Keller}, Ben and {Wadsley}, James},
        title = "{Dwarf diversity in $Λ$CDM with baryons}",
      journal = {arXiv e-prints},
         year = 2025,
        month = oct,
          eid = {arXiv:2510.11800},
        pages = {arXiv:2510.11800},
          doi = {10.48550/arXiv.2510.11800},
archivePrefix = {arXiv},
       eprint = {2510.11800},
 primaryClass = {astro-ph.GA},
       adsurl = {https://ui.adsabs.harvard.edu/abs/2025arXiv251011800C}
}

@ARTICLE{Riggs_24,
       author = {{Riggs}, Claire L. and {Brooks}, Alyson M. and {Munshi}, Ferah and {Christensen}, Charlotte R. and {Cohen}, Roger E. and {Quinn}, Thomas R. and {Wadsley}, James},
        title = "{Testable predictions of outside-in age gradients in dwarf galaxies of all types}",
      journal = {arXiv e-prints},
         year = 2024,
        month = aug,
          eid = {arXiv:2408.10379},
        pages = {arXiv:2408.10379},
          doi = {10.48550/arXiv.2408.10379},
archivePrefix = {arXiv},
       eprint = {2408.10379},
 primaryClass = {astro-ph.GA},
       adsurl = {https://ui.adsabs.harvard.edu/abs/2024arXiv240810379R}
}

@ARTICLE{Ma_16,
       author = {{Ma}, Xiangcheng and {Hopkins}, Philip F. and {Faucher-Gigu{\`e}re}, Claude-Andr{\'e} and {Zolman}, Nick and {Muratov}, Alexander L. and {Kere{\v{s}}}, Du{\v{s}}an and {Quataert}, Eliot},
        title = "{The origin and evolution of the galaxy mass-metallicity relation}",
      journal = {\mnras},
         year = 2016,
        month = feb,
       volume = {456},
       number = {2},
        pages = {2140-2156},
          doi = {10.1093/mnras/stv2659},
archivePrefix = {arXiv},
       eprint = {1504.02097},
 primaryClass = {astro-ph.GA},
       adsurl = {https://ui.adsabs.harvard.edu/abs/2016MNRAS.456.2140M}
}

@ARTICLE{Tremmel2017,
       author = {{Tremmel}, M. and {Karcher}, M. and {Governato}, F. and {Volonteri}, M. and {Quinn}, T.~R. and {Pontzen}, A. and {Anderson}, L. and {Bellovary}, J.},
        title = "{The Romulus cosmological simulations: a physical approach to the formation, dynamics and accretion models of SMBHs}",
      journal = {\mnras},
         year = 2017,
        month = sep,
       volume = {470},
       number = {1},
        pages = {1121-1139},
          doi = {10.1093/mnras/stx1160},
archivePrefix = {arXiv},
       eprint = {1607.02151},
 primaryClass = {astro-ph.GA},
       adsurl = {https://ui.adsabs.harvard.edu/abs/2017MNRAS.470.1121T}
}

@ARTICLE{BryanNorman_98,
       author = {{Bryan}, Greg L. and {Norman}, Michael L.},
        title = "{Statistical Properties of X-Ray Clusters: Analytic and Numerical Comparisons}",
      journal = {\apj},
         year = 1998,
        month = mar,
       volume = {495},
       number = {1},
        pages = {80-99},
          doi = {10.1086/305262},
archivePrefix = {arXiv},
       eprint = {astro-ph/9710107},
 primaryClass = {astro-ph},
       adsurl = {https://ui.adsabs.harvard.edu/abs/1998ApJ...495...80B}
}

@ARTICLE{Kravtsov_18,
       author = {{Kravtsov}, A.~V. and {Vikhlinin}, A.~A. and {Meshcheryakov}, A.~V.},
        title = "{Stellar Mass{\textemdash}Halo Mass Relation and Star Formation Efficiency in High-Mass Halos}",
      journal = {Astronomy Letters},
         year = 2018,
        month = jan,
       volume = {44},
       number = {1},
        pages = {8-34},
          doi = {10.1134/S1063773717120015},
archivePrefix = {arXiv},
       eprint = {1401.7329},
 primaryClass = {astro-ph.CO},
       adsurl = {https://ui.adsabs.harvard.edu/abs/2018AstL...44....8K}
}

@ARTICLE{Tremmel2015,
       author = {{Tremmel}, M. and {Governato}, F. and {Volonteri}, M. and {Quinn}, T.~R.},
        title = "{Off the beaten path: a new approach to realistically model the orbital decay of supermassive black holes in galaxy formation simulations}",
      journal = {\mnras},
         year = 2015,
        month = aug,
       volume = {451},
       number = {2},
        pages = {1868-1874},
          doi = {10.1093/mnras/stv1060},
archivePrefix = {arXiv},
       eprint = {1501.07609},
 primaryClass = {astro-ph.GA},
       adsurl = {https://ui.adsabs.harvard.edu/abs/2015MNRAS.451.1868T}
}

@ARTICLE{WoosleyWeaver_95,
       author = {{Woosley}, S.~E. and {Weaver}, Thomas A.},
        title = "{The Evolution and Explosion of Massive Stars. II. Explosive Hydrodynamics and Nucleosynthesis}",
      journal = {\apjs},
         year = 1995,
        month = nov,
       volume = {101},
        pages = {181},
          doi = {10.1086/192237},
       adsurl = {https://ui.adsabs.harvard.edu/abs/1995ApJS..101..181W}
}

@ARTICLE{Starburst99,
       author = {{Leitherer}, Claus and {Schaerer}, Daniel and {Goldader}, Jeffrey D. and {Delgado}, Rosa M. Gonz{\'a}lez and {Robert}, Carmelle and {Kune}, Denis Foo and {de Mello}, Du{\'\i}lia F. and {Devost}, Daniel and {Heckman}, Timothy M.},
        title = "{Starburst99: Synthesis Models for Galaxies with Active Star Formation}",
      journal = {\apjs},
         year = 1999,
        month = jul,
       volume = {123},
       number = {1},
        pages = {3-40},
          doi = {10.1086/313233},
archivePrefix = {arXiv},
       eprint = {astro-ph/9902334},
 primaryClass = {astro-ph},
       adsurl = {https://ui.adsabs.harvard.edu/abs/1999ApJS..123....3L}
}

@ARTICLE{Gandhi_22,
       author = {{Gandhi}, Pratik J. and {Wetzel}, Andrew and {Hopkins}, Philip F. and {Shappee}, Benjamin J. and {Wheeler}, Coral and {Faucher-Gigu{\`e}re}, Claude-Andr{\'e}},
        title = "{Exploring metallicity-dependent rates of Type Ia supernovae and their impact on galaxy formation}",
      journal = {\mnras},
         year = 2022,
        month = oct,
       volume = {516},
       number = {2},
        pages = {1941-1958},
          doi = {10.1093/mnras/stac2228},
archivePrefix = {arXiv},
       eprint = {2202.10477},
 primaryClass = {astro-ph.GA},
       adsurl = {https://ui.adsabs.harvard.edu/abs/2022MNRAS.516.1941G}
}

@ARTICLE{Nomoto_06,
       author = {{Nomoto}, Ken'ichi and {Tominaga}, Nozomu and {Umeda}, Hideyuki and {Kobayashi}, Chiaki and {Maeda}, Keiichi},
        title = "{Nucleosynthesis yields of core-collapse supernovae and hypernovae, and galactic chemical evolution}",
      journal = {\nphysa},
         year = 2006,
        month = oct,
       volume = {777},
        pages = {424-458},
          doi = {10.1016/j.nuclphysa.2006.05.008},
archivePrefix = {arXiv},
       eprint = {astro-ph/0605725},
 primaryClass = {astro-ph},
       adsurl = {https://ui.adsabs.harvard.edu/abs/2006NuPhA.777..424N}
}

@ARTICLE{Mannucci_06,
       author = {{Mannucci}, F. and {Della Valle}, M. and {Panagia}, N.},
        title = "{Two populations of progenitors for Type Ia supernovae?}",
      journal = {\mnras},
         year = 2006,
        month = aug,
       volume = {370},
       number = {2},
        pages = {773-783},
          doi = {10.1111/j.1365-2966.2006.10501.x},
archivePrefix = {arXiv},
       eprint = {astro-ph/0510315},
 primaryClass = {astro-ph},
       adsurl = {https://ui.adsabs.harvard.edu/abs/2006MNRAS.370..773M}
}

@ARTICLE{Iwamoto_99,
       author = {{Iwamoto}, Koichi and {Brachwitz}, Franziska and {Nomoto}, Ken'ICHI and {Kishimoto}, Nobuhiro and {Umeda}, Hideyuki and {Hix}, W. Raphael and {Thielemann}, Friedrich-Karl},
        title = "{Nucleosynthesis in Chandrasekhar Mass Models for Type IA Supernovae and Constraints on Progenitor Systems and Burning-Front Propagation}",
      journal = {\apjs},
         year = 1999,
        month = dec,
       volume = {125},
       number = {2},
        pages = {439-462},
          doi = {10.1086/313278},
archivePrefix = {arXiv},
       eprint = {astro-ph/0002337},
 primaryClass = {astro-ph},
       adsurl = {https://ui.adsabs.harvard.edu/abs/1999ApJS..125..439I}
}

@ARTICLE{Hoek_97,
       author = {{van den Hoek}, L.~B. and {Groenewegen}, M.~A.~T.},
        title = "{New theoretical yields of intermediate mass stars}",
      journal = {\aaps},
         year = 1997,
        month = jun,
       volume = {123},
        pages = {305-328},
          doi = {10.1051/aas:1997162},
       adsurl = {https://ui.adsabs.harvard.edu/abs/1997A&AS..123..305V}
}

@ARTICLE{Marigo_01,
       author = {{Marigo}, P.},
        title = "{Chemical yields from low- and intermediate-mass stars: Model predictions and basic observational constraints}",
      journal = {\aap},
         year = 2001,
        month = apr,
       volume = {370},
        pages = {194-217},
          doi = {10.1051/0004-6361:20000247},
archivePrefix = {arXiv},
       eprint = {astro-ph/0012181},
 primaryClass = {astro-ph},
       adsurl = {https://ui.adsabs.harvard.edu/abs/2001A&A...370..194M}
}

@ARTICLE{Izzard_04,
       author = {{Izzard}, Robert G. and {Tout}, Christopher A. and {Karakas}, Amanda I. and {Pols}, Onno R.},
        title = "{A new synthetic model for asymptotic giant branch stars}",
      journal = {\mnras},
         year = 2004,
        month = may,
       volume = {350},
       number = {2},
        pages = {407-426},
          doi = {10.1111/j.1365-2966.2004.07446.x},
archivePrefix = {arXiv},
       eprint = {astro-ph/0402403},
 primaryClass = {astro-ph},
       adsurl = {https://ui.adsabs.harvard.edu/abs/2004MNRAS.350..407I}
}

@ARTICLE{Cloudy_Ferland17,
   author = {{Ferland}, G.~J. and {Chatzikos}, M. and {Guzm{\'a}n}, F. and {Lykins}, M.~L. and {van Hoof}, P.~A.~M. and {Williams}, R.~J.~R. and {Abel}, N.~P. and {Badnell}, N.~R. and {Keenan}, F.~P. and {Porter}, R.~L. and {Stancil}, P.~C.},
    title = "{The 2017 Release Cloudy}",
  journal = {\rmxaa},
archivePrefix = "arXiv",
   eprint = {1705.10877},
     year = 2017,
    month = oct,
   volume = 53,
    pages = {385-438},
   adsurl = {http://adsabs.harvard.edu/abs/2017RMxAA..53..385F}}

@ARTICLE{Taira_25,
       author = {{Taira}, Elias and {Kopenhafer}, Claire and {O'Shea}, Brian W. and {Manning}, Alexis and {Fuhrman}, Evelyn and {Peeples}, Molly S. and {Tumlinson}, Jason and {Smith}, Britton D.},
        title = "{Impacts of the Metagalactic Ultraviolet Background on Circumgalactic Medium Absorption Systems}",
      journal = {\apj},
         year = 2025,
        month = oct,
       volume = {991},
       number = {2},
          eid = {221},
        pages = {221},
          doi = {10.3847/1538-4357/adfc4e},
archivePrefix = {arXiv},
       eprint = {2503.11775},
 primaryClass = {astro-ph.GA},
       adsurl = {https://ui.adsabs.harvard.edu/abs/2025ApJ...991..221T}
}

@ARTICLE{Puchwein19,
       author = {{Puchwein}, Ewald and {Haardt}, Francesco and {Haehnelt}, Martin G. and {Madau}, Piero},
        title = "{Consistent modelling of the meta-galactic UV background and the thermal/ionization history of the intergalactic medium}",
      journal = {\mnras},
         year = 2019,
        month = may,
       volume = {485},
       number = {1},
        pages = {47-68},
          doi = {10.1093/mnras/stz222},
archivePrefix = {arXiv},
       eprint = {1801.04931},
 primaryClass = {astro-ph.GA},
       adsurl = {https://ui.adsabs.harvard.edu/abs/2019MNRAS.485...47P}
}

@ARTICLE{Dekel_Silk86,
       author = {{Dekel}, A. and {Silk}, J.},
        title = "{The Origin of Dwarf Galaxies, Cold Dark Matter, and Biased Galaxy Formation}",
      journal = {\apj},
         year = 1986,
        month = apr,
       volume = {303},
        pages = {39},
          doi = {10.1086/164050},
       adsurl = {https://ui.adsabs.harvard.edu/abs/1986ApJ...303...39D}
}

@ARTICLE{Somerville_Dave15,
       author = {{Somerville}, Rachel S. and {Dav{\'e}}, Romeel},
        title = "{Physical Models of Galaxy Formation in a Cosmological Framework}",
      journal = {\araa},
         year = 2015,
        month = aug,
       volume = {53},
        pages = {51-113},
          doi = {10.1146/annurev-astro-082812-140951},
archivePrefix = {arXiv},
       eprint = {1412.2712},
 primaryClass = {astro-ph.GA},
       adsurl = {https://ui.adsabs.harvard.edu/abs/2015ARA&A..53...51S}
}

@ARTICLE{TumCGMreview,
       author = {{Tumlinson}, Jason and {Peeples}, Molly S. and {Werk}, Jessica K.},
        title = "{The Circumgalactic Medium}",
      journal = {\araa},
         year = 2017,
        month = aug,
       volume = {55},
       number = {1},
        pages = {389-432},
          doi = {10.1146/annurev-astro-091916-055240},
archivePrefix = {arXiv},
       eprint = {1709.09180},
 primaryClass = {astro-ph.GA},
       adsurl = {https://ui.adsabs.harvard.edu/abs/2017ARA&A..55..389T}
}

@ARTICLE{FaucherGigure_Oh23,
       author = {{Faucher-Gigu{\`e}re}, Claude-Andr{\'e} and {Oh}, S. Peng},
        title = "{Key Physical Processes in the Circumgalactic Medium}",
      journal = {\araa},
         year = 2023,
        month = aug,
       volume = {61},
        pages = {131-195},
          doi = {10.1146/annurev-astro-052920-125203},
archivePrefix = {arXiv},
       eprint = {2301.10253},
 primaryClass = {astro-ph.GA},
       adsurl = {https://ui.adsabs.harvard.edu/abs/2023ARA&A..61..131F}
}

@ARTICLE{Rey_25,
       author = {{Rey}, Maxime and {Blaizot}, J{\'e}r{\'e}my and {Kimm}, Taysun and {Rosdahl}, Joakim and {Michel-Dansac}, L{\'e}o},
        title = "{ARCHITECTS I: impact of subgrid physics on the simulated properties of the circumgalactic medium}",
      journal = {\mnras},
         year = 2025,
        month = oct,
       volume = {543},
       number = {1},
        pages = {12-27},
          doi = {10.1093/mnras/staf1372},
archivePrefix = {arXiv},
       eprint = {2602.13392},
 primaryClass = {astro-ph.GA},
       adsurl = {https://ui.adsabs.harvard.edu/abs/2025MNRAS.543...12R}
}

@ARTICLE{2026arXiv260213394R,
       author = {{Rey}, Maxime and {Blaizot}, J{\'e}r{\'e}my and {Kimm}, Taysun and {Rosdahl}, Joakim and {Michel-Dansac}, L{\'e}o and {Mauerhofer}, Valentin},
        title = "{ARCHITECTS II: Impact of subgrid physics on the observable properties of the circumgalactic medium}",
      journal = {arXiv e-prints},
         year = 2026,
        month = feb,
          eid = {arXiv:2602.13394},
        pages = {arXiv:2602.13394},
          doi = {10.48550/arXiv.2602.13394},
archivePrefix = {arXiv},
       eprint = {2602.13394},
 primaryClass = {astro-ph.GA},
       adsurl = {https://ui.adsabs.harvard.edu/abs/2026arXiv260213394R}
}

@ARTICLE{Smith_17,
       author = {{Smith}, Britton D. and {Bryan}, Greg L. and {Glover}, Simon C.~O. and {Goldbaum}, Nathan J. and {Turk}, Matthew J. and {Regan}, John and {Wise}, John H. and {Schive}, Hsi-Yu and {Abel}, Tom and {Emerick}, Andrew and {O'Shea}, Brian W. and {Anninos}, Peter and {Hummels}, Cameron B. and {Khochfar}, Sadegh},
        title = "{GRACKLE: a chemistry and cooling library for astrophysics}",
      journal = {\mnras},
         year = 2017,
        month = apr,
       volume = {466},
       number = {2},
        pages = {2217-2234},
          doi = {10.1093/mnras/stw3291},
archivePrefix = {arXiv},
       eprint = {1610.09591},
 primaryClass = {astro-ph.CO},
       adsurl = {https://ui.adsabs.harvard.edu/abs/2017MNRAS.466.2217S}
}

@ARTICLE{Bieri_26,
       author = {{Bieri}, Rebekka and {Pakmor}, R{\"u}diger and {van de Voort}, Freeke and {Talbot}, Rosie Y. and {Werhahn}, Maria and {Pfrommer}, Christoph and {Springel}, Volker},
        title = "{Unveiling the impact of cosmic rays on the disc sizes and outflows from dwarf scales to galaxy groups}",
      journal = {\mnras},
         year = 2026,
        month = apr,
       volume = {547},
       number = {2},
          eid = {stag216},
        pages = {stag216},
          doi = {10.1093/mnras/stag216},
archivePrefix = {arXiv},
       eprint = {2509.07124},
 primaryClass = {astro-ph.GA},
       adsurl = {https://ui.adsabs.harvard.edu/abs/2026MNRAS.547ag216B}
}

@ARTICLE{Iyer_20,
       author = {{Iyer}, Kartheik G. and {Tacchella}, Sandro and {Genel}, Shy and {Hayward}, Christopher C. and {Hernquist}, Lars and {Brooks}, Alyson M. and {Caplar}, Neven and {Dav{\'e}}, Romeel and {Diemer}, Benedikt and {Forbes}, John C. and {Gawiser}, Eric and {Somerville}, Rachel S. and {Starkenburg}, Tjitske K.},
        title = "{The diversity and variability of star formation histories in models of galaxy evolution}",
      journal = {\mnras},
         year = 2020,
        month = oct,
       volume = {498},
       number = {1},
        pages = {430-463},
          doi = {10.1093/mnras/staa2150},
archivePrefix = {arXiv},
       eprint = {2007.07916},
 primaryClass = {astro-ph.GA},
       adsurl = {https://ui.adsabs.harvard.edu/abs/2020MNRAS.498..430I}
}

@ARTICLE{Tremonti_04,
       author = {{Tremonti}, Christy A. and {Heckman}, Timothy M. and {Kauffmann}, Guinevere and {Brinchmann}, Jarle and {Charlot}, St{\'e}phane and {White}, Simon D.~M. and {Seibert}, Mark and {Peng}, Eric W. and {Schlegel}, David J. and {Uomoto}, Alan and {Fukugita}, Masataka and {Brinkmann}, Jon},
        title = "{The Origin of the Mass-Metallicity Relation: Insights from 53,000 Star-forming Galaxies in the Sloan Digital Sky Survey}",
      journal = {\apj},
         year = 2004,
        month = oct,
       volume = {613},
       number = {2},
        pages = {898-913},
          doi = {10.1086/423264},
archivePrefix = {arXiv},
       eprint = {astro-ph/0405537},
 primaryClass = {astro-ph},
       adsurl = {https://ui.adsabs.harvard.edu/abs/2004ApJ...613..898T}
}

@ARTICLE{2026arXiv260123264E,
       author = {{Engelhardt}, Anna and {Munshi}, Ferah and {Peter}, Annika H.~G. and {Nadler}, Ethan O. and {Cruz}, Akaxia and {Brooks}, Alyson M. and {Carton Zeng}, Zhichao and {Quinn}, Thomas R. and {Keith}, Blake},
        title = "{MARVELously Dark: the gravothermal evolution of dwarf halos in velocity-dependent SIDM}",
      journal = {arXiv e-prints},
         year = 2026,
        month = jan,
          eid = {arXiv:2601.23264},
        pages = {arXiv:2601.23264},
          doi = {10.48550/arXiv.2601.23264},
archivePrefix = {arXiv},
       eprint = {2601.23264},
 primaryClass = {astro-ph.GA},
       adsurl = {https://ui.adsabs.harvard.edu/abs/2026arXiv260123264E}
}

@ARTICLE{2025arXiv250819396B,
       author = {{Baumschlager}, Bernhard and {Shen}, Sijing and {Wadsley}, James W. and {Keller}, Benjamin and {Wissing}, Robert and {Mayer}, Lucio and {Madau}, Piero and {Munshi}, Ferah and {Brooks}, Alyson},
        title = "{The Seven Dwarfs illuminated. The impact of radiation on dwarf galaxies and their circumgalactic medium}",
      journal = {arXiv e-prints},
         year = 2025,
        month = aug,
          eid = {arXiv:2508.19396},
        pages = {arXiv:2508.19396},
          doi = {10.48550/arXiv.2508.19396},
archivePrefix = {arXiv},
       eprint = {2508.19396},
 primaryClass = {astro-ph.GA},
       adsurl = {https://ui.adsabs.harvard.edu/abs/2025arXiv250819396B}
}

@ARTICLE{Nahar_99,
       author = {{Nahar}, Sultana N.},
        title = "{Electron-Ion Recombination Rate Coefficients, Photoionization Cross Sections, and Ionization Fractions for Astrophysically Abundant Elements. II. Oxygen Ions}",
      journal = {\apjs},
         year = 1999,
        month = jan,
       volume = {120},
       number = {1},
        pages = {131-145},
          doi = {10.1086/313173},
       adsurl = {https://ui.adsabs.harvard.edu/abs/1999ApJS..120..131N}
}

@ARTICLE{2025arXiv251026875G,
       author = {{Geda}, Robel and {Cruz}, Akaxia and {Wright}, Anna C. and {Greene}, Jenny E. and {Brooks}, Alyson and {Quinn}, Thomas and {Wadsley}, James and {Keller}, Ben},
        title = "{Disk Formation and the Size-sSFR Relation of Dwarf Galaxies}",
      journal = {arXiv e-prints},
         year = 2025,
        month = oct,
          eid = {arXiv:2510.26875},
        pages = {arXiv:2510.26875},
          doi = {10.48550/arXiv.2510.26875},
archivePrefix = {arXiv},
       eprint = {2510.26875},
 primaryClass = {astro-ph.GA},
       adsurl = {https://ui.adsabs.harvard.edu/abs/2025arXiv251026875G}
}

@ARTICLE{Hopkins20,
       author = {{Hopkins}, Philip F. and {Chan}, T.~K. and {Garrison-Kimmel}, Shea and {Ji}, Suoqing and {Su}, Kung-Yi and {Hummels}, Cameron B. and {Kere{\v{s}}}, Du{\v{s}}an and {Quataert}, Eliot and {Faucher-Gigu{\`e}re}, Claude-Andr{\'e}},
        title = "{But what about...: cosmic rays, magnetic fields, conduction, and viscosity in galaxy formation}",
      journal = {\mnras},
         year = 2020,
        month = mar,
       volume = {492},
       number = {3},
        pages = {3465-3498},
          doi = {10.1093/mnras/stz3321},
archivePrefix = {arXiv},
       eprint = {1905.04321},
 primaryClass = {astro-ph.GA},
       adsurl = {https://ui.adsabs.harvard.edu/abs/2020MNRAS.492.3465H}
}

@ARTICLE{Buck_20,
       author = {{Buck}, Tobias and {Pfrommer}, Christoph and {Pakmor}, R{\"u}diger and {Grand}, Robert J.~J. and {Springel}, Volker},
        title = "{The effects of cosmic rays on the formation of Milky Way-mass galaxies in a cosmological context}",
      journal = {\mnras},
         year = 2020,
        month = sep,
       volume = {497},
       number = {2},
        pages = {1712-1737},
          doi = {10.1093/mnras/staa1960},
archivePrefix = {arXiv},
       eprint = {1911.00019},
 primaryClass = {astro-ph.GA},
       adsurl = {https://ui.adsabs.harvard.edu/abs/2020MNRAS.497.1712B}
}

@ARTICLE{Corbelli_14,
       author = {{Corbelli}, Edvige and {Thilker}, David and {Zibetti}, Stefano and {Giovanardi}, Carlo and {Salucci}, Paolo},
        title = "{Dynamical signatures of a {\ensuremath{\Lambda}}CDM-halo and the distribution of the baryons in M 33}",
      journal = {\aap},
         year = 2014,
        month = dec,
       volume = {572},
          eid = {A23},
        pages = {A23},
          doi = {10.1051/0004-6361/201424033},
archivePrefix = {arXiv},
       eprint = {1409.2665},
 primaryClass = {astro-ph.GA},
       adsurl = {https://ui.adsabs.harvard.edu/abs/2014A&A...572A..23C}
}

@ARTICLE{Li_23,
       author = {{Li}, Jiadong and {Liu}, Chao and {Zhang}, Zhi-Yu and {Tian}, Hao and {Fu}, Xiaoting and {Li}, Jiao and {Yan}, Zhi-Qiang},
        title = "{Stellar initial mass function varies with metallicity and time}",
      journal = {\nat},
         year = 2023,
        month = jan,
       volume = {613},
       number = {7944},
        pages = {460-462},
          doi = {10.1038/s41586-022-05488-1},
archivePrefix = {arXiv},
       eprint = {2301.07029},
 primaryClass = {astro-ph.GA},
       adsurl = {https://ui.adsabs.harvard.edu/abs/2023Natur.613..460L}
}

@ARTICLE{Geha13,
       author = {{Geha}, Marla and {Brown}, Thomas M. and {Tumlinson}, Jason and {Kalirai}, Jason S. and {Simon}, Joshua D. and {Kirby}, Evan N. and {VandenBerg}, Don A. and {Mu{\~n}oz}, Ricardo R. and {Avila}, Roberto J. and {Guhathakurta}, Puragra and {Ferguson}, Henry C.},
        title = "{The Stellar Initial Mass Function of Ultra-faint Dwarf Galaxies: Evidence for IMF Variations with Galactic Environment}",
      journal = {\apj},
         year = 2013,
        month = jul,
       volume = {771},
       number = {1},
          eid = {29},
        pages = {29},
          doi = {10.1088/0004-637X/771/1/29},
archivePrefix = {arXiv},
       eprint = {1304.7769},
 primaryClass = {astro-ph.CO},
       adsurl = {https://ui.adsabs.harvard.edu/abs/2013ApJ...771...29G}
}

@ARTICLE{Baumschlager_24,
       author = {{Baumschlager}, Bernhard and {Shen}, Sijing and {Wadsley}, James W.},
        title = "{Spectral reconstruction for radiation hydrodynamic simulations of galaxy evolution}",
      journal = {\aap},
         year = 2024,
        month = nov,
       volume = {691},
          eid = {A219},
        pages = {A219},
          doi = {10.1051/0004-6361/202348164},
archivePrefix = {arXiv},
       eprint = {2310.16902},
 primaryClass = {astro-ph.GA},
       adsurl = {https://ui.adsabs.harvard.edu/abs/2024A&A...691A.219B}
}

@ARTICLE{Garza_24,
       author = {{Garza}, Samantha L. and {Werk}, Jessica K. and {Oppenheimer}, Benjamin D. and {Tchernyshyov}, Kirill and {Sanchez}, N. Nicole and {Faerman}, Yakov and {Rubin}, Kate H.~R. and {Bentz}, Misty C. and {Davies}, Jonathan J. and {Burchett}, Joseph N. and {Crain}, Robert A. and {Prochaska}, J. Xavier},
        title = "{The COS-Holes Survey: Connecting Galaxy Black Hole Mass with the State of the CGM}",
      journal = {\apj},
         year = 2024,
        month = aug,
       volume = {970},
       number = {2},
          eid = {115},
        pages = {115},
          doi = {10.3847/1538-4357/ad4ecc},
archivePrefix = {arXiv},
       eprint = {2405.20316},
 primaryClass = {astro-ph.GA},
       adsurl = {https://ui.adsabs.harvard.edu/abs/2024ApJ...970..115G}
}

@ARTICLE{Hopkins_18SNe, 
       author = {{Hopkins}, Philip F. and {Wetzel}, Andrew and {Kere{\v{s}}}, Du{\v{s}}an and {Faucher-Gigu{\`e}re}, Claude-Andr{\'e} and {Quataert}, Eliot and {Boylan-Kolchin}, Michael and {Murray}, Norman and {Hayward}, Christopher C. and {El-Badry}, Kareem},
        title = "{How to model supernovae in simulations of star and galaxy formation}",
      journal = {\mnras},
         year = 2018,
        month = jun,
       volume = {477},
       number = {2},
        pages = {1578-1603},
          doi = {10.1093/mnras/sty674},
archivePrefix = {arXiv},
       eprint = {1707.07010},
 primaryClass = {astro-ph.GA},
       adsurl = {https://ui.adsabs.harvard.edu/abs/2018MNRAS.477.1578H}
}

@ARTICLE{Sanchez_24,
       author = {{Sanchez}, N. Nicole and {Werk}, Jessica K. and {Christensen}, Charlotte and {Telford}, O. Grace and {Quinn}, Thomas R. and {Tremmel}, Michael and {Mead}, Jennifer and {Sharma}, Ray S. and {Brooks}, Alyson M.},
        title = "{The Scatter Matters: Circumgalactic Metal Content in the Context of the M─{\ensuremath{\sigma}} Relation}",
      journal = {\apj},
         year = 2024,
        month = jun,
       volume = {967},
       number = {2},
          eid = {100},
        pages = {100},
          doi = {10.3847/1538-4357/ad39eb},
archivePrefix = {arXiv},
       eprint = {2305.07672},
 primaryClass = {astro-ph.GA},
       adsurl = {https://ui.adsabs.harvard.edu/abs/2024ApJ...967..100S}
}

@ARTICLE{Hummels19,
       author = {{Hummels}, Cameron B. and {Smith}, Britton D. and {Hopkins}, Philip F. and {O'Shea}, Brian W. and {Silvia}, Devin W. and {Werk}, Jessica K. and {Lehner}, Nicolas and {Wise}, John H. and {Collins}, David C. and {Butsky}, Iryna S.},
        title = "{The Impact of Enhanced Halo Resolution on the Simulated Circumgalactic Medium}",
      journal = {\apj},
         year = 2019,
        month = sep,
       volume = {882},
       number = {2},
          eid = {156},
        pages = {156},
          doi = {10.3847/1538-4357/ab378f},
archivePrefix = {arXiv},
       eprint = {1811.12410},
 primaryClass = {astro-ph.GA},
       adsurl = {https://ui.adsabs.harvard.edu/abs/2019ApJ...882..156H}
}

@ARTICLE{Lucchini26,
       author = {{Lucchini}, Scott and {Abramson}, Cecilia and {Hummels}, Cameron and {Conroy}, Charlie and {Hernquist}, Lars and {Smith}, Aaron},
        title = "{ENhanced Galactic Atmospheres With Arepo: Resolving the CGM at 200 pc with the ENGAWA Simulations}",
      journal = {arXiv e-prints},
         year = 2026,
        month = mar,
          eid = {arXiv:2603.05584},
        pages = {arXiv:2603.05584},
          doi = {10.48550/arXiv.2603.05584},
archivePrefix = {arXiv},
       eprint = {2603.05584},
 primaryClass = {astro-ph.GA},
       adsurl = {https://ui.adsabs.harvard.edu/abs/2026arXiv260305584L}
}

@ARTICLE{Hummels_13,
       author = {{Hummels}, Cameron B. and {Bryan}, Greg L. and {Smith}, Britton D. and {Turk}, Matthew J.},
        title = "{Constraints on hydrodynamical subgrid models from quasar absorption line studies of the simulated circumgalactic medium}",
      journal = {\mnras},
         year = 2013,
        month = apr,
       volume = {430},
       number = {3},
        pages = {1548-1565},
          doi = {10.1093/mnras/sts702},
archivePrefix = {arXiv},
       eprint = {1212.2965},
 primaryClass = {astro-ph.GA},
       adsurl = {https://ui.adsabs.harvard.edu/abs/2013MNRAS.430.1548H}
}

@ARTICLE{Benincasa_16,
       author = {{Benincasa}, S.~M. and {Wadsley}, J. and {Couchman}, H.~M.~P. and {Keller}, B.~W.},
        title = "{The anatomy of a star-forming galaxy: pressure-driven regulation of star formation in simulated galaxies}",
      journal = {\mnras},
         year = 2016,
        month = nov,
       volume = {462},
       number = {3},
        pages = {3053-3068},
          doi = {10.1093/mnras/stw1741},
archivePrefix = {arXiv},
       eprint = {1607.05795},
 primaryClass = {astro-ph.GA},
       adsurl = {https://ui.adsabs.harvard.edu/abs/2016MNRAS.462.3053B}
}

@ARTICLE{Christensen_14,
       author = {{Christensen}, C.~R. and {Governato}, F. and {Quinn}, T. and {Brooks}, A.~M. and {Shen}, S. and {McCleary}, J. and {Fisher}, D.~B. and {Wadsley}, J.},
        title = "{The effect of models of the interstellar media on the central mass distribution of galaxies}",
      journal = {\mnras},
         year = 2014,
        month = may,
       volume = {440},
       number = {3},
        pages = {2843-2859},
          doi = {10.1093/mnras/stu399},
archivePrefix = {arXiv},
       eprint = {1211.0326},
 primaryClass = {astro-ph.CO},
       adsurl = {https://ui.adsabs.harvard.edu/abs/2014MNRAS.440.2843C}
}

@ARTICLE{Hopkins_14_SF,
       author = {{Hopkins}, Philip F. and {Kere{\v{s}}}, Du{\v{s}}an and {O{\~n}orbe}, Jos{\'e} and {Faucher-Gigu{\`e}re}, Claude-Andr{\'e} and {Quataert}, Eliot and {Murray}, Norman and {Bullock}, James S.},
        title = "{Galaxies on FIRE (Feedback In Realistic Environments): stellar feedback explains cosmologically inefficient star formation}",
      journal = {\mnras},
         year = 2014,
        month = nov,
       volume = {445},
       number = {1},
        pages = {581-603},
          doi = {10.1093/mnras/stu1738},
archivePrefix = {arXiv},
       eprint = {1311.2073},
 primaryClass = {astro-ph.CO},
       adsurl = {https://ui.adsabs.harvard.edu/abs/2014MNRAS.445..581H}
}
\bibliographystyle{aasjournalv7}

\appendix
\section{Stellar Mass Dependence for Mishra+24 Observations}\label{apx:mstardep}

In Figure \ref{fig:CompObs_Mass}, we scale the observationally derived mass of \OVI\, in \citetalias{Mishra_24} with stellar mass. The dependence on stellar mass is due to the dependence on $R_{vir}$ for the mass of \OVI\, ($M_{\rm OVI}$) following Equation \ref{eqn:m24movi}. 
The calculation for mass was performed for two annuli: $R_{\rm inner}=0, ~R_{\rm outer}=R_{vir}$ and  $R_{\rm inner}=R_{vir}, ~R_{\rm outer}=2R_{vir}$. 
\citetalias{Mishra_24} estimates a galaxy's $R_{vir}$ by estimating a halo mass ($M_{vir}$) from the observed \Mstar, assuming a stellar mass to halo mass relation from \citet{Kravtsov_18}, and finding the virial radius for a given halo mass based on the relation defined in \citet{BryanNorman_98}. From this, we determine a $R_{vir}\propto M_*$ relation and ultimately the $M_{\rm OVI}^{M24} (M_*)$ scaling following,

\begin{equation}
    \begin{split}
    M_* \propto M_{vir}^{\beta} \\
    R_{vir} \propto M_{vir}^{1/3} \rightarrow  R_{vir} \propto M_{*}^{\frac{1}{3\beta}} \\
    \end{split}
\end{equation}

\begin{equation}
    M_{\rm OVI}^{M24} (M_*) = M_{\rm OVI}^{M24} \times\biggl(\frac{M_*}{10^{8.3}M_{\odot}} \biggr)^{2(\frac{1}{3\beta})}
\end{equation}
The factor of 2 in the exponent comes from Equation \ref{eqn:m24movi}, where mass depends on $R^2$.  We adopt $\beta=1.8$ based on \citet{Kravtsov_18}.

\section{Uniformly Increasing CGM OVI Reservoirs}\label{apx:IncreasedNOVI}

\begin{figure*}
    \centering
    \includegraphics[width=1\linewidth]{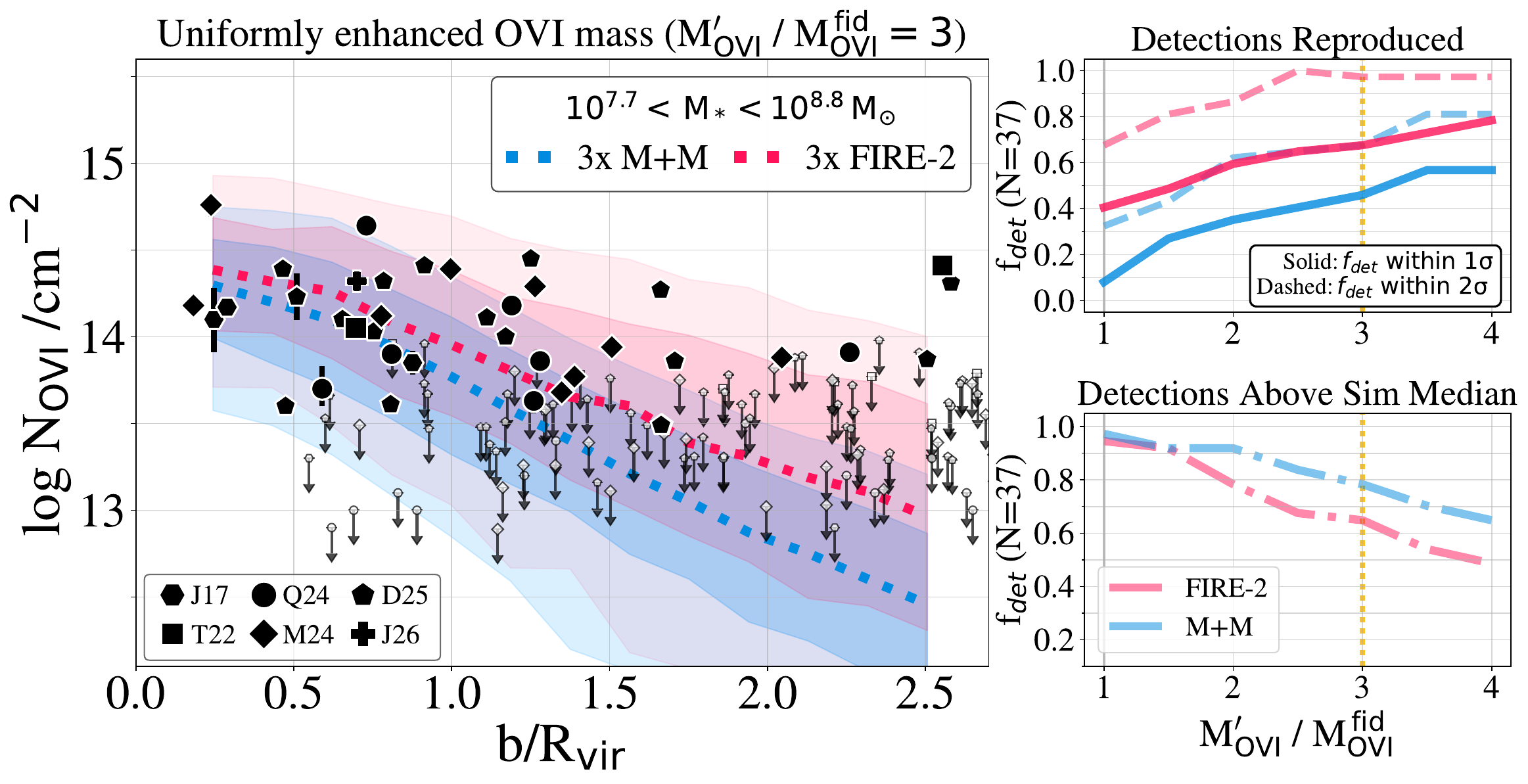}
    \caption{\OVI\, column densities (\NOVI) for $10^{7.7}< M_*<10^{8.8} M_{\odot}$ galaxies (\textit{left panel}). Simulated values represent the resulting column densities after increasing the total simulated \OVI\, mass in the CGM by a factor of 3. Median \NOVI\, values and 16th-84th percentile ranges are shown as dotted lines and shaded bands, and are calculated within $0.2 b/R_{vir}$ wide bins. Observational data from \citet{Johnson_17} (J17), \citet{Tchernyshyov_22} (T22), \citet{Qu_24} (Q24), \citet{Mishra_24} (M24), \citet{Dutta_25_col} (D25), \citet{Johnson26} (J26) are shown as hexagons, squares, circles, diamonds, pentagons, and crosses, respectively. Observationally derived detections of \OVI\, are shown as solid markers, while non-detections are shown as open-faced markers with a downward-pointing arrow. \textit{Top right panel:} Fraction of observed detections that lie within $1\sigma$ (solid line) and $2\sigma$ of the simulated medians (dashed line) as a function of the uniform total \MOVI\, enhancement, such that $M^{\prime}_{\rm OVI}/M^{\rm fid}_{\rm OVI}=1$ represents the fiducial \NOVI\, distributions. \textit{Bottom right panel:} Fraction of observed detections that lie above the simulated medians. A vertical yellow dotted line is drawn to represent the uniform enhancement shown in the left panel.}
    \label{fig:increasedNOVI}
\end{figure*}

In this Appendix, we briefly explore the agreement current simulations can achieve with column density observations, under a uniform increase in \MOVI. The column density of a single line of sight is simply

\begin{equation}
    N_{OVI} = \int n_{OVI} ~dl = \int \frac{\rho_{OVI}}{m_{OVI}} ~dl, 
\end{equation}

where $n_{OVI}$ and $\rho_{OVI}$ is the number and mass density of \OVI, respectively, and $m_{OVI}$ is the mass of a single \OVI\, nucleus. If the \OVI\, mass of the CGM is uniformly increased everywhere, this translates to an equal increase in a given sightline's \NOVI. 

The left panel Figure \ref{fig:increasedNOVI} shows the resulting simulated \NOVI\, distribution after this total enhancement in \MOVI\, by a factor of $3$. Compared to Figure \ref{fig:CompObs_NOVI}, Figure \ref{fig:increasedNOVI} simply shows the entire distribution increased by a factor of 3, preserving the scatter and slope that the \NOVI\, distribution of each simulation exhibits. This increase in \NOVI\, results in roughly $70\%$ ($100\%$) and $50\%$ ($70\%$) of observed detections lying within $1\sigma$ ($2\sigma$) for FIRE-2 and \MM, respectively (shown as the solid lines in the upper right panel in  Figure \ref{fig:increasedNOVI}).  Although a $2\times$ increase in \NOVI\, yields relatively similar fits in terms of the fraction of detections within $1-2\sigma$, this increase still results in $80-90\%$ of detections lying above the simulated medians (shown as the dash-dotted line in the bottom right panel in  Figure \ref{fig:increasedNOVI}). We consider a $2\times$ increase to be still underpredicted since the vast majority of detections are systematically higher than the medians, while a $3\times$ increase results in fewer detections lying above the median ($60-80\%$). 

We emphasize that the decision between a $2\times$ versus $3\times$ increase is not motivated by robust statistics. This section is not aimed at discerning the most statistically robust fit to the observations, nor does it consider the large number of non-detections in this comparison. Rather, this section is meant to demonstrate the resulting \NOVI\, given an idealized and uniform increase in \MOVI\, to the degree discussed in the text. An increase to this degree would provide stronger agreement with observations, in particular for FIRE-2, which has a shallower slope, allowing the suite to reproduce the column densities at large impact parameters. 

As shown in Section \ref{subsubsec:phasevary}, modifications to the \OVI\, ionization efficiency cannot achieve a $3\times$ increase in \MOVI, thus more oxygen mass is likely required in the CGM. Yet, increasing the CGM oxygen mass would, at the very least, require modified feedback that transports the oxygen from the ISM to the CGM, if not additionally increased oxygen production. Modifications to feedback may likely change the radial distribution of oxygen mass and therefore the slope of \NOVI\, profiles. As discussed in Section \ref{subsec:quantgap}, enhanced metallicity in the CGM may also affect inflow and outflows. Therefore, the assumption that an increase in \MOVI\, to this magnitude will necessarily preserve the same radial distribution is highly idealized. Nonetheless, this section provides an approximate benchmark for how well \NOVI\, values will agree with observations given an increase in \MOVI\, as suggested by Figure \ref{fig:CompObs_MassRatios}.

\end{document}